\documentclass[ba]{imsart}

\RequirePackage{amsthm,amsmath,amsfonts,amssymb}
\RequirePackage[numbers]{natbib}
\RequirePackage[colorlinks,citecolor=blue,urlcolor=blue,backref=page,backref=page]{hyperref}
\RequirePackage{graphicx}

\usepackage{amsmath}
\usepackage{graphicx}%
\usepackage{amsfonts}%
\usepackage{amssymb}
\usepackage{overpic}
\usepackage[noend]{algpseudocode}
\usepackage{subcaption}
\usepackage{rotating}
\usepackage{epstopdf}
\usepackage{url}
\usepackage{xr}
\usepackage{caption}
\usepackage{color}
\usepackage{comment}
\usepackage{enumerate}
\usepackage{mathtools}
\usepackage{multirow}
\usepackage{pgfplots}
\usepackage{natbib}
\usepackage{pst-solides3d}
\usepackage{bbm}
\usepackage{float}
\usepackage{wrapfig}
\usepackage{bm}
\usepackage[plain,noend]{algorithm2e}

\pubyear{2024}
\volume{TBA}
\issue{TBA}
\firstpage{1}
\lastpage{1}

\startlocaldefs
\theoremstyle{plain}

\newtheorem{theorem}{Theorem}[section]
\newtheorem{lemma}[theorem]{Lemma}
\theoremstyle{definition}

\theoremstyle{remark}

\newcommand{\bC}{ {\boldsymbol C} }

\newcommand{\be}{ {\boldsymbol e} }
\newcommand{\bE}{ {\boldsymbol E} }
\newcommand{\boldf}{ {\boldsymbol f} }

\newcommand{\bG}{ {\boldsymbol G} }

\newcommand{\bI}{ {\boldsymbol I} }

\newcommand{\boldm}{ {\boldsymbol m} }

\newcommand{\bo}{ {\boldsymbol o} }

\newcommand{\bp}{ {\boldsymbol p} }
\newcommand{\bP}{ {\boldsymbol P} }

\newcommand{\bQ}{ {\boldsymbol Q} }
\newcommand{\br}{ {\boldsymbol r} }
\newcommand{\bR}{ {\boldsymbol R} }

\newcommand{\bu}{ {\boldsymbol u} }

\newcommand{\bv}{ {\boldsymbol v} }
\newcommand{\bV}{ {\boldsymbol V} }
\newcommand{\bw}{ {\boldsymbol w} }

\newcommand{\bx}{ {\boldsymbol x} }
\newcommand{\bX}{ {\boldsymbol X} }
\newcommand{\by}{ {\boldsymbol y} }

\newcommand{\bz}{ {\boldsymbol z} }

\newcommand{\bDelta}  { {\boldsymbol \Delta} }

\newcommand{\bet}     { {\boldsymbol \eta} }

\newcommand{\bmu}     { {\boldsymbol \mu} }

\newcommand{\bSigma}  { {\boldsymbol \Sigma} }

\newcommand{\bphi}    { {\boldsymbol \phi} }

\endlocaldefs

\begin{document}

\begin{frontmatter}
\title{\bf Data Sketching and Stacking: A Confluence of Two Strategies for Predictive Inference in Nonlinear High-Dimensional Regression}
\runtitle{Data Sketching and Stacking}

\begin{aug}
\author[A]{\fnms{Samuel}~\snm{Gailliot}\ead[label=e1]{samuel.gailliot@stat.tamu.edu}},
\author[A]{\fnms{Niladri}~\snm{Kal}\ead[label=e4]{niladrik@tamu.edu}},
\author[B]{\fnms{Roger D.}~\snm{Peng}\ead[label=e3]{roger.peng@austin.utexas.edu}}
\and
\author[A]{\fnms{Rajarshi}~\snm{Guhaniyogi}\ead[label=e2]{rajguhaniyogi@stat.tamu.edu}}\thanks{
    Rajarshi Guhaniyogi acknowledges funding from National Science Foundation through DMS-2220840 and DMS-2210672; and funding from National Institute of Health Award Number R01NS131604}
\address[A]{Department of Statistics,
Texas A\&M University\printead[presep={,\ }]{e1,e4,e2}}

\address[B]{Department of Statistics and Data Sciences, University of Texas at Austin\printead[presep={,\ }]{e3}}
\runauthor{Gailliot et al.}
\end{aug}

\begin{abstract}
This article focuses on drawing computationally-efficient predictive inference from Gaussian process (GP) regressions with a large number of predictors when the response is conditionally
independent of the predictors given the projection to a noisy low dimensional manifold. Bayesian estimation of the regression relationship using Markov chain Monte Carlo and subsequent predictive inference is computationally prohibitive and may lead to inferential inaccuracies since accurate variable selection is essentially impossible in such high-dimensional GP regressions. As an alternative, this article proposes a strategy to sketch the high-dimensional predictor vector with a carefully constructed sketching matrix, before fitting a GP with the scalar outcome and the sketched predictor vector to draw predictive inference. The analysis is performed in parallel with many different sketching matrices and smoothing parameters in different processors, and the predictive inferences are combined using \emph{Bayesian predictive stacking}. Since posterior predictive distribution in each processor is analytically tractable, the algorithm allows bypassing the robustness issues due to convergence and mixing of MCMC chains, leading to fast implementation with very large number of predictors. 
Simulation studies show superior performance of the proposed approach with a wide variety of competitors. The approach outperforms competitors in drawing point prediction with predictive uncertainties of outdoor air pollution from satellite images.
\end{abstract}



\end{frontmatter}

\section{Introduction}\label{sec:intro}
We focus on the problem of drawing predictive inference of a random variable from a high-dimensional predictor vector using ``sketching'' of the predictor vector when it truly lies on a low-dimensional noisy unknown manifold. In recent years, there has been a growing literature on ``data sketching,'' which involves sketching or compressing the original data before analysis 
\citep{ halko2011finding, mahoney2011randomized, woodruff2014sketching, guhaniyogi2015bayesian, guhaniyogi2016compressed}. However, our approach differs from the existing data sketching literature in two key aspects. Firstly, while most data sketching approaches aim to reduce the number of data samples, our approach is distinct in that it maintains the same number of samples but instead reduces the dimensionality of the predictor vector. Secondly, the majority of research in data sketching focuses on performance evaluation of ordinary and high-dimensional penalized regression methods with sketched data \citep{zhang2013recovering,  dobriban2018new, drineas2011faster, ahfock2017statistical, huang2018near}, with only a few recent articles considering application of data sketching in Bayesian high-dimensional linear and non-linear regressions \citep{guhaniyogi2021sketching,guhaniyogi2015bayesian,guhaniyogi2016compressed}. In contrast, our approach demonstrates that when the predictor vector resides on a noisy unknown low-dimensional manifold, and the response has a nonlinear relationship with the predictors through the low-dimensional manifold coordinates, an exact nonparametric Bayesian framework can be employed—without the need for iterative sampling methods like Markov chain Monte Carlo. This framework enables efficient and accurate predictions in ultra-high dimensions by effectively utilizing   Gaussian process regression with predictors compressed through data sketching, and stacking of predictive distributions \citep{yao2018using}.


We consider a regression framework with an outcome $y\in\mathcal{Y}\subseteq\mathbb{R}$ and a predictor vector $\bx=(x_1,...,x_p)^T$ when $\bx$ resides on a noisy unknown manifold, i.e., $\bx=\bphi(\bo)+\bet$, where $\bo=(o_1,...,o_d)^T$ is $d$-dimensional co-ordinates for a manifold $\mathcal{O}\subseteq\mathbb{R}^p$, $\bphi(\cdot):\mathbb{R}^d\rightarrow\mathbb{R}^p$ is a mapping function such that $\bphi(\bo)\in\mathcal{O}$ and $\bet$ is $p$-dimensional noise. Often the complex dependence between $y$ and $\bx$ is encoded
via co-ordinates of the low-dimensional manifold, i.e., 
\begin{align}
y=h(\bo)+\epsilon, 
\end{align}
where $h$ is a complex function encoding the true relationship between the response and co-ordinates of the manifold and $\epsilon$ is the error. Since the manifold $\mathcal{O}$ is unobserved, the co-ordinates $\bo$ is typically unknown. Hence, the common practice is to estimate complex dependencies between $y$ and $\bx$ through a non-linear regression model given by,
\begin{align}
y=f(\bx)+\epsilon,    
\end{align}
where $f$ is an unknown regression function and $\epsilon$ is the residual. When dealing with high-dimensional predictors, Gaussian process (GP) priors with an automatic relevance determination (ARD) kernel are commonly used to estimate the underlying function $f$ with sufficient sparsity assumption in the relationship between $y$ and $\bx$ \citep{zhao2018variational, jensen2021scalable}. The estimated $f$ is then employed to predict the response variable. However, when the number of predictors reaches the order of a few thousand, estimation of $f$ with GP-ARD framework is often inaccurate, leading to unsatisfactory predictive inference. This article proposes an alternative approach that exclusively focuses on drawing predictive inference on the response variable $y$, including both point prediction and uncertainty estimation, using GP regression. We review below a list of existing strategies to draw predictive inference on $y$ in non-linear regressions before introducing our approach.


In the literature, a significant line of work follows a two-stage approach for dealing with high-dimensional predictors in non-linear manifold regression tasks. In this approach, the first stage involves constructing a lower-dimensional representation of the high-dimensional predictors using manifold learning techniques. Some popularly employed parsimonious manifold learning algorithms include Isomap \citep{tenenbaum2000global}, Diffusion Maps \citep{coifman2006diffusion}, and Laplacian eigenmap \citep{belkin2003laplacian}. These algorithms enable the reduction of dimensionality while preserving the essential characteristics of the data. Additionally, there are model-based approaches that estimate the unknown Riemannian manifold structure within the predictor vector. These methods utilize techniques such as local PCA \citep{weingessel2000local, arias2017spectral} or geometric multiresolution analysis \citep{maggioni2016multiscale}, and, more recently, spherical basis functions \citep{li2022efficient}. Non-linear regression models in the second stage are based on these projected predictors in lower-dimensions.
However, it is important to note that such two-stage approaches rely on learning the manifold structure embedded in the high-dimensional predictors. While this can be valuable for understanding the underlying data structure, it adds unnecessary computational burden when the primary focus is on prediction rather than inference.

An alternative line of research focuses on estimating the unknown function $f$ using tree-based approaches or deep neural network methods. Tree-based approaches, such as CART \citep{denison1998bayesian}, BART \citep{chipman2010bart}, and random forest \citep{breiman2001random} are based on finding the best splitting attribute, which can become less efficient as the number of predictors ($p$) increases. While there is a growing literature on variable selection within tree-based methods, such as BART \citep{bleich2014variable} and its variants \citep{liu2021variable}, estimating the true regression function with a large number of predictors ($p$ of the order of thousands) can pose challenges.
Deep neural networks with variable selection architecture \citep{ dinh2020consistent} are also not ideal as they lack predictive uncertainty and struggle to handle the high-dimensional predictor space efficiently.

Bayesian modeling approaches are naturally appealing when the focus is on quantifying predictive uncertainty. To this end, the more traditional Bayesian models simultaneously learn the mapping to the lower-dimensional subspace along with the regression function in the coordinates on this subspace. These approaches range 
from Gaussian process latent variable models (GP-LVMs) \citep{lawrence2005probabilistic,titsias2010bayesian} for probabilistic nonlinear principle component analysis to mixture of factor models \citep{chen2010compressive}. However, such methods pose daunting computational challenges with even moderately large $p$ and sample size due to learning the number and distribution of latent variables, as well as the mapping functions, while maintaining identifiability restrictions.

To enhance the time efficiency of the aforementioned approaches, pre-processing steps are often employed, and two popular pre-processing methods are predictor screening and projection. The predictor screening approach identifies predictors that exhibit the strongest marginal association with the response variable. By selecting the predictors with the highest marginal association, this approach aims to reduce the dimensionality of the problem. predictor screening methods are generally straightforward to implement, and offer asymptotic guarantees of selecting a superset of important predictors \citep{chen2018error}. On the other hand, projection approaches aim to construct lower-dimensional predictor vectors by combining the original high-dimensional predictors. One common method in projection approaches is to construct a few principal components (PCs) from the original $p$-dimensional predictor vector. 

A naive implementation of the above pre-processing steps is unappealing to our scenario. For example, in a non-parametric regression with a large number of correlated predictors and low signal-to-noise ratio, it may be important to choose a conservative threshold for screening, which limits the scope of dimension reduction at this stage. On the other hand, construction of PCs are agnostic to the relationship between the response and the predictor vector. Instead, we propose an approach that first employs variable screening \citep{chen2018error} with a conservative threshold to identify a large subset of predictors, typically a few thousand, having the highest non-linear marginal association with the response. After variable screening, the screened predictor vector is further compressed using a short and fat random sketching matrix \citep{mahoney2011randomized,drineas2012fast}. This matrix has a small number of rows ($m$) and entries that are independently and identically drawn from a normal distribution. 
Predictive inference proceeds by fitting a non-parametric Gaussian process (GP) regression model \citep{williams2006gaussian,gramacy2020surrogates} to the scalar outcome and the $m$-dimensional sketched predictor vector after fixing values for the weakly identifiable tuning parameters within the covariance kernel of the Gaussian processes. The posterior predictive distribution corresponding to a choice of tuning parameters and random sketching matrix comes in a closed form without the need to implement MCMC sampling, so that one can obtain the predictive distribution extremely rapidly even in problems with huge numbers of predictors. To reduce the sensitivity of predictive inference to the choice of the sketching matrix and the tuning parameters in GP regression, the model is fit with multiple different choices of the sketching matrix and parameters. The predictive inferences obtained from such choices are then aggregated using Bayesian predictive stacking \citep{yao2018using} to improve accuracy and robustness. When predictor vector lies on a low-dimensional unknown manifold and the response has a non-linear relationship with the manifold co-ordinates, the proposed approach has provable guarantee of drawing optimal predictive inference, as argued in this article.

In regressions involving high-dimensional predictors and a large sample size, \cite{guhaniyogi2021sketching} propose an approach orthogonal to ours which exploits random sketching matrices to reduce the sample size rather than the number of predictors. A few approaches closely related to ours develop theoretical bound on predictive accuracy when high-dimensional predictors are sketched with random matrices \citep{guhaniyogi2015bayesian,guhaniyogi2016compressed,thanei2017random}. These articles include random linear combinations of many unimportant predictors, diminishing signal in the analysis, which results in less than satisfactory predictive performance with massive-dimensional predictors. Addressing this issue, the most closely related to our approach is \cite{mukhopadhyay2020targeted}, which proposes novel constructions of projection matrices tailored to deliver more accurate predictive inference. This approach aim to overcome the challenges associated with sketching high-dimensional predictors and improve predictive performance. However, our approach offers two key advantages compared to \cite{mukhopadhyay2020targeted} \textbf{First}, while \cite{mukhopadhyay2020targeted} is designed for high-dimensional parametric linear regression under predictor sparsity, our work addresses nonlinear GP regression with predictors on a noisy manifold, which presents a more complex form of sparsity. \textbf{Second}, \cite{mukhopadhyay2020targeted} manages sensitivity to sketching matrices using Bayesian model averaging (BMA) \cite{raftery1997bayesian}, rather than stacking used in our approach. While Bayesian model averaging \citep{raftery1997bayesian} is most popularly used for aggregating inference from multiple models, it is less suited to the stacking procedure in our settings. To see this, assume that there are $S$ candidate models $\mathcal{M}=\{\mathcal{M}_1,...,\mathcal{M}_S\}$.  Bayesian model comparison falls into one of three settings: (i) $\mathcal{M}$-closed where a true data generating model exists and is included in $\mathcal{M}$; (ii) $\mathcal{M}$-complete where a true model exists but is not included in $\mathcal{M}$; and (iii) $\mathcal{M}$-open where we do not assume the existence of a true data generating model. Although Bayesian model averaging has the advantage of asymptotically identifying the true data generating model in the first setting, predictive stacking has advantages in the $\mathcal{M}$-complete and $\mathcal{M}$-open settings  \cite{le2017bayes, yao2018using,pavlyshenko2020using,yao2022bayesian,yao2022stacking}. Given that the true model may not be included in the class of fitted GP regression models with sketched predictors, predictive stacking offers substantial advantages over model averaging.

The rest of the article proceeds as follows. Section~\ref{sec:air pollution} discusses the motivating dataset on outdoor air pollution and satellite imagery. Section~\ref{sec:stacked_GP} proposes the model and computational approach for predictive inference in manifold regression with large number of predictors. Section~\ref{sec:simulations} offers empirical evaluation of the proposed approach along with its competitors for simulation studies. Section~\ref{sec:real_data} investigates the proposed approach in drawing predictive inference of outdoor air pollution concentration from satellite images. Finally, Section~\ref{sec:conclusion} concludes the article with an eye towards future work.

\subsection{Outdoor Air Pollution Application}\label{sec:air pollution}

As a motivation for the development of our methodology, we consider the problem of predicting outdoor air pollution concentrations across the United States. Outdoor air pollution in the U.S. is measured using a network of ground-based monitors managed by local air quality agencies and the U.S. Environmental Protection Agency~\citep{epa:1996} (EPA). While the combined network of monitors consists of thousands of locations, the spatial coverage of the network is actually quite sparse, leaving many areas of the country without any ground-level data~\citep{apte2017high}. To address the sparsity of the network, there have been efforts to deploy low-cost sensors across urban areas to fill the gaps. While such approaches have promise, they are still experimental and ad hoc in nature, and the sensors themselves can sometimes introduce new measurement problems~\citep{heffernan2023dynamic}.

Remote sensing techniques, which use satellite imagery to predict ground-level concentrations of outdoor air pollution have the potential to address the spatial coverage problem because of their constant monitoring of the entire planet. Traditional approaches have employed aerosol optical depth as a proxy for such pollutants as fine particulate matter~\citep{paciorek2008spatiotemporal}, or PM2.5. 
In recent years, there has been a revolution in the deployment of satellite constellations, where hundreds of smaller inexpensive satellites orbit the Earth, providing constant coverage of all areas~\citep{team2017planet}. Furthermore, these satellites have much higher resolution. Given the recent emergence of data from satellite constellations, there is still a question of how best to use them for the purpose of predicting ground-level air pollution concentrations. For this application, we focus on predicting fine particulate matter pollution (PM2.5) from multi-band satellite images. For ground-truth information we use the EPA's network of monitors to provide PM2.5 measurements. The combination of high-resolution spatial and temporal coverage of the entire U.S. with novel statistical predictive approaches has the potential to dramatically increase the monitoring of outdoor air pollution and its subsequent health effects.

\section{Our Approach: Stacked Gaussian Process Regression}\label{sec:stacked_GP}

Let $\mathcal{D}_n=\{(\bx_i^T,y_i):i=1,...,n\}$ be a dataset containing $n$ observations each with a $p$-variate predictor $\bx_i=(x_{i,1},...,x_{i,p})^T$ and a scalar-valued response $y_i$. We assume $n$ is moderately large and $p$ is large. The predictor vector $\bx_i$ lies on an unknown noisy manifold $\mathcal{O}\subseteq\mathbb{R}^p$ with $d$-dimensional latent co-ordinates $\bo_i$ (i.e., $\bx_i=\bphi(\bo_i)+\bet_i$, $\bphi(\bo_i)\in\mathcal{O}$). We assume a nonlinear regression relationship between $y_i$ and $\bx_i$, and approximate the density of $y_i$ by sketching the high-dimensional predictor vector $\bx_i$ to lower-dimensions using a sketching matrix $\bP_n$ as follows
\begin{align}\label{eq:fitted_model}
y_i=f(\bP_n\bx_i)+\epsilon_i,\:\:\epsilon_i\stackrel{i.i.d.}{\sim} N(0,\xi^2),
\end{align}
with $\xi^2$ as the noise variance and $f(\cdot)$ as an unknown continuous function in the Holder class of smoothness $s$. Discussion on the choice of the sketching matrix $\bP_n\in\mathbb{R}^{m\times p}$ is provided in Section~\ref{Sec:proj}. 

\subsection{Choice of the Sketching Matrix}\label{Sec:proj}
The sketching matrix $\bP_n \in \mathbb{R}^{m \times p}$ embeds $p$-dimensional predictors $\bx_i$ into $m$ dimensions while not throwing away excessive amounts of information. The most popular linear embedding is obtained from the singular value decomposition (SVD) of 
$\bX=[\bx_1:\cdots:\bx_n]^T$, but they are problematic to estimate when $p>>n$. In contrast, random sketching matrices are often used to embed the high-dimensional predictors to a random subspace, and appropriate choices of the random matrices allow distances between samples to be approximately preserved \citep{li2017restricted}.

The direct application of sketching matrices on high-dimensional predictors is unappealing, as it constructs random linear combinations of many unimportant predictors, diminishing the signal in the analysis. As an alternative strategy, we first reduce the number of predictors to a more manageable size using a screening procedure before applying a sketching matrix that forms random linear combinations from this selected set. To identify these predictors, this article performs nonparametric B-spline regression of $y_i$ onto each component of 
$\bx_i$ separately. The order of importance of the predictors is determined, in descending order, by the residual sum of squares of the marginal nonparametric regressions. predictors with a residual sum of squares greater than a user-defined threshold are considered important predictors with the highest marginal association with the response. We adopt a conservative threshold following \cite{fan2014nonparametric} to select a large superset of important predictors, which allows for joint contributions of predictors in explaining the response. While variable screening is not strictly necessary it is recommended by the authors as it is unhelpful to allow a large number of uninformative predictors to contaminate the sketched design matrix. We emphasize that variable screening is not the main contribution of the article and that any alternative screening well designed to select predictors can also be incorporated into our framework.

Let $\mathcal{I}$ correspond to the indices of the predictors chosen by marginal screening and $\bar{\mathcal{I}}$ be the indices of the predictors screened out through this procedure, such that $\mathcal{I}\cup\bar{\mathcal{I}}=\{1,...,p\}$. Let $\bE_n$ be a permutation matrix such that
$\bE_n\bx=(\bx_{\mathcal{I}}^T,\bx_{\bar{\mathcal{I}}}^T)^T$. We construct a matrix $\bR_n=[\bR_{n,1}:\bR_{n,2}]$ where $\bR_{n,2}={\boldsymbol 0}_{m\times (p-|\mathcal{I}|)}$  and $\bR_{n,1}$ is an $m\times |\mathcal{I}|$ matrix constructed by first drawing entries independently from $N(0,1)$, and then orthonormalizing the rows using Gram--Schmidt orthogonalization, following the literature on random sketching matrices \citep{baraniuk2008simple}. The resulting sketching matrix is given by $\bP_n=\bR_n\bE_n$.

\subsection{Prior, Posterior and Posterior Predictive Distributions}\label{sec:GP}

Following a Bayesian approach, we assign a zero-centered Gaussian process prior on the unknown regression function $f(\cdot)$, denoted  by $f(\cdot)\sim GP(0,\sigma^2\delta_{\theta})$. Here, $\delta_{\theta}$ corresponds to an exponential correlation kernel $\delta_{\theta}(\bx_i,\bx_j)=\exp(-\theta||\bx_i-\bx_j||)$ involving the length-scale parameter $\theta$. The parameter $\sigma^2$ is the signal variance parameter and $||\cdot||$ denotes the Euclidean norm. 
A significant finding by \cite{yang2016bayesian} establishes that when predictors $\bx_i$ lie on a $d$-dimensional manifold $\mathcal{O}$, GP regression approximates the true regression function at  the minimax optimal rate of l. The de-noised compressed predictors $\bP_n\bx_i$ exhibit a higher concentration around the manifold compared to the original predictors $\bx_i$. With this enhanced concentration, the theory presented by \cite{yang2016bayesian} suggests that an appropriate GP prior can yield excellent performance. In addition to denoising, the compression of the high-dimensional predictor vector has a major advantage in avoiding the estimation of a geodesic distance along the unknown manifold $\mathcal{O}$ between any two predictor vectors $\bx_i$ and $\bx_{i'}$.


In practical applications, utilizing the recommended prior distributions on $\theta$ and $\sigma^2$ from \cite{yang2016bayesian} requires computationally expensive Markov chain Monte Carlo (MCMC) sampling. In particular, the posterior computation of $\theta$ typically entails meticulous tuning, especially when dealing with high-dimensional predictors, imposing a significant computational burden. This article introduces an alternative approach that enables exact predictive inference from the model, entirely bypassing the MCMC algorithm in model estimation. The details of the strategy are elaborated below.

Denote $\boldf=(f(\bP_n\bx_1),...,f(\bP_n\bx_n))^T$ as the vector consisting of the function $f$ evaluated at the sketched predictors $\bP_n\bx_1$,...,$\bP_n\bx_n$ and $\bC$ as an $n\times n$ covariance matrix with $(i,j)$th entry  $\delta_{\theta}(\bP_n \bx_i, \bP_n \bx_j)$. With $\by=(y_1,...,y_n)^T$ as the response vector, a customary Bayesian hierarchical model is constructed as
$\by|\boldf,\xi^2\sim N(\boldf,\xi^2\bI),\:\:\:
(\boldf | \xi^2)\sim N(0, \xi^2\psi^2\bC),\:\:\:\: \pi(\xi^2)\propto \frac{1}{\xi^2}$,
where we fix the length-scale parameter $\theta$ and the signal-to-noise variance ratio $\psi^2=\frac{\sigma^2}{\xi^2}$. This ensures closed-form conjugate marginal posterior predictive distributions.
More specifically, the marginal posterior distribution of $\xi^2$, given the projection matrix $\bP_n$, $\theta$, $\psi^2$ and $\mathcal{D}_n$, is inverse gamma with parameters $a=n/2$ and $b=\by^T(\psi^2\bC+\bI)^{-1}\by/2$. The marginal posterior distribution of $\boldf$, given $\bP_n$, $\psi^2$, $\theta$ and $\mathcal{D}_n$, follows a scaled $n$-variate $t$ distribution with degrees of freedom $n$, location $\bmu_t$ and scale matrix $\bSigma_t$, denoted by $t_n(\bmu_t,\bSigma_t)$, where 
$\bmu_t=(\bI+\bC^{-1}/\psi^2)^{-1}\by,\:\:\:\bSigma_t=(2b/n)(\bI+\bC^{-1}/\psi^2)^{-1}.$
Consider prediction for the response at $n_{new}$ data points with corresponding covariates $\tilde{\bx_1},...,\tilde{\bx}_{n_{new}}$. Let $\bC_{new,new}$ and $\bC_{new,old}$ denote $n_{new}\times n_{new}$ and $n_{new}\times n$ matrices with $(i,j)$th elements $\delta_{\theta}(\bP_n\tilde{\bx}_i,\bP_n\tilde{\bx}_j)$ and $\delta_{\theta}(\bP_n\tilde{\bx}_i,\bP_n\bx_j)$, respectively.
The posterior predictive distribution of the response $\tilde{\by}_{new}=(\tilde{y}_1,...,\tilde{y}_{n_{new}})^T$ given $\tilde{\bx_1},...,\tilde{\bx}_{n_{new}}$, $\bP_n$, $\theta$, $\psi^2$ and $\mathcal{D}_n$, marginalizing out $(\boldf, \xi^2)$, follows a scaled $n_{new}$-variate t-distribution $t_{n_{new}}(\tilde{\bmu}_t,\tilde{\bSigma}_t)$, where 
\begin{align}\label{eq:posterior_predictive_dist}
\tilde{\bmu}_t &=\psi^2\bC_{new,old}(\bI+\psi^2\bC)^{-1}\by\nonumber\\
\tilde{\bSigma}_t &=(2b/n)\left[\bI+\psi^2\bC_{new,new}-\psi^4\bC_{new,old}(\bI+\psi^2\bC)^{-1}\bC_{new,old}^T\right].
\end{align}
Since the posterior predictive distribution is available in closed form, Bayesian inference can proceed from exact posterior samples.

This tractability is only possible if the length-scale parameter $\theta$ and the signal-to-noise variance ratio $\psi^2$
are fixed. While it is possible to estimate their full posterior distributions through expensive Markov chain Monte Carlo (MCMC) sampling, these parameters are inconsistently estimable for the general Matern class of correlation functions \citep{zhang2004inconsistent} often resulting in poorer convergence. Therefore, for the chosen sketching matrix $\bP_n$, we obtain $(\theta, \psi^2)$ such that
\begin{align}
\mbox{max}_{\theta,\psi^2} f(\theta, \psi^2 | \bP_n, \by) &\propto \mbox{max}_{\theta,\psi^2}\frac{1}{|\psi^2\bC + \bI|^{\frac{1}{2}}} \frac{2^{\frac{n}{2}}\Gamma(\frac{n}{2})}{\left[\by'(\psi^2\bC + \bI)^{-1}\by\right]^{\frac{n}{2}}(\sqrt{2\pi})^{n} }
\label{eq:SNR_post}
\end{align}
Our approach
will conduct exact predictive inference using the closed form predictive distribution in (\ref{eq:posterior_predictive_dist}) and stack the predictive inference over different fixed values of $\{\bP_n, \theta,\psi^2\}$.




\subsubsection{Stacking of Predictive Distributions}
\begin{figure}
\centering
\includegraphics[width=0.5\textwidth]{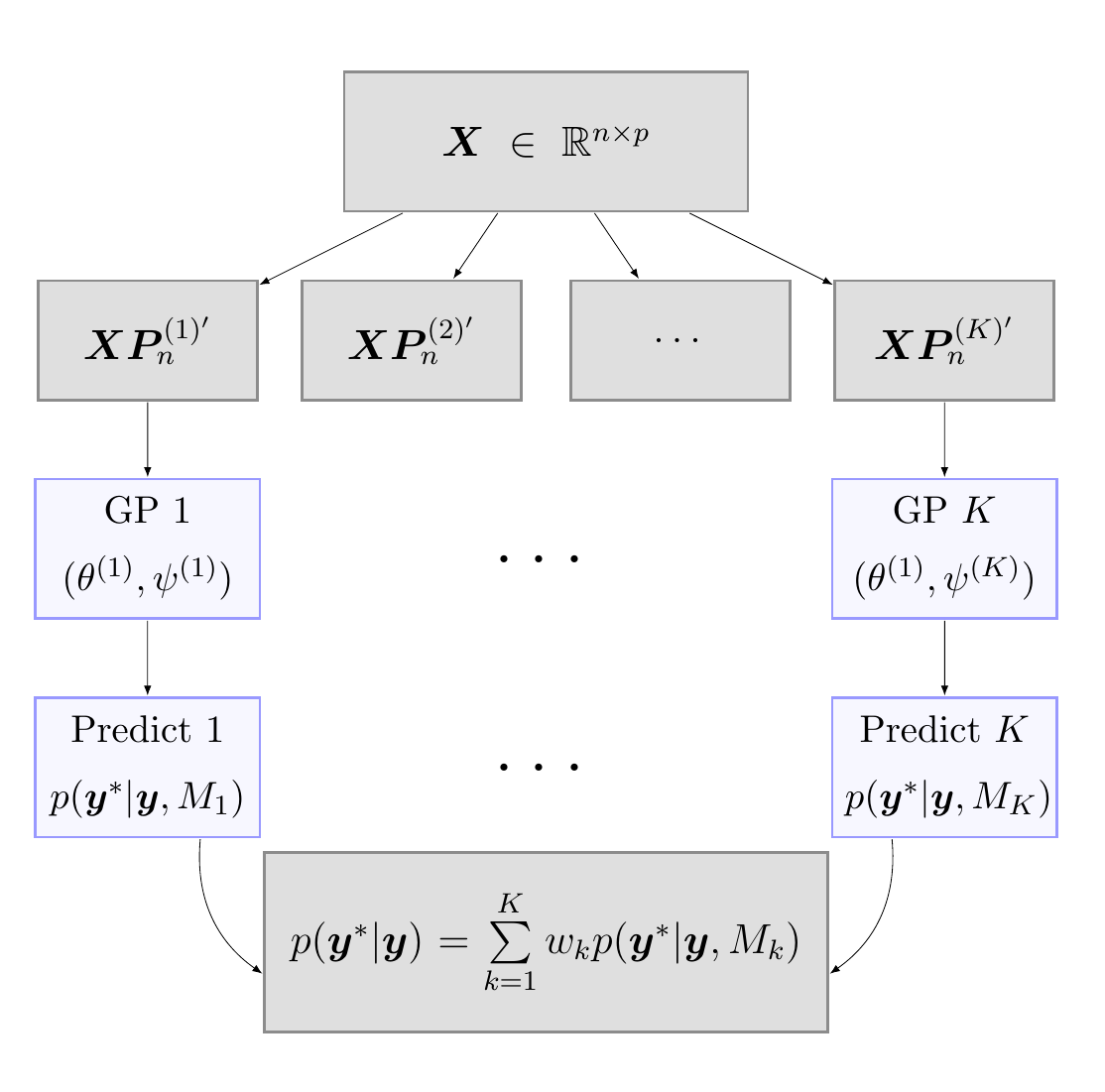}
\caption{Flowchart representing the Sketched Gaussian process regression framework.}\label{fig:Model_Diagram}
\end{figure}

Let $\mathcal{M}_k$ represent the fitted model (\ref{eq:fitted_model}) with ${\bP_n^{(k)}, \theta^{(k)},\psi^{2(k)}}$ for $k=1,...,K$. While the sketching matrix $\bP_n^{(k)}$ is randomly generated for each $k$, $\theta^{(k)}$ and $\psi^{2(k)}$ are obtained following equation (\ref{eq:SNR_post}) for the choice of $\bP_n^{(k)}$. Employing the generalized Bayesian stacking framework proposed by \cite{yao2018using}, we implement a stacking procedure over the predictive distribution obtained from each $\mathcal{M}_k$. Let $p(\tilde{\by}_{new}|\by, \mathcal{M}_k)$ denote the predictive distribution under model $\mathcal{M}_k$, and $p_t(\tilde{\by}_{new}| \by)$ denote the true predictive distribution. Our objective is to determine the distribution in the convex hull $\mathcal{C} = \{\sum_{k = 1}^K w_k p(\cdot | \by,\mathcal{M}_k): w_k \in \mathcal{S}_1^K\}$, where $\mathcal{S}_1^K = \{\bw \in [0,1]^K: \sum_{k=1}^K w_k = 1\}$, that is optimal with respect to some proper scoring function. Using the logarithmic score, which corresponds to the KL divergence, we seek to find the vector $\tilde{\bw} = (\tilde{w}_1, \ldots, \tilde{w}_K)$ such that
\begin{equation}\label{eq:stacking_approx}
 \tilde{\bw}=   \max_{\bw \in \mathcal{S}_1^K} \frac{1}{n} \sum_{i = 1}^n \log\left( \sum_{k = 1}^K w_k p_{k,-i}(y_i)\right),
\end{equation}
where $\by_{-i}=(y_j:j\neq i,\:j=1,...,n)^T$ and $p_{k,-i}(y_i) = p(y_i|\by_{-i}, \mathcal{M}_k)$ has a closed from univariate $t$-distribution with parameters $\tilde{\mu}_{-i}^{(k)}$ and $\tilde{\Sigma}_{-i}^{(k)}$ obtained using the formula for posterior predictive distribution given in equation (\ref{eq:posterior_predictive_dist}). In practice, calculating the predictive densities $p_{k,-i}(y_i)$ one at a time is computationally expensive as the calculation of $\tilde{\Sigma}_{-i}^{(k)}$ requires inverting an $(n-1)\times(n-1)$ matrix for every $k=1,...,K$ and $i=1,...,n$. To avoid this, we randomly split the data into $S=10$ disjoint folds of approximately equal size, $(\by_{(1)}, \bX_{(1)}), \ldots, (\by_{(S)}, \bX_{(S)})$, and compute 
$\by_{(s)}|\by_{(1)},...,\by_{(s-1)},\by_{(s+1)},...,\by_{(S)}$ for every $s=1,...,S$, which follows a multivariate t-distribution with parameters  obtained using equation (\ref{eq:posterior_predictive_dist}). If the $i$th sample belongs to the $s$th fold, we will replace $p_{k,-i}(y_i)$ in (\ref{eq:stacking_approx}) by $p_{k,(s)}(y_i)$, where $p_{k,(s)}(y_i)$ represents the marginal distribution of $y_i$ from $\by_{(s)}|\by_{(1)},...,\by_{(s-1)},\by_{(s+1)},...,\by_{(S)}$. This strategy requires inverting an $(n-n/S)\times (n-n/S)$ matrix only $S$ times (assuming that all folds are of equal size), which leads to substantial computational benefits.
No analytical solution to this non-convex constrained optimization problem in (\ref{eq:stacking_approx}) is available, but first and second derivatives are easily obtained to construct an iterative optimizer.
The optimal distribution provides a pseudo posterior predictive distribution 
given by $\tilde{p}(\tilde{\by}_{new}|\by)=\sum_{k=1}^K\tilde{w}_k t_{n_{new}}(\tilde{\bmu}_t^{(k)},\tilde{\bSigma}_t^{(k)})$, where $\tilde{\bmu}_t^{(k)}$ and $\tilde{\bSigma}_t^{(k)}$ are obtained from equation (\ref{eq:posterior_predictive_dist}) by evaluating $\tilde{\bmu}_t$ and $\tilde{\bSigma}_t$ at $\bP_n^{(k)}, \theta^{(k)}, \psi^{(k)}$. The pseudo posterior predictive distribution is further used to draw point prediction and 95\% predictive interval to quantify predictive uncertainty. Figure~\ref{fig:Model_Diagram} offers a flowchart outlining the proposed framework.

While Bayesian model averaging (BMA) is a common method for combining multiple distributions, its applicability in our context is limited for several reasons. Firstly, the fitted models (\ref{eq:fitted_model}) with randomly sketched predictors are likely to deviate from the true model, placing us outside the $\mathcal{M}$-closed setting where BMA is optimal. Additionally, stacking is designed to determine weights for optimal prediction, whereas asymptotically, BMA assigns full weight to the ``best'' single model closest in KL divergence to the true model \citep{yao2022bayesian}. However, when the true model lies outside the space of fitted models, it may be more advantageous to leverage multiple models in predictive inference. Subsequent empirical experiments demonstrate stacking as a powerful tool for drawing posterior predictive inference in our setting.
 
 
\subsection{Theoretical Analysis}
This section presents the theoretical foundation that explains the strong practical performance of the proposed method. In this context, when the predictor vector lies on a noisy manifold, i.e., 
$\bx=\bphi(\bo)+\bet $, it follows that $\bP_n\bx=\bP_n\bphi(\bo)+\bP_n\bet $ Two key arguments support the claim that data sketching leads to desirable inference:
\begin{enumerate}[(A)]
\item When predictors lie exactly on a manifold, our proposed approach of GP regression with multiple sketched predictors followed by stacking achieves asymptotically accurate estimation of the true regression function. This shows that using the sketched predictors $\{\bP_n\bphi(\bo_i)\}_{i=1}^n$ in our proposed approach will result in accurate asymptotic estimation of the regression function.
\item Sketching the noise vector 
$\bet$ with compression matrices reduces the impact of noise in 
$\bx$ on predictive accuracy.
\end{enumerate}
To support (A), we first introduce the following definition and notations.\\
\textbf{Definition:} Let $\mathcal{H}_1$ and $\mathcal{H}_2$ be two manifolds. A differentiable map $g:\mathcal{H}_1\rightarrow\mathcal{H}_2$ is called
a diffeomorphism if it is a bijection and its inverse $g^{-1}:\mathcal{H}_2\rightarrow\mathcal{H}_1$ is differentiable. 

 Let $f_k(\cdot)$ represent the fitted regression function for model $\mathcal{M}_k$, which corresponds to the fitted model
 (\ref{eq:fitted_model})  with parameters ${\bP_n^{(k)}, \theta^{(k)},\psi^{2(k)}}$ for $k=1,...,K$. Let $f_0(\cdot)$ denote the true regression function. We define the distance between the stacked regression function and the true regression function under a fixed design of covariates lying on a manifold as 
$\Delta_\phi(f_0)=\frac{1}{n}\sum_{i=1}^n(\sum_{k=1}^K w_kf_k(\bphi(\bo_i))-f_0(\bphi(\bo_i)))^2$. When the covariate design is random, this distance is expressed as
$\Delta_\phi(f_0)=\int(\sum_{k=1}^K w_kf_k(\bphi(\bo))-f_0(\bphi(\bo)))^2 dG$, where $G(\cdot)$ is the marginal distribution of the predictors. Let $\Pi(\cdot|\mbox{Data})$ represent the posterior distribution of the model parameters. We will also denote $C^s(\mathcal{O})$ as the Banach space of functions on $\mathcal{O}$ with Holder smoothness $s$.
The following result shows (A), the proof of which can be found in the Appendix.

\begin{lemma}\label{lem:clean-manifold}
Let $\Delta_\phi^{(k)}(f_0)=\frac{1}{n}\sum_{i=1}^n(f_k(\phi(\bo_i))-f_0(\phi(\bo_i)))^2$ for fixed design and $\Delta_\phi^{(k)}(f_0)=\int(f_k(\phi(\bo))-f_0(\phi(\bo)))^2 dG$ for random design of covariates. Assume that the stacking weights $w_k$, $1\leq k\leq K$, are all bounded. Further assume that $\mathcal O\subset \mathbb R^p$ is a compact $C^\gamma$ $d$-dimensional manifold satisfying the regularity conditions of Theorem 3.1 in \cite{baraniuk2009random}, and that $f_0\in C^s(\mathcal O)$ for some $s\le \min\{2,\gamma-1\}$. Then for any $\zeta>0$
\[
\Pi\left(\Delta_\phi(f_0)>\zeta 
\,\middle|\,
\mathcal D\right)\rightarrow 0, \text{ as }n\rightarrow\infty.
\]
\end{lemma}
To support (B), the following lemma is stated. The proof of the lemma can be found in Appendix.

\begin{lemma}
Let $\bP_n^{(k)} = [\bP_{\mathcal I}^{(k)} :\; \mathbf{0}]$ with $\bP_{\mathcal I}^{(k)} \in \mathbb R^{m \times |\mathcal I|}$ obtained by Gram--Schmidt orthonormalization of a Gaussian matrix with independent $N(0,1)$ entries. Assume the off-manifold perturbation $\bet_i = [\bet_{i,\mathcal I}:\; \bet_{i, \tilde{\mathcal I}}] \in \mathbb R^p$ is independent of $\bP_n^{(k)}$ with $\|\bet_i\| = O_P(1)$. Then each component of the projected noise satisfies:
\[
\bp_{j}^{(k)} \bet_{i, \mathcal I} = O_P(|\mathcal I|^{-1/2}),
\]
where $\bp_{j}^{(k)}$ denotes the $j$-th row of $\bP_{\mathcal I}^{(k)}$.
\end{lemma}

Thus, each compressed coordinate contains an attenuated contribution of the off-manifold noise. At the vector level, the same calculation gives
\[
\mathbb E_{\bP_n}\left[
\|\bP_n^{(k)}\bet_i\|^2
\,\middle|\,\bet_i
\right]
=
\frac{m}{|\mathcal I|}\|\bet_{i,\mathcal I}\|^2.
\]
Therefore, under the assumption \(\|\bet_i\|=O_P(1)\),
\[
\|\bP_n^{(k)}\bet_i\|
=
O_P\left(\sqrt{\frac{m}{|\mathcal I|}}\right).
\]
In particular, if \(m/|\mathcal I|\to0\), then, for every fixed \(i\),
\[
\|\bP_n^{(k)}\bet_i\|=o_P(1).
\]

Since
\[
\bP_n^{(k)}\bx_i = \bP_n^{(k)}\bphi(\bo_i)+\bP_n^{(k)}\bet_i,
\]
the noisy sketched predictors are asymptotically close to the corresponding clean sketched predictors, in the sense that the projected noise contribution vanishes coordinatewise as $|\mathcal I|\to\infty$. A uniform version follows under the additional condition
\[
\max_{1\le i\le n}\|\bet_i\|=o_P(1).
\]
Since \(\bP_n^{(k)}\) has orthonormal rows, it is non-expansive, and hence
\[
\max_{1\le i\le n}
\|\bP_n^{(k)}\bet_i\|
\le
\max_{1\le i\le n}
\|\bet_i\|
=
o_P(1).
\]
Consequently, defining
\[
\bz_i^{(k)}=\bP_n^{(k)}\bx_i,
\qquad
\bu_i^{(k)}=\bP_n^{(k)}\bphi(\bo_i),
\qquad
\br_i^{(k)}=\bP_n^{(k)}\bet_i
\]
we obtain
\[
\delta_n = \max_{1\le i\le n} \|\br_i^{(k)}\| = 
\max_{1\le i\le n} \|\bz_i^{(k)}-\bu_i^{(k)}\| =
o_P(1).
\]

The condition \(\delta_n=o_P(1)\) controls the size of the perturbation in the sFketched predictors. To translate this perturbation control into a posterior stability statement, we also need a mild regularity condition on the clean sketched design. 

While our empirical analysis employs the exponential kernel for the Gaussian Process, we adopt the squared exponential kernel to establish the following theoretical results. This approach aligns with established practices in the literature on Bayesian manifold regression using GP \citep{yang2016bayesian}. Since the squared exponential kernel decays rapidly with distance, it is enough to require that the effective row sums of the kernel matrix remain bounded. The clean sketched predictors are allowed to range over a growing region, but they should not cluster too strongly. In particular, we assume that
\[
A_n^{(k)}
:=
\max_{1\le i\le n}
\sum_{j=1}^n
\|\bu_i^{(k)}-\bu_j^{(k)}\|
e^{-(\theta^{(k)}/2)\|\bu_i^{(k)}-\bu_j^{(k)}\|^2}
=
O_P(1),
\]
and
\[
B_n
:=
\max_{1\le i\le n}
\sum_{j=1}^n
e^{-(\theta^{(k)}/2)\|\bu_i^{(k)}-\bu_j^{(k)}\|^2}
=
O_P(1).
\]
These assumptions hold, for example, when the clean sketched predictors are
sufficiently separated in an increasing domain, so that each point has only a
uniformly bounded effective number of neighbors under the squared exponential
kernel.

\begin{lemma}[Posterior stability under sketched perturbations]\label{lem:posterior-stability}
Let \(\Pi_Z^{(k)}(\cdot\mid \mathcal D)\) and
\(\Pi_U^{(k)}(\cdot\mid \mathcal D)\) denote the Gaussian-process posteriors
obtained using the noisy sketched predictors
\(\{\bz_i^{(k)}\}_{i=1}^n\) and the clean sketched predictors
\(\{\bu_i^{(k)}\}_{i=1}^n\), respectively. Suppose the covariance kernel is
squared exponential, $\delta_\theta(\bx_i,\bx_j)=\exp(-\theta\|\bx_i-\bx_j\|^2)$, with fixed \(\theta>0\). If
\[
\delta_n^{(k)}
:=
\max_{1\le i\le n}\|\bz_i^{(k)}-\bu_i^{(k)}\|
=
o_P(1),
\qquad
A_n^{(k)}=O_P(1),
\qquad
B_n^{(k)}=O_P(1),
\]
then, for every \(\zeta>0\),
\[
\Pi\left(
\frac1n
\sum_{i=1}^n
\{f_Z^{(k)}(\bz_i^{(k)})-f_U^{(k)}(\bu_i^{(k)})\}^2
>
\zeta
\,\middle|\,
\mathcal D
\right)
\to 0
\]
in \(P_0\)-probability, under the Gaussian coupling of the two posterior draws.
\end{lemma}

Thus, for each sketch \(k\), replacing the clean sketched predictors $\left\{\bP_n^{(k)}\bphi(\bo_i)\right\}_{i=1}^n$ by the noisy sketched predictors $\left\{\bP_n^{(k)}\bx_i\right\}_{i=1}^n$ has an asymptotically negligible effect on the corresponding GP posterior. We now combine this posterior stability with the clean-manifold consistency result in Lemma~\ref{lem:clean-manifold} to obtain posterior consistency for the proposed stacked procedure based on the observed noisy predictors.
\begin{theorem}
For \(k=1,\ldots,K\), define
$
\Delta_x^{(k)}(f_0)
=
\frac{1}{n}\sum_{i=1}^n
\left\{
f_k\!\left(\bx_i\right)
-
f_0(\bx_i)
\right\}^2$
for fixed design, and
$\Delta_x^{(k)}(f_0)
=
\int
\left\{
f_k\!\left(\bx\right)
-
f_0(\bx)
\right\}^2\,dG(\bx)$
for random design. Define
\[
\Delta_x(f_0)
=
\frac{1}{n}\sum_{i=1}^n
\left[
\sum_{k=1}^K
w_k f_k\!\left(\bx_i\right)
-
f_0(\bx_i)
\right]^2
\]
for fixed design, with the analogous definition in the random-design case. Suppose that the conditions of Lemma~\ref{lem:posterior-stability} hold for each \(k=1,\ldots,K\). Assume further that \(f_0\) is Lipschitz. Then, for every \(\zeta>0\),
\[
\Pi\left(
\Delta_x(f_0)>\zeta
\,\middle|\,
\mathcal{D}
\right)
\to 0,
\qquad n\to\infty.
\]
\end{theorem}


\section{Simulation Study}\label{sec:simulations}
We assess the performance of Sketched Gaussian Process (SkGP) regression through a range of simulation scenarios, varying the structure of the underlying manifold ($\mathcal{O}$), predictor dimension ($p$), and predictor noise level ($\tau^2$). In each case, we compare the out-of-sample predictive accuracy of SkGP to several benchmark methods: standard Gaussian Process (GP) regression with unsketched predictors (which has optimal prediction rate guarantees when predictors lie exactly on a manifold \citep{yang2016bayesian}), Bayesian Additive Regression Trees (BART) \citep{chipman2010bart}, Random Forests (RF) \citep{breiman2001random}, Bayesian Adaptive Smoothing Splines (BASS) and deep neural networks (NN). We also include sketched versions of BART and RF, termed Sketched BART (SkBART) and Sketched Random Forest (SkRF), which accelerate computation by projecting predictors through a single random matrix. Additionally, we construct a competitor, Compressed GP (CompGP). CompGP differs from SkGP by compressing the full unscreened predictor set to dimension $m$, fitting separate GPs with compressed predictors, and using Bayesian model averaging (BMA) to aggregate predictions. The comparison with CompGP highlights the advantages of both stacking over BMA and predictor screening.

Except for CompGP, all methods are applied to a screened subset of $|\mathcal{I}|=1000$ predictors with the highest marginal association to the response. For consistency, we use $m=60$ as the default sketching dimension. While \cite{mukhopadhyay2020targeted} is a related method, results are not presented as it addresses only ordinary linear regression, making CompGP a more relevant comparator. Detailed sensitivity analyses for different choices of $|\mathcal{I}|$ and $m$ are provided.

\subsection{Simulated Data Generation} 
\begin{figure}
  \subcaptionbox{\label{fig:g1}}{\includegraphics[width=0.46\textwidth]{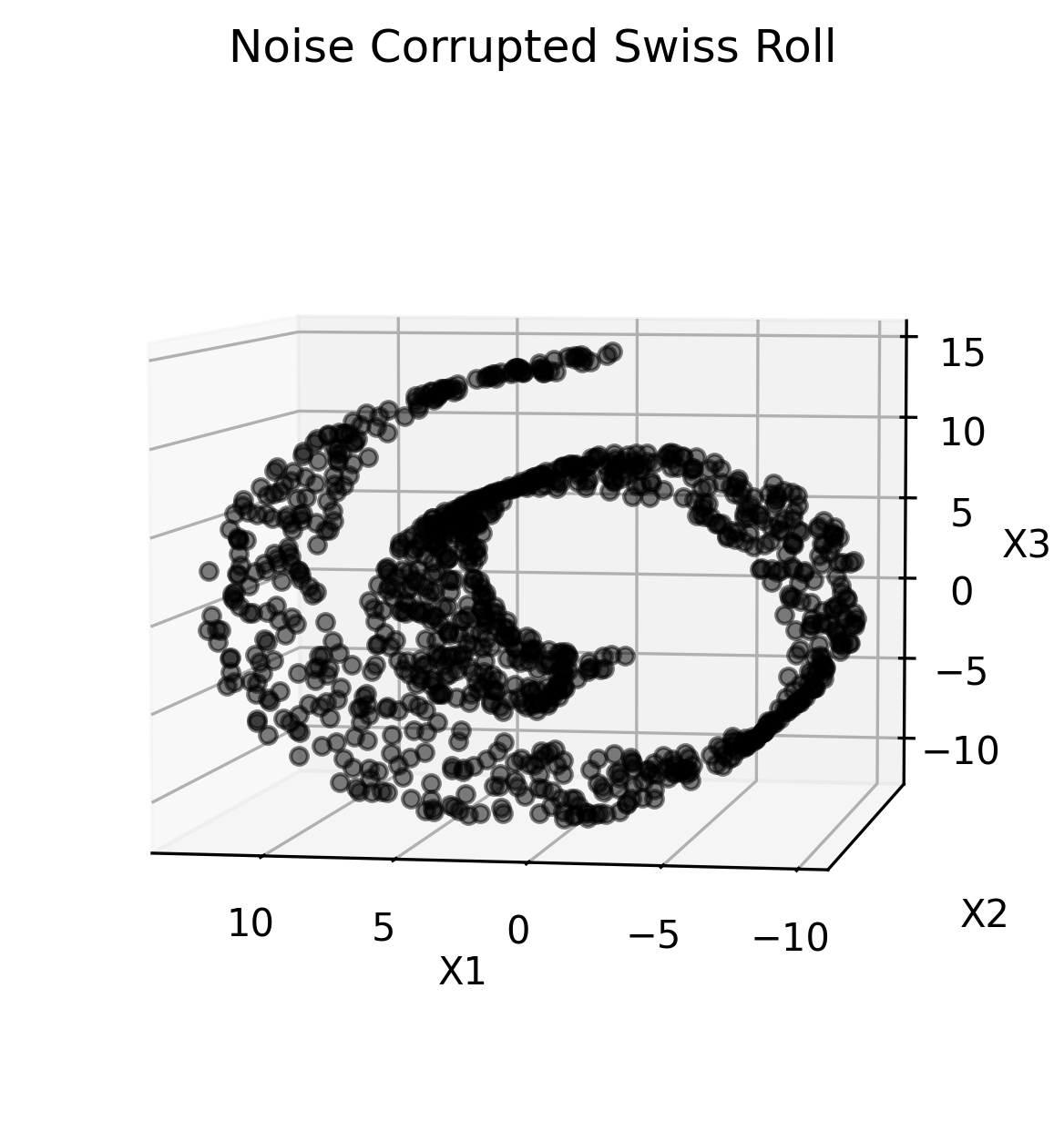}}\hfill%
  \subcaptionbox{\label{fig:g2}}{\includegraphics[width=0.46\textwidth]{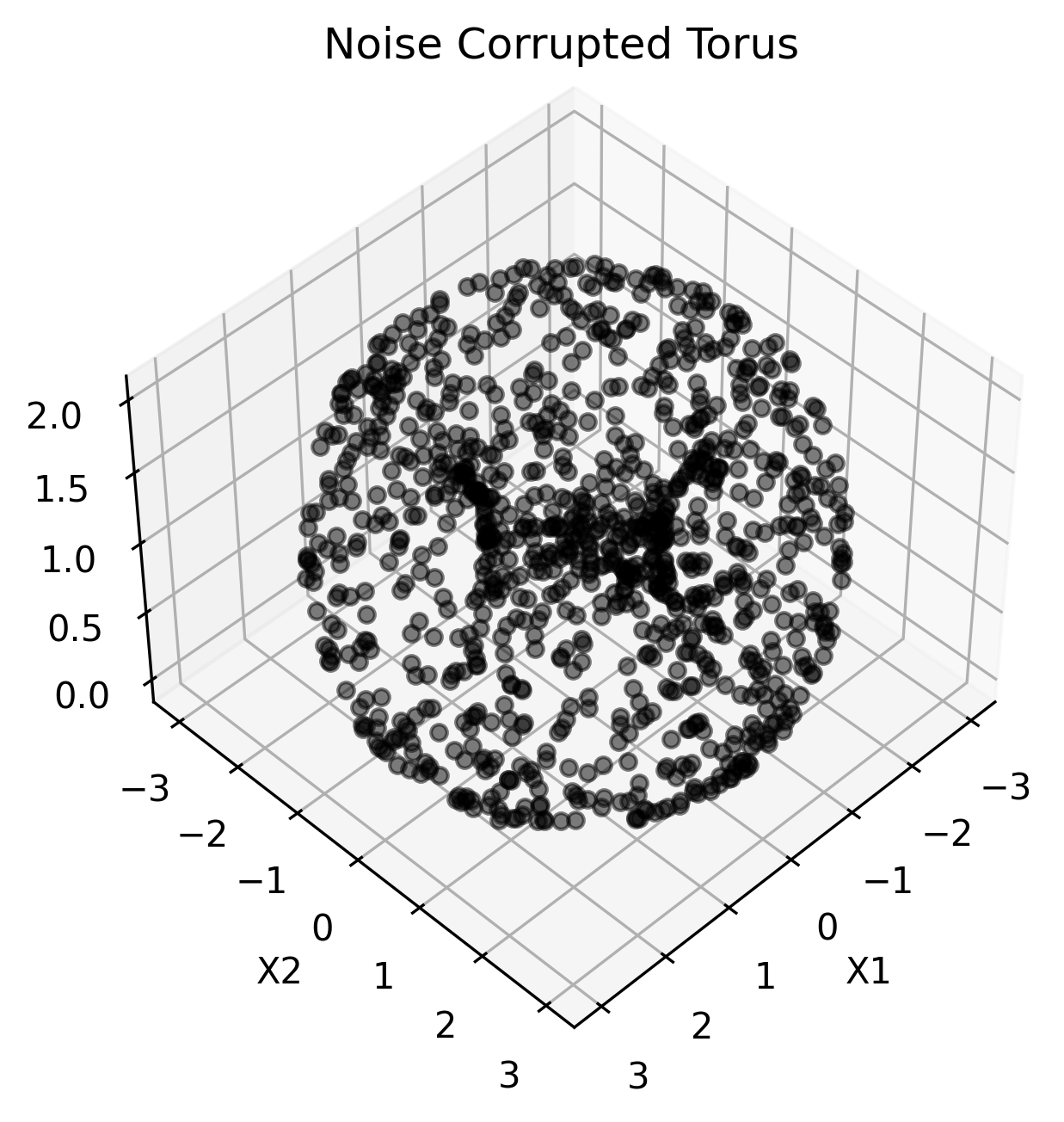}} \\
  \caption{Manifolds embedded in noisy high-dimensional predictors.}
\end{figure}

During data simulation, we explore specific scenarios where the response distribution follows a nonlinear function of $d$-dimensional coordinates for a manifold $\mathcal{O}\subseteq\mathbb{R}^p$, embedded in a high-dimensional ambient space. Two distinct choices for $\mathcal{O}$ and their corresponding response distributions are simulated.\\
\underline{\textbf{$\mathcal{O}$ is a swiss roll and $d=2$.}} For the swiss roll, we sample manifold coordinates, $o_1 \sim U(\frac{3\pi}{2}, \frac{9\pi}{2})$, $o_2\sim U(0, 3)$. A high dimensional predictor $\bx = (x_1, \ldots, x_p)$ is simulated according to $x_1 = o_1 \cos(o_1) + \eta_1,\; x_2 = o_2+\eta_2,\; x_3 = o_1 \sin(o_1) + \eta_3, \; x_i = \eta_i, \; i \geq 4$. The response $y$ have a non-linear relationship with these predictors and is simulated following,
\begin{align}
 y = \sin(0.5\pi o_1) + o_2^2 + \epsilon,\:\:\epsilon \sim N(0,0.02^2),
\end{align}

where $\eta_1, \ldots, \eta_p \sim N(0, \tau^2)$. Notably, $\bx$ and $y$ are conditionally independent given $o_1,o_2$ which is the low-dimensional signal manifold. In particular, $\bx$ lives on a (noise corrupted) swiss roll embedded in a $p$-dimensional ambient space (see Figure~\ref{fig:g1}), but y is only a function of coordinates along
the swiss roll $\mathcal{O}$.\\
\underline{\textbf{$\mathcal{O}$ is a torus and $d=3$.}} For the torus, we consider $x_1=o_1+\eta_1, x_2=o_2+\eta_2$ and $x_3=o_3+\eta_3$ where $o_1,o_2,o_3$ lie on a three dimensional torus with interior radius $1$ and exterior radius $3$ (see Figure~\ref{fig:g2}), such that $(3-\sqrt{o_1^2+o_2^2})^2+o_3^2=1$, and set $x_i=\eta_i$ for $i\geq 4$. The predictor noise $\eta_1,...,\eta_n$ are generated i.i.d. from $N(0, \tau^2)$. The response is generated as,
\begin{align}
y = o_{2}^2 + \sin(0.5\pi o_{3}) + \epsilon,\; \epsilon \sim N(0, 0.1^2).
\end{align}

The geodesic distance between two points on both a swiss roll and a torus can substantially differ from their Euclidean distance in the ambient space $\mathbb{R}^p$. The swiss roll, in particular, poses a challenging setup for SkGP, as points on $\mathcal{O}$ that are close in a Euclidean sense can be quite far in a geodesic sense.

To assess the impact of the number of predictors ($p$) and noise levels of the predictors ($\tau^2$) on the performance of the competitors, various simulation scenarios are considered by varying $p=2000, 10000$
and $\tau^2=0.01,0.03,0.05,0.1$. For each of these simulation scenarios, 50 datasets are generated, and metrics such as mean squared prediction error (MSPE), coverage, and lengths of 95\% predictive intervals (PI) are calculated across all replicates. All simulations use $n=500$ training samples and $n_{new}=100$ test samples for prediction. Our focus is on computational challenges when the
number of predictors $p$ is large, which is why $n$ is kept moderate. However, our approach can
be efficiently scaled to larger $n$ using modern GP approximations, and we offer a discussion on this in Section~\ref{sec:conclusion}.

\subsection{Point prediction}
Tables~\ref{MSPE_swissroll} and \ref{MSPE_torus} display the MSPE averaged over $50$ replications for all the competing methods in the swiss roll and torus examples, respectively. Values in parentheses represent the standard error of MSPE over $50$ replicates.

Both Tables~\ref{MSPE_swissroll} and \ref{MSPE_torus} show that incorporating randomly sketched predictors into the GP model within the SkGP framework yields strong predictive performance, significantly surpassing the performance of the neural network. For both $p=2000$ and $p=10000$, when the manifold is affected by low noise, SkGP significantly outperforms GP, BART, and RF with pre-screened but unsketched predictors. While SkBART emerges as the second-best performer in scenarios with very low noise in the manifold ($\tau^2=0.01$), its performance declines notably with an increase in the noise level in the predictors.
In comparison, SkGP effectively mitigates the impact of noise in the predictors, but there exists a tipping point (depending on the structure of the underlying manifold $\mathcal{O}$ and sample size $n$) where noise distorts the manifold excessively, causing SkGP to perform similarly to other competitors. This is observed in the MSPE values corresponding to $\tau^2=0.1$. Among the sketched competitors, SkRF and CompGP exhibit notably inferior performance compared to both SkGP and SkBART in all simulation examples. While theoretically the performance of SkGP should remain similar for both $p=2000$ and $p=10000$ when all predictors lie exactly on a low-dimensional manifold, in practice, we observe a significant decline in the performance of SkGP with an increase in $p$. This is attributed to the substantial impact of noise corruption on the manifold, influencing predictive performance.

\begin{table}[h!]
\begin{tabular}{|ll|llll|}
\hline
Swiss Roll &      & \multicolumn{4}{c|}{Noise} \\ \cline{3-6}
          &      & \multicolumn{1}{l|}{0.01} & \multicolumn{1}{l|}{0.03} & \multicolumn{1}{l|}{0.05} & 0.10 \\ \hline
\multicolumn{1}{|l|}{p = 2000} & SkGP   & \multicolumn{1}{l|}{\textbf{0.24 (0.035)}} & \multicolumn{1}{l|}{\textbf{0.40 (0.034)}} & \multicolumn{1}{l|}{0.65 (0.038)} & 1.45 (0.035) \\
\multicolumn{1}{|l|}{}         & GP     & \multicolumn{1}{l|}{1.06 (0.051)} & \multicolumn{1}{l|}{1.19 (0.041)} & \multicolumn{1}{l|}{1.35 (0.050)} & 2.06 (0.036) \\
\multicolumn{1}{|l|}{}         & CompGP & \multicolumn{1}{l|}{1.22 (0.047)} & \multicolumn{1}{l|}{2.13 (0.063)} & \multicolumn{1}{l|}{3.33 (0.088)} & 5.42 (0.110) \\
\multicolumn{1}{|l|}{}         & BART   & \multicolumn{1}{l|}{0.39 (0.014)} & \multicolumn{1}{l|}{0.41 (0.017)} & \multicolumn{1}{l|}{\textbf{0.46 (0.022)}} & \textbf{0.55 (0.017)} \\
\multicolumn{1}{|l|}{}         & SkBART & \multicolumn{1}{l|}{0.36 (0.025)} & \multicolumn{1}{l|}{1.01 (0.033)} & \multicolumn{1}{l|}{1.90 (0.056)} & 3.92 (0.094) \\
\multicolumn{1}{|l|}{}         & RF     & \multicolumn{1}{l|}{5.99 (0.123)} & \multicolumn{1}{l|}{6.33 (0.128)} & \multicolumn{1}{l|}{6.15 (0.143)} & 6.23 (0.084) \\
\multicolumn{1}{|l|}{}         & SkRF   & \multicolumn{1}{l|}{0.80 (0.056)} & \multicolumn{1}{l|}{1.87 (0.068)} & \multicolumn{1}{l|}{3.08 (0.106)} & 5.16 (0.095) \\
\multicolumn{1}{|l|}{}         & LASSO  & \multicolumn{1}{l|}{0.95 (0.020)} & \multicolumn{1}{l|}{0.98 (0.018)} & \multicolumn{1}{l|}{0.96 (0.022)} &1.09 (0.023) \\
\multicolumn{1}{|l|}{}         & BASS   & \multicolumn{1}{l|}{2.83 (0.440)} & \multicolumn{1}{l|}{2.20 (0.390)} & \multicolumn{1}{l|}{2.12 (0.401)} & 2.85 (0.449) \\
\multicolumn{1}{|l|}{}         & NN     & \multicolumn{1}{l|}{0.58 (0.019)} & \multicolumn{1}{l|}{1.17 (0.038)} & \multicolumn{1}{l|}{1.90 (0.051)} & 4.17 (0.100) \\ \hline

\multicolumn{1}{|l|}{p = 10{,}000} & SkGP   & \multicolumn{1}{l|}{\textbf{0.28 (0.027)}} & \multicolumn{1}{l|}{\textbf{0.79 (0.025)}} & \multicolumn{1}{l|}{1.51 (0.043)} & 3.94 (0.064) \\
\multicolumn{1}{|l|}{}           & GP     & \multicolumn{1}{l|}{1.10 (0.030)} & \multicolumn{1}{l|}{1.61 (0.044)} & \multicolumn{1}{l|}{2.12 (0.058)} & 3.78 (0.054) \\
\multicolumn{1}{|l|}{}           & CompGP & \multicolumn{1}{l|}{1.20 (0.038)} & \multicolumn{1}{l|}{2.12 (0.057)} & \multicolumn{1}{l|}{3.30 (0.093)} & 5.47 (0.090) \\
\multicolumn{1}{|l|}{}           & BART   & \multicolumn{1}{l|}{0.84 (0.025)} & \multicolumn{1}{l|}{0.90 (0.039)} & \multicolumn{1}{l|}{\textbf{0.95 (0.034)}} & 1.11 (0.035) \\
\multicolumn{1}{|l|}{}           & SkBART & \multicolumn{1}{l|}{0.78 (0.030)} & \multicolumn{1}{l|}{2.35 (0.070)} & \multicolumn{1}{l|}{4.04 (0.098)} & 6.52 (0.112) \\
\multicolumn{1}{|l|}{}           & RF     & \multicolumn{1}{l|}{6.72 (0.129)} & \multicolumn{1}{l|}{7.13 (0.139)} & \multicolumn{1}{l|}{6.94 (0.146)} & 7.05 (0.094) \\
\multicolumn{1}{|l|}{}           & SkRF   & \multicolumn{1}{l|}{1.53 (0.056)} & \multicolumn{1}{l|}{3.85 (0.106)} & \multicolumn{1}{l|}{5.30 (0.127)} & 6.89 (0.101) \\
\multicolumn{1}{|l|}{}           & LASSO  & \multicolumn{1}{l|}{0.94 (0.020)} & \multicolumn{1}{l|}{0.97 (0.018)} & \multicolumn{1}{l|}{0.96 (0.022)} & \textbf{1.08 (0.021)} \\
\multicolumn{1}{|l|}{}           & BASS   & \multicolumn{1}{l|}{5.72 (0.419)} & \multicolumn{1}{l|}{6.27 (0.457)} & \multicolumn{1}{l|}{6.07 (0.447)} & 5.78 (0.450) \\
\multicolumn{1}{|l|}{}           & NN     & \multicolumn{1}{l|}{1.05 (0.024)} & \multicolumn{1}{l|}{3.00 (0.112)} & \multicolumn{1}{l|}{5.06 (0.134)} & 8.31 (0.193) \\ \hline
\end{tabular}
\caption{Averaged Mean squared Prediction Error (MSPE) over 50 replications are shown for the competing models in Swiss Roll example. Standard errors are presented within parenthesis.}
\label{MSPE_swissroll}
\end{table}

\begin{table}[h!]
\begin{tabular}{|ll|llll|}
\hline
Torus &      & \multicolumn{4}{c|}{Noise} \\ \cline{3-6}
          &      & \multicolumn{1}{l|}{0.01} & \multicolumn{1}{l|}{0.03} & \multicolumn{1}{l|}{0.05} & 0.10 \\ \hline
\multicolumn{1}{|l|}{p = 2000} & SkGP   & \multicolumn{1}{l|}{\textbf{0.05 (0.005)}} & \multicolumn{1}{l|}{\textbf{0.08 (0.005)}} & \multicolumn{1}{l|}{0.12 (0.008)} & 0.25 (0.009) \\
\multicolumn{1}{|l|}{}         & GP     & \multicolumn{1}{l|}{0.06 (0.006)} & \multicolumn{1}{l|}{0.09 (0.006)} & \multicolumn{1}{l|}{\textbf{0.09 (0.009)}} & 0.40 (0.009) \\
\multicolumn{1}{|l|}{}         & CompGP & \multicolumn{1}{l|}{0.07 (0.006)} & \multicolumn{1}{l|}{0.26 (0.009)} & \multicolumn{1}{l|}{0.57 (0.015)} & 0.92 (0.019) \\
\multicolumn{1}{|l|}{}         & BART   & \multicolumn{1}{l|}{0.08 (0.007)} & \multicolumn{1}{l|}{0.10 (0.007)} & \multicolumn{1}{l|}{\textbf{0.09 (0.006)}} & \textbf{0.09 (0.005)} \\
\multicolumn{1}{|l|}{}         & SkBART & \multicolumn{1}{l|}{0.06 (0.007)} & \multicolumn{1}{l|}{0.15 (0.008)} & \multicolumn{1}{l|}{0.27 (0.011)} & 0.76 (0.020) \\
\multicolumn{1}{|l|}{}         & RF     & \multicolumn{1}{l|}{0.85 (0.021)} & \multicolumn{1}{l|}{0.91 (0.016)} & \multicolumn{1}{l|}{0.87 (0.018)} & 0.85 (0.019) \\
\multicolumn{1}{|l|}{}         & SkRF   & \multicolumn{1}{l|}{0.07 (0.006)} & \multicolumn{1}{l|}{0.21 (0.009)} & \multicolumn{1}{l|}{0.38 (0.013)} & 0.74 (0.018) \\
\multicolumn{1}{|l|}{}         & LASSO  & \multicolumn{1}{l|}{1.19 (0.031)} & \multicolumn{1}{l|}{1.27 (0.022)} & \multicolumn{1}{l|}{1.23 (0.023)} & 1.23 (0.026) \\
\multicolumn{1}{|l|}{}         & BASS   & \multicolumn{1}{l|}{0.75 (0.050)} & \multicolumn{1}{l|}{0.65 (0.054)} & \multicolumn{1}{l|}{0.65 (0.058)} & 0.61 (0.054) \\
\multicolumn{1}{|l|}{}         & NN     & \multicolumn{1}{l|}{0.08 (0.007)} & \multicolumn{1}{l|}{0.23 (0.010)} & \multicolumn{1}{l|}{0.38 (0.011)} & 1.01 (0.028) \\ \hline

\multicolumn{1}{|l|}{p = 10{,}000} & SkGP   & \multicolumn{1}{l|}{\textbf{0.05 (0.005)}} & \multicolumn{1}{l|}{\textbf{0.13 (0.006)}} & \multicolumn{1}{l|}{\textbf{0.32 (0.010)}} & 0.95 (0.021) \\
\multicolumn{1}{|l|}{}           & GP     & \multicolumn{1}{l|}{0.06 (0.006)} & \multicolumn{1}{l|}{0.16 (0.007)} & \multicolumn{1}{l|}{0.48 (0.012)} & 0.88 (0.019) \\
\multicolumn{1}{|l|}{}           & CompGP & \multicolumn{1}{l|}{0.08 (0.006)} & \multicolumn{1}{l|}{0.26 (0.009)} & \multicolumn{1}{l|}{0.56 (0.014)} & 0.93 (0.020) \\
\multicolumn{1}{|l|}{}           & BART   & \multicolumn{1}{l|}{0.36 (0.018)} & \multicolumn{1}{l|}{0.40 (0.013)} & \multicolumn{1}{l|}{0.39 (0.015)} & \textbf{0.38 (0.016)} \\
\multicolumn{1}{|l|}{}           & SkBART & \multicolumn{1}{l|}{0.10 (0.006)} & \multicolumn{1}{l|}{0.45 (0.015)} & \multicolumn{1}{l|}{0.85 (0.023)} & 0.99 (0.021) \\
\multicolumn{1}{|l|}{}           & RF     & \multicolumn{1}{l|}{0.92 (0.022)} & \multicolumn{1}{l|}{0.97 (0.017)} & \multicolumn{1}{l|}{0.95 (0.019)} & 0.92 (0.020) \\
\multicolumn{1}{|l|}{}           & SkRF   & \multicolumn{1}{l|}{0.13 (0.006)} & \multicolumn{1}{l|}{0.54 (0.013)} & \multicolumn{1}{l|}{0.81 (0.020)} & 0.95 (0.021) \\
\multicolumn{1}{|l|}{}           & LASSO  & \multicolumn{1}{l|}{0.89 (0.023)} & \multicolumn{1}{l|}{0.96 (0.016)} & \multicolumn{1}{l|}{0.92 (0.020)} & 0.90 (0.021) \\
\multicolumn{1}{|l|}{}           & BASS   & \multicolumn{1}{l|}{0.88 (0.032)} & \multicolumn{1}{l|}{0.97 (0.034)} & \multicolumn{1}{l|}{0.93 (0.028)} & 0.89 (0.035) \\
\multicolumn{1}{|l|}{}           & NN     & \multicolumn{1}{l|}{0.19 (0.009)} & \multicolumn{1}{l|}{0.72 (0.024)} & \multicolumn{1}{l|}{1.08 (0.022)} & 1.27 (0.026) \\ \hline
\end{tabular}
\caption{Averaged Mean squared Prediction Error (MSPE) over 50 replications are shown for the competing models in Torus example. Standard errors are presented within parenthesis.}
\label{MSPE_torus}
\end{table}
\newpage
\subsection{Predictive Uncertainty}
\begin{figure}[h!]
\centering
\subfloat[][$\tau^2 = 0.01$]{\includegraphics[height=4.5cm]{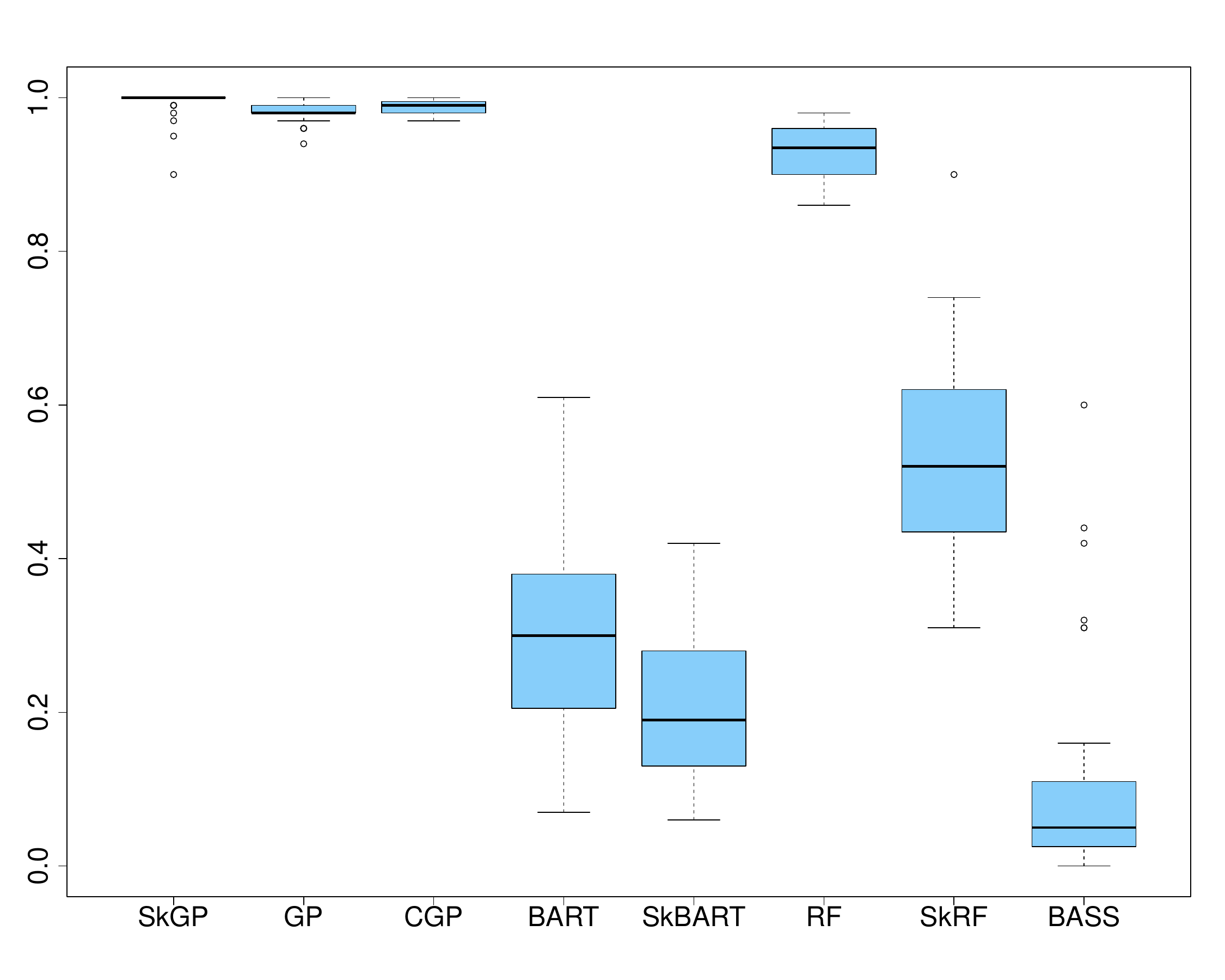}}
\subfloat[][$\tau^2 = 0.03$]{\includegraphics[height=4.5cm]{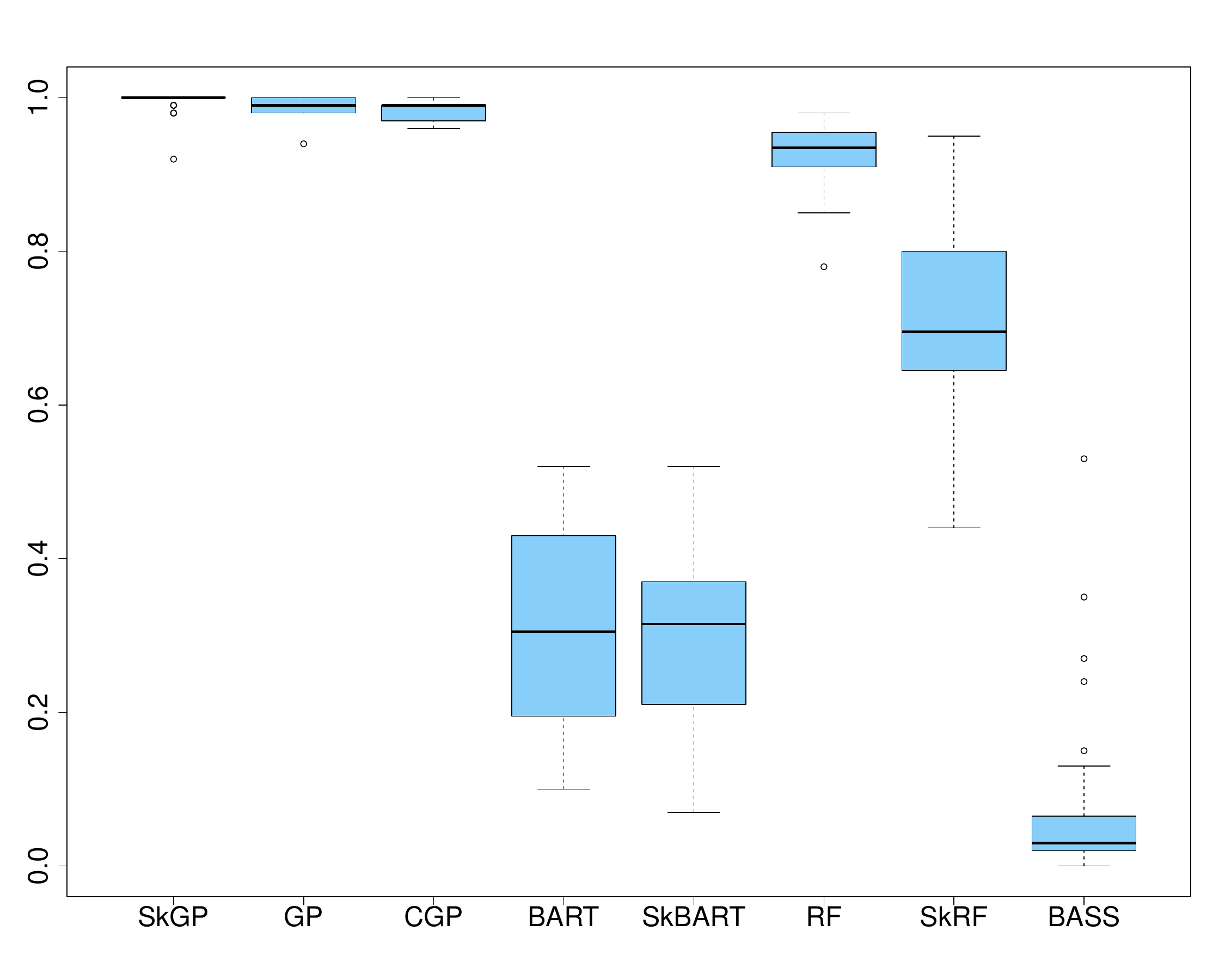}}\par\medskip
\subfloat[][$\tau^2 = 0.05$]{\includegraphics[height=4.5cm]{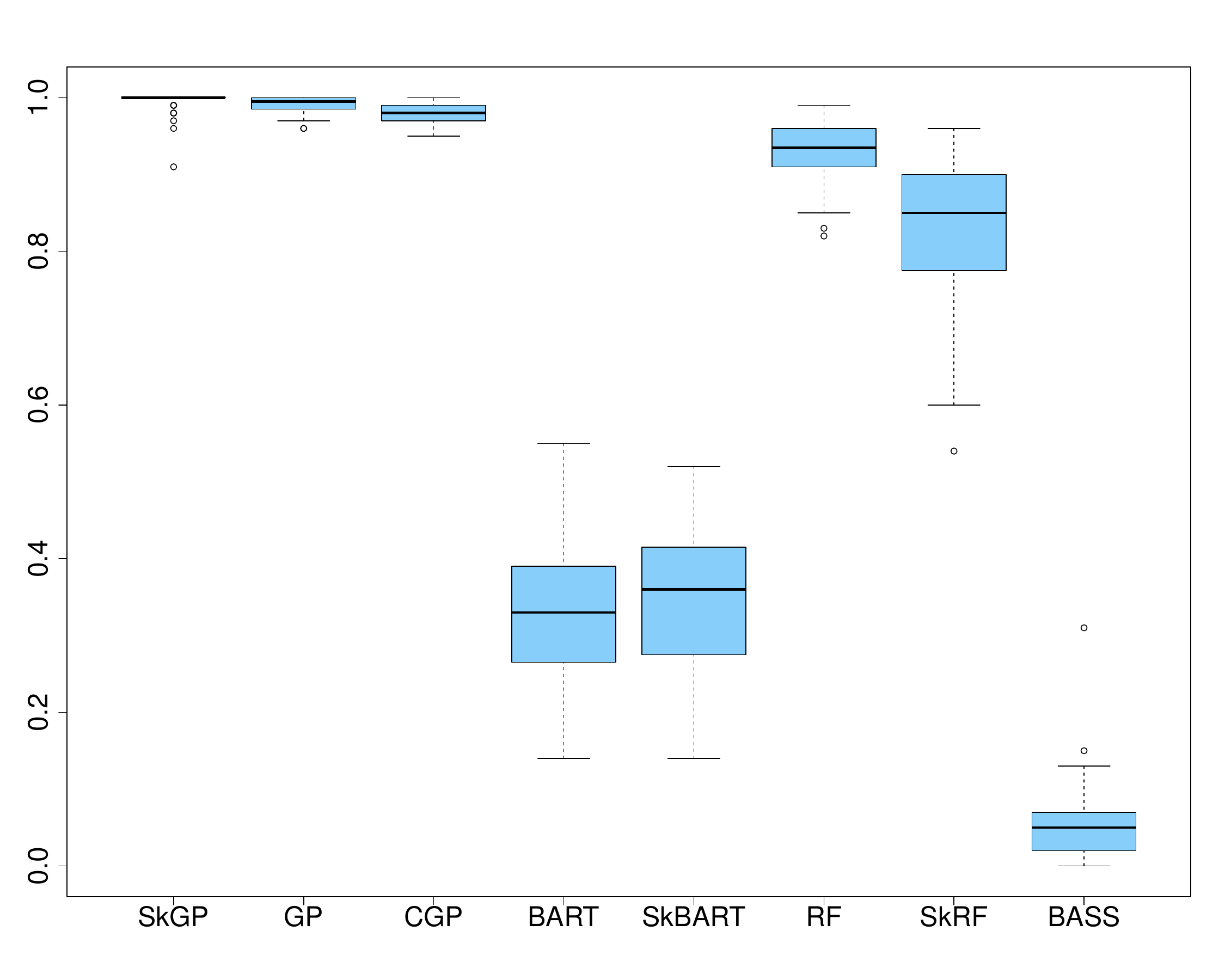}}
\subfloat[][$\tau^2 = 0.1$]{\includegraphics[height = 4.5cm]{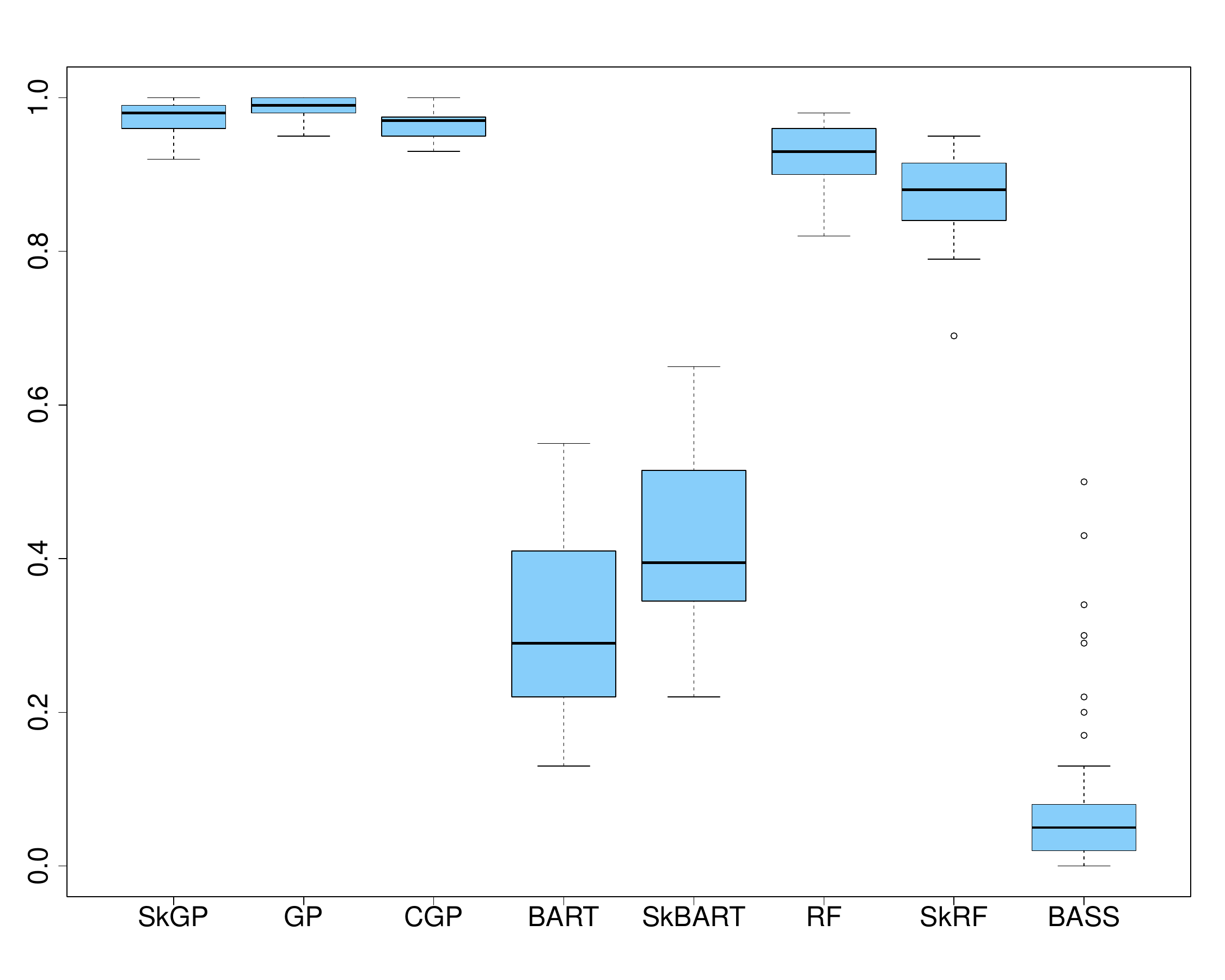}}
\caption{Coverage of 95\% predictive interval for Swiss Roll Simulations for different noise levels $\tau^2$ in the manifold containing predictors in the case of $p=2000$.}\label{fig:swiss_roll_coverage}
\end{figure}

\begin{figure}[h!]
\centering
\subfloat[][$\tau^2 = 0.01$]{\includegraphics[height=4.5cm]{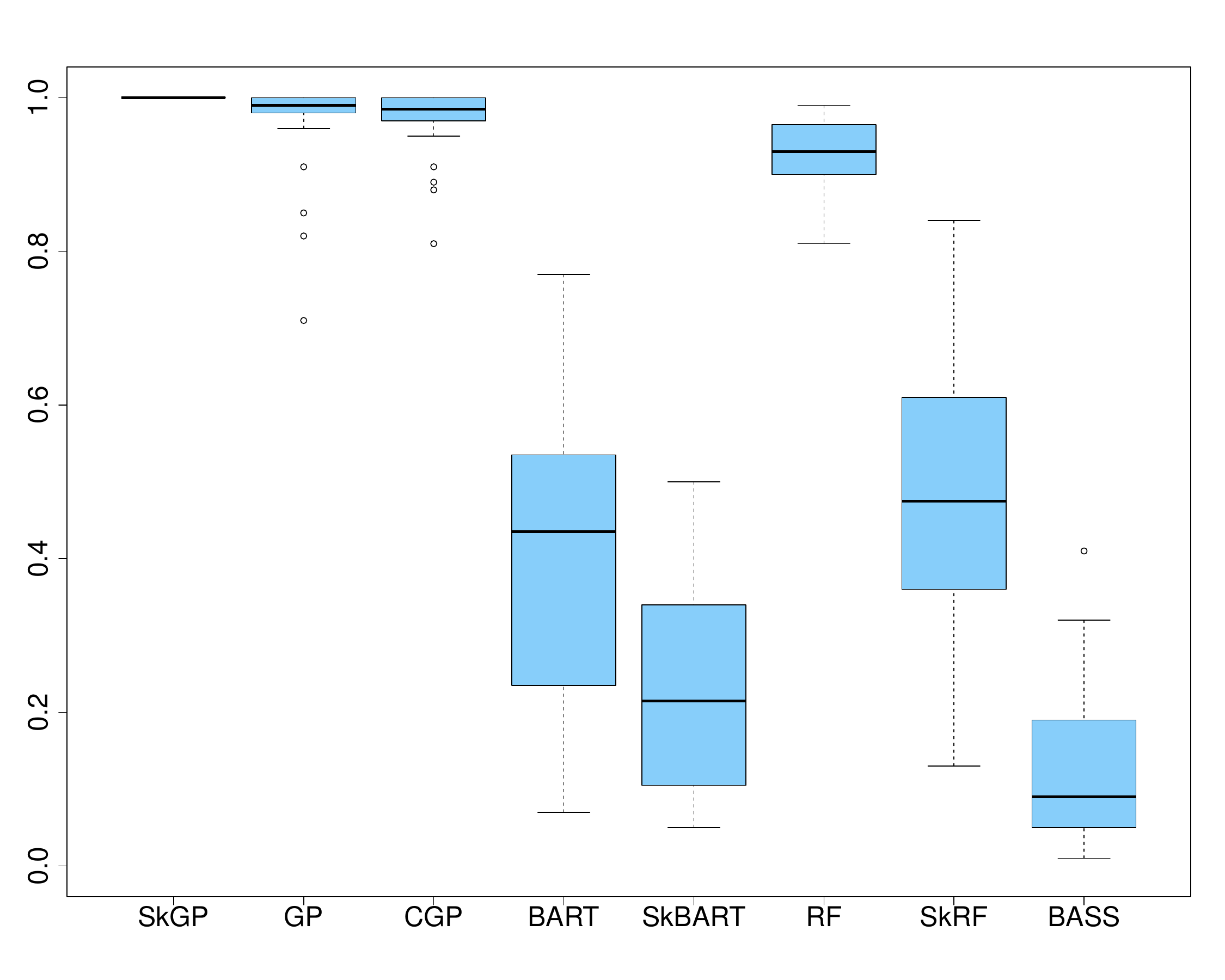}}
\subfloat[][$\tau^2 = 0.03$]{\includegraphics[height=4.5cm]{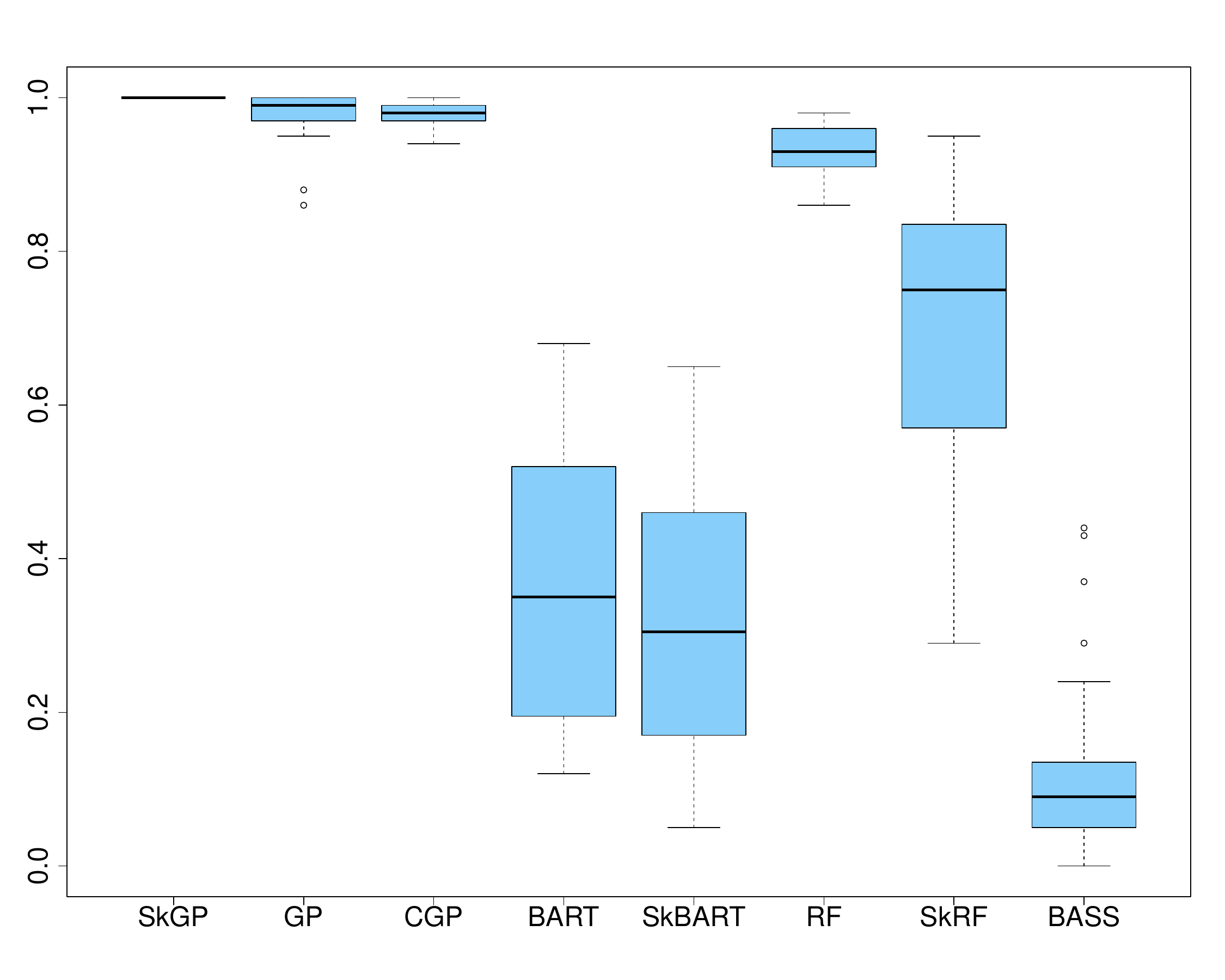}}\par\medskip
\subfloat[][$\tau^2 = 0.05$]{\includegraphics[height=4.5cm]{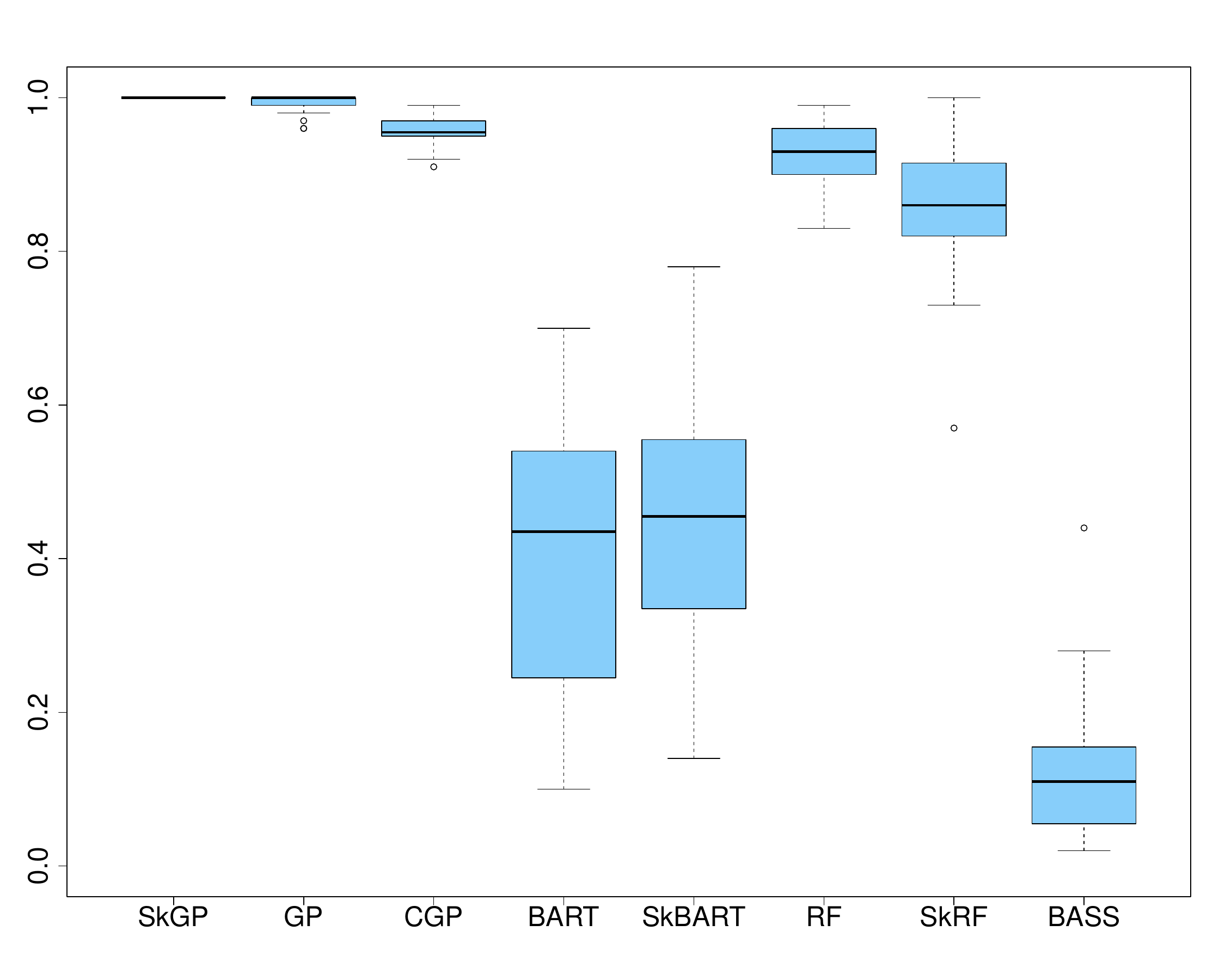}}
\subfloat[][$\tau^2 = 0.1$]{\includegraphics[height = 4.5cm]{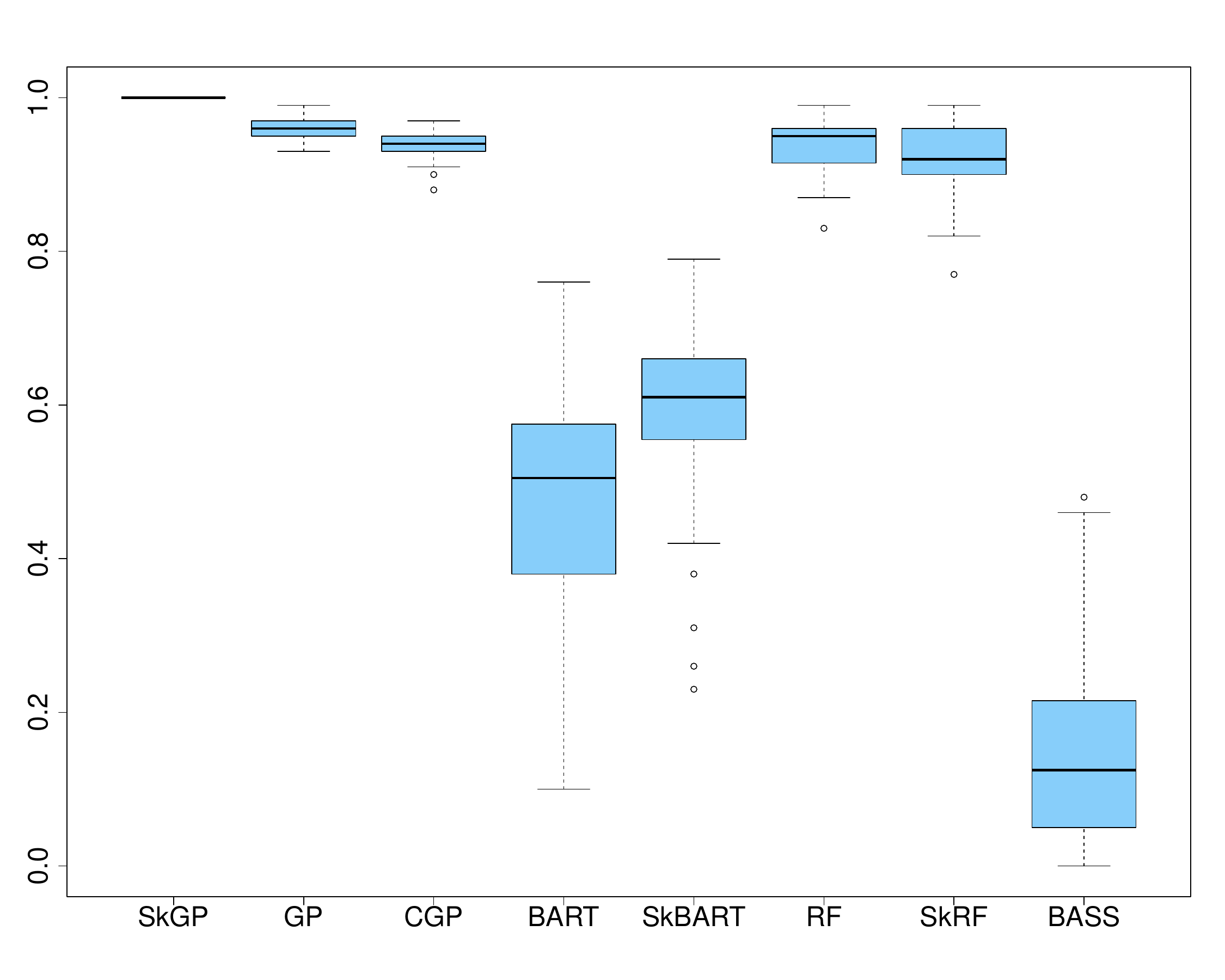}}
\caption{Coverage of 95\% predictive interval for Torus Simulations for different noise levels $\tau^2$ in the manifold containing predictors in the case of $p=2000$.}\label{fig:torus_coverage}
\end{figure}

\begin{figure}[h!]
  \subcaptionbox{Swiss Roll, $p = 2000$}{\includegraphics[width=0.4\textwidth]{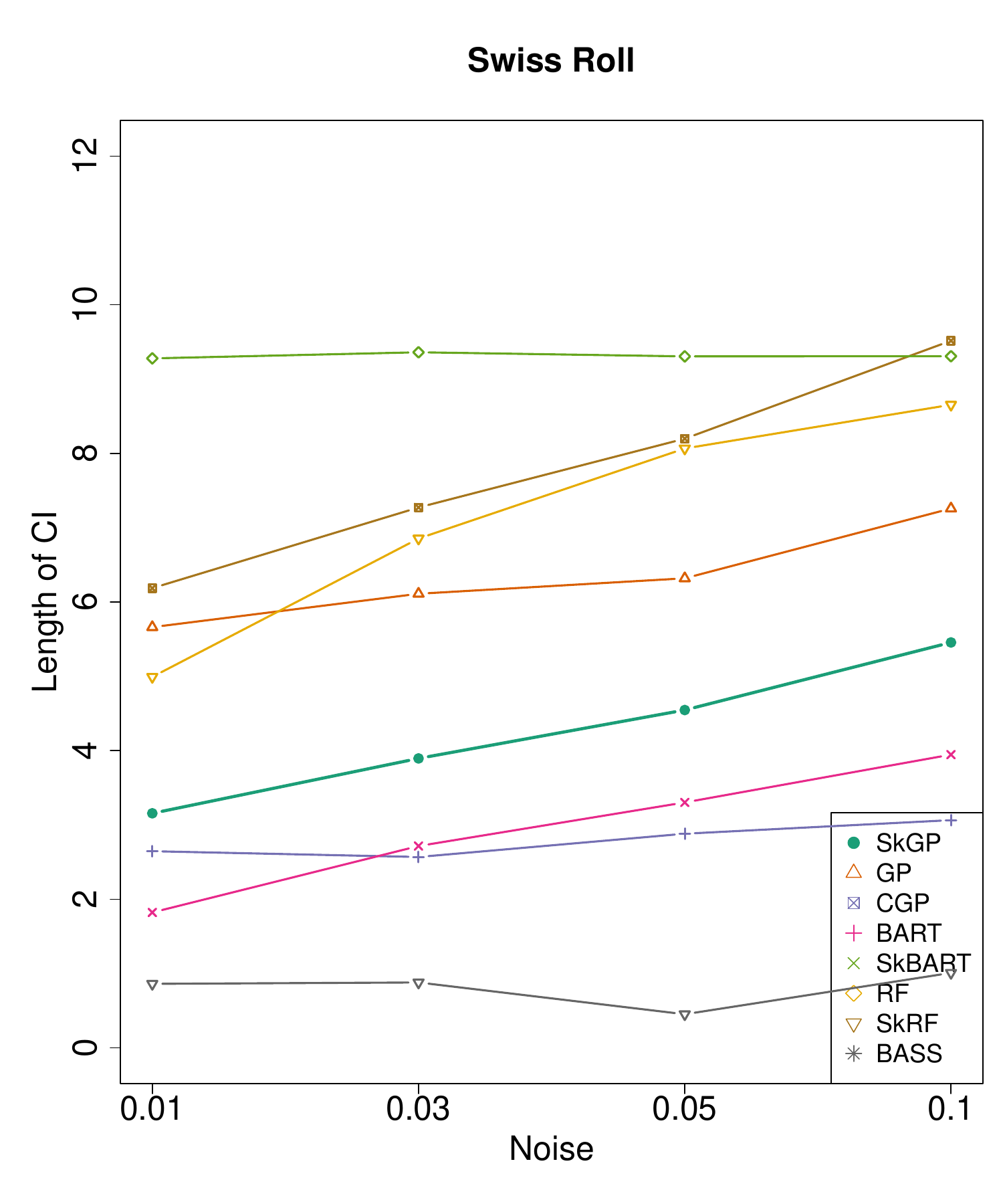}}\hfill%
  \subcaptionbox{Torus, $p = 2000$}{\includegraphics[width=0.4\textwidth]{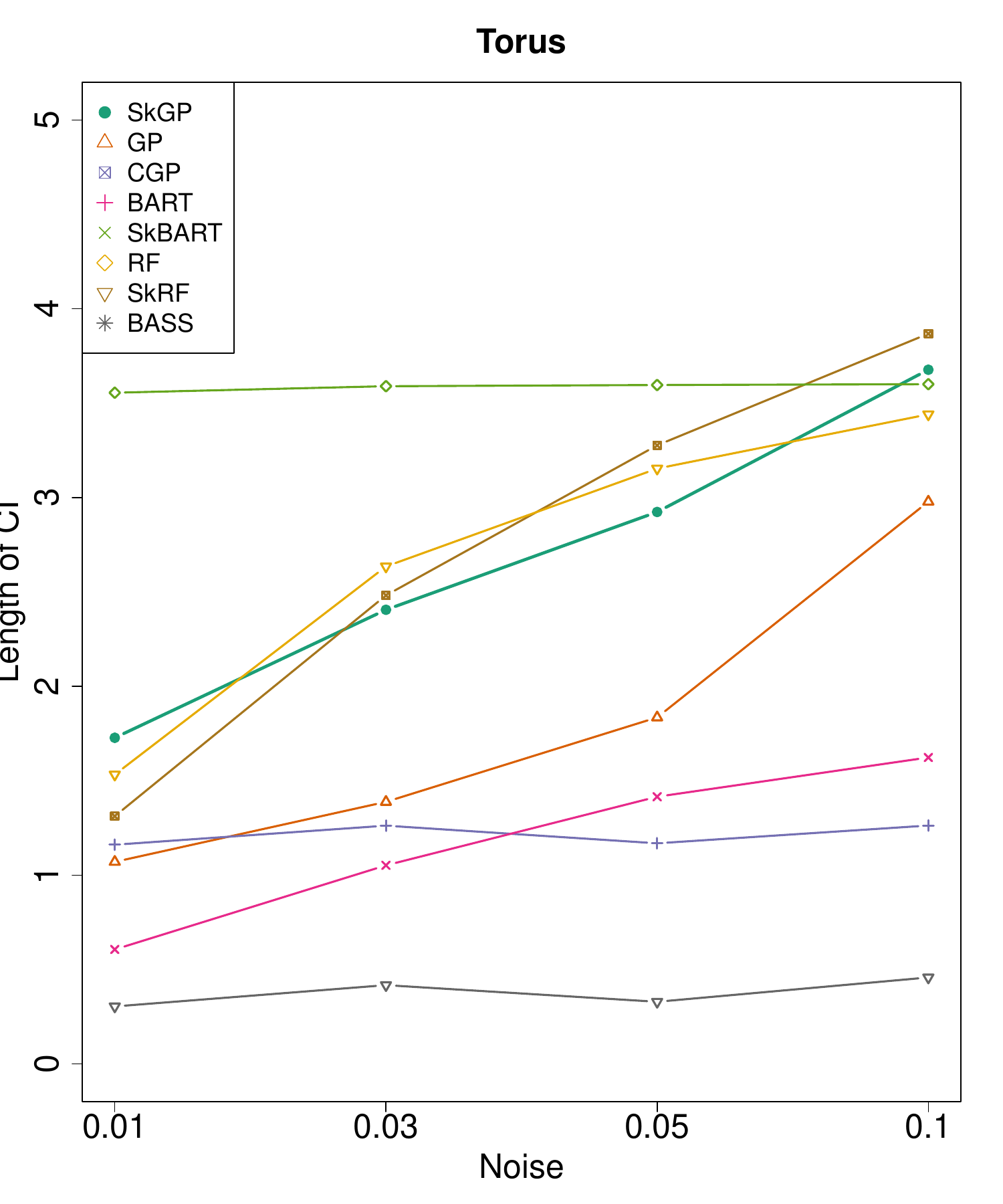}} \\
  \caption{Length of 95\% predictive intervals for all competitors in both swiss roll and torus simulations for $p=2000$.}\label{fig:length_PI}
\end{figure}

To evaluate quality of predictive uncertainty, we calculate the coverage and length of 95\% predictive intervals (PI) for SkGP and competitors. While frequentist methods, like SkRF and RF, do not inherently provide coverage probabilities with point estimates, we employ a two-stage plug-in approach for them: (i) estimate the regression function in the first stage, and (ii) construct 95\% PI based on the normal distribution centered on the predictive mean from the regression model, with variance equal to the estimated variance in the residuals. Coverage probability boxplots over $50$ replications for $p=2000$ in the swiss roll example and torus example are presented in Figures~\ref{fig:swiss_roll_coverage} and \ref{fig:torus_coverage}, respectively.
Figure~\ref{fig:length_PI} displays the median lengths of the 95\% PI for all competitors for $p=2000$ in both the swiss roll and torus examples.
Results for predictive uncertainties for $p=10000$ are presented in Section 1 of the supplementary file due to space constraint.

Across all simulations, SkGP's 95\% PI coverage is near the nominal level, with intervals widening as noise increases (higher $\tau^2$). BART and SkBART show poor coverage with narrower PIs. RF and SkRF also exhibit undercoverage ($\sim$ 80\%) with wider PIs than SkGP. In the swiss roll example, GP's coverage is similar to SkGP but with intervals about twice as wide. In the torus example, GP and SkGP perform similarly for $p=2000$, but for $p=10000$, SkGP produces narrower PIs than GP while maintaining similar coverage as noise increases (see Section 1 of the supplementary file). Overall, SkGP provides more precise predictive uncertainty compared to its competitors.


\subsection{Computation Time}
The main goal of developing SkGP was to improve computational scalability in large $p$ settings. For specific choices of $\bP_n$, $\theta$, and $\psi^2$, the computation time is driven by two key factors: (a) inverting an $n \times n$ matrix, which has a complexity of $n^3$, and (b) multiplying an $m \times |\mathcal{I}|$ matrix with an $|\mathcal{I}| \times n$ matrix, with a complexity of $mn|\mathcal{I}|$. Since the posterior predictive distribution is available in closed form, these computations are performed once per choice of $\bP_n$, $\theta$, and $\psi^2$. Parallelization across different CPUs will handle various parameter choices efficiently. The stacking step involves inverting $S$ matrices of dimension $(n - n/S) \times (n - n/S)$ with a complexity of $S(n - n/S)^3$. Given that the focus is on moderate $n$, all computations are highly efficient, enabling fast SkGP calculations.

\begin{figure}[h!]
  \subcaptionbox{Vary $p$, fix $m=60$\label{fig:a}}{\includegraphics[width=0.4\textwidth]{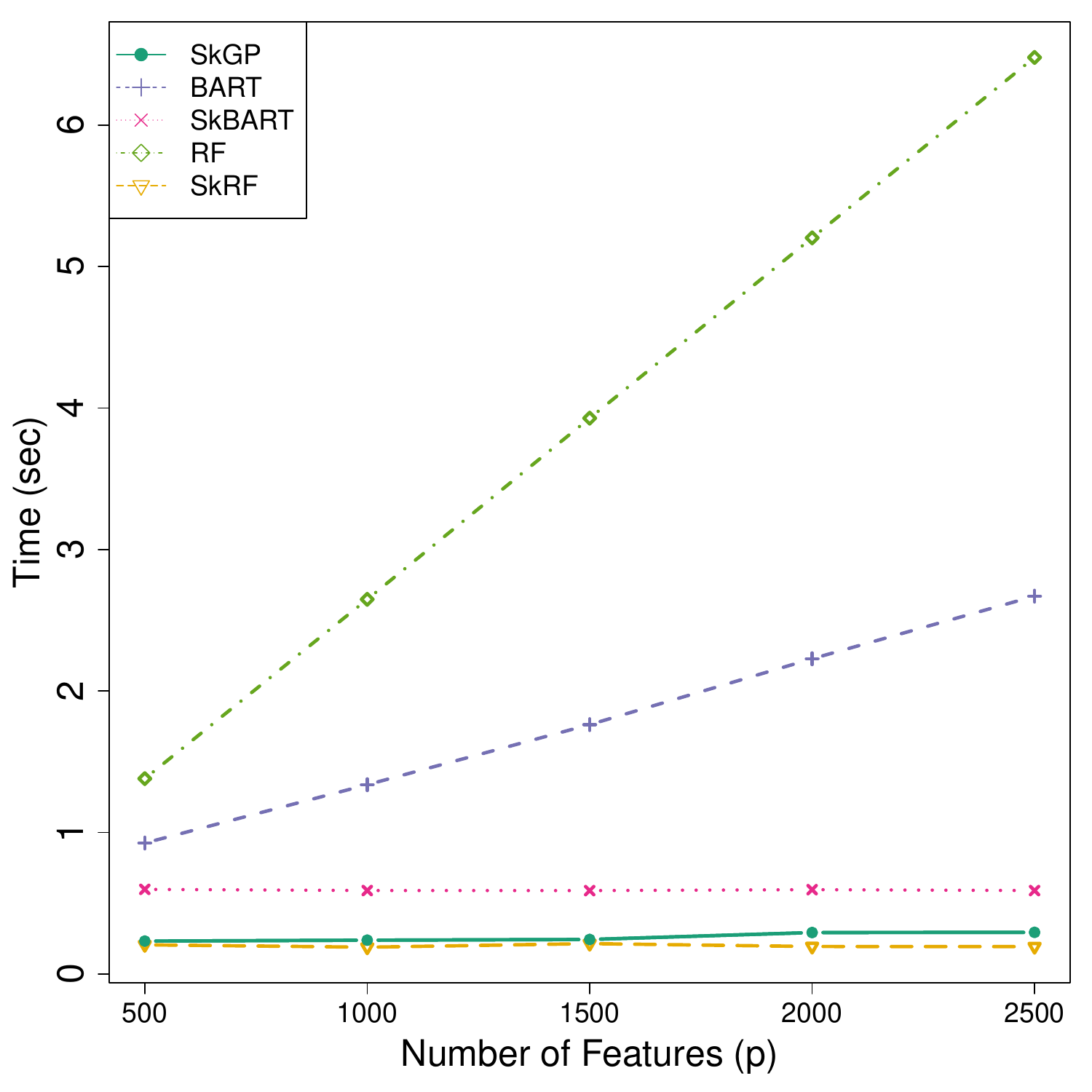}}\hfill%
  \subcaptionbox{Vary $m$, fix $|\mathcal{I}|=1000$\label{fig:b}}{\includegraphics[width=0.4\textwidth]{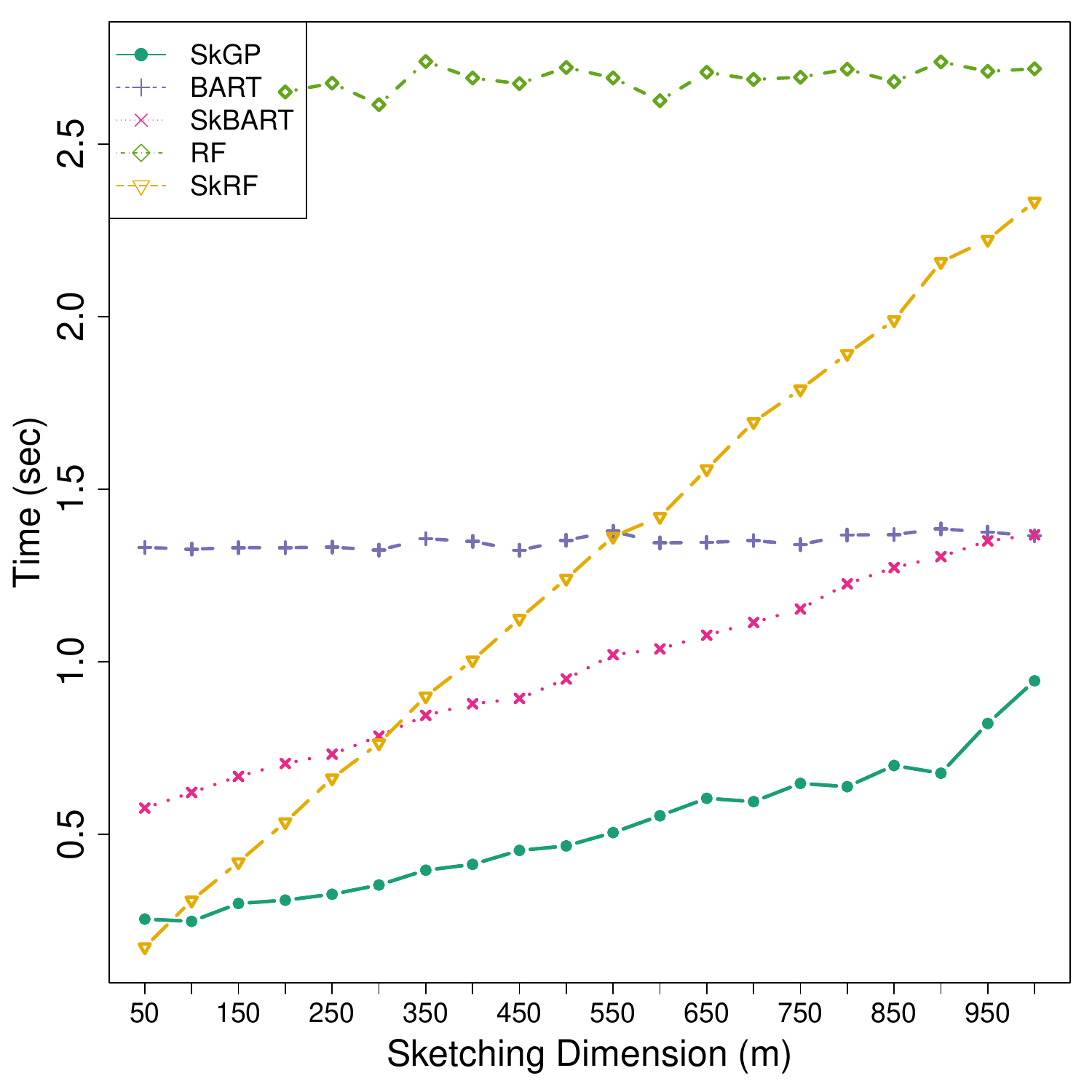}} \\
  \caption{The left panel shows the computation time for competitors by fixing the sketching dimension ($m$), while varying the number of predictors. The right panel shows the computation time for competitors by fixing the number of screened predictors $|\mathcal{I}|=1000$, while varying $m.$}
\end{figure}

Figure~\ref{fig:a} demonstrates that as the number of predictors increases with a fixed sketching dimension ($m = 60$), computation times for non-sketched methods grow linearly, while sketched methods remain constant, depending only on the number of screened predictors $|\mathcal{I}|$. Figure~\ref{fig:b} shows that as the sketching dimension increases with a fixed number of screened predictors ($|\mathcal{I}| = 1000$), computation times for sketched methods increase linearly. Importantly, SkGP achieves computation times comparable to frequentist approaches while enabling principled Bayesian predictive inference.

\subsection{Sensitivity to the choice of $m$, $|\mathcal{I}|$ and $K$}
\begin{figure}[h!]
\centering
\subfloat{\includegraphics[height=4.5cm]{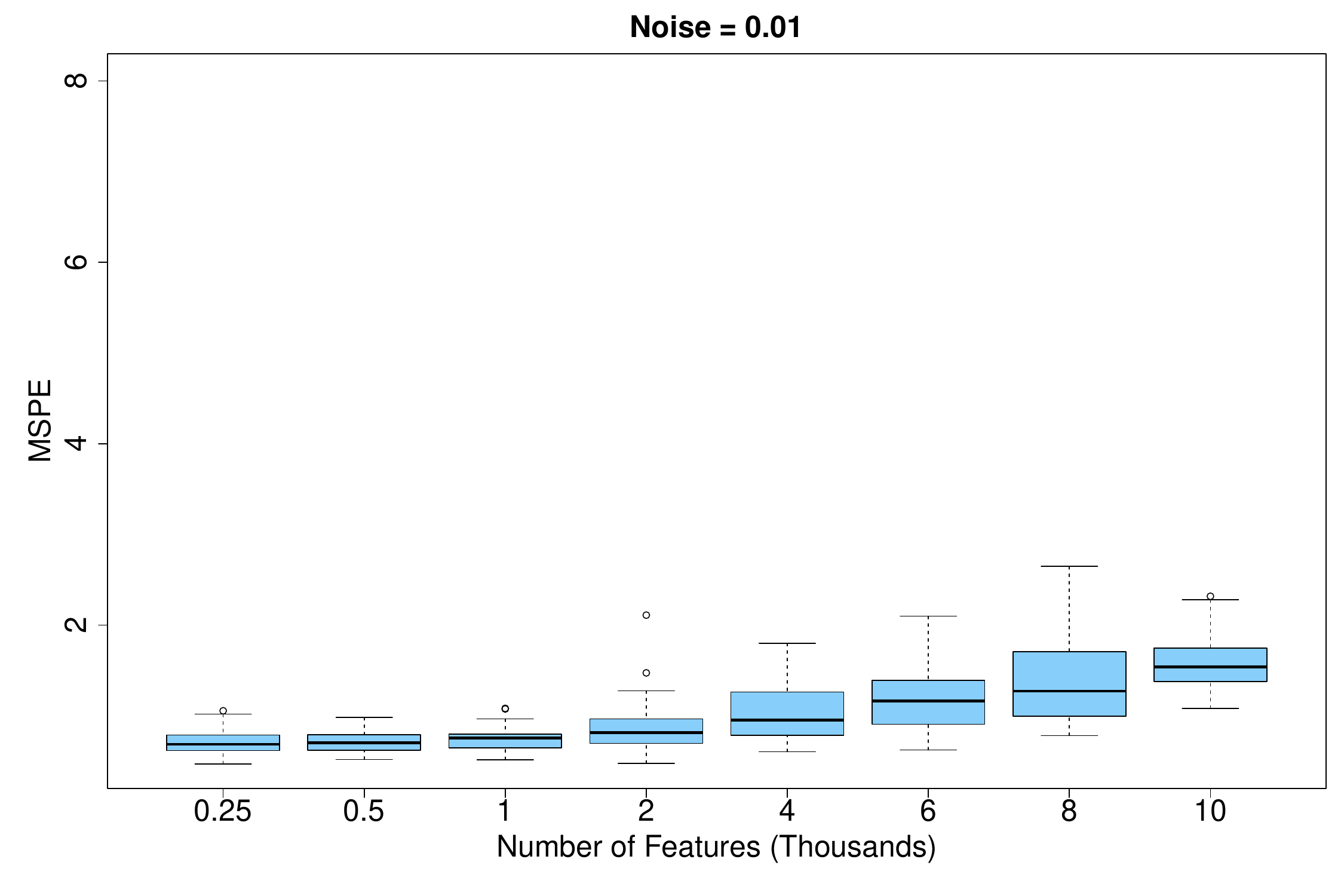}}
\subfloat{\includegraphics[height=4.5cm]{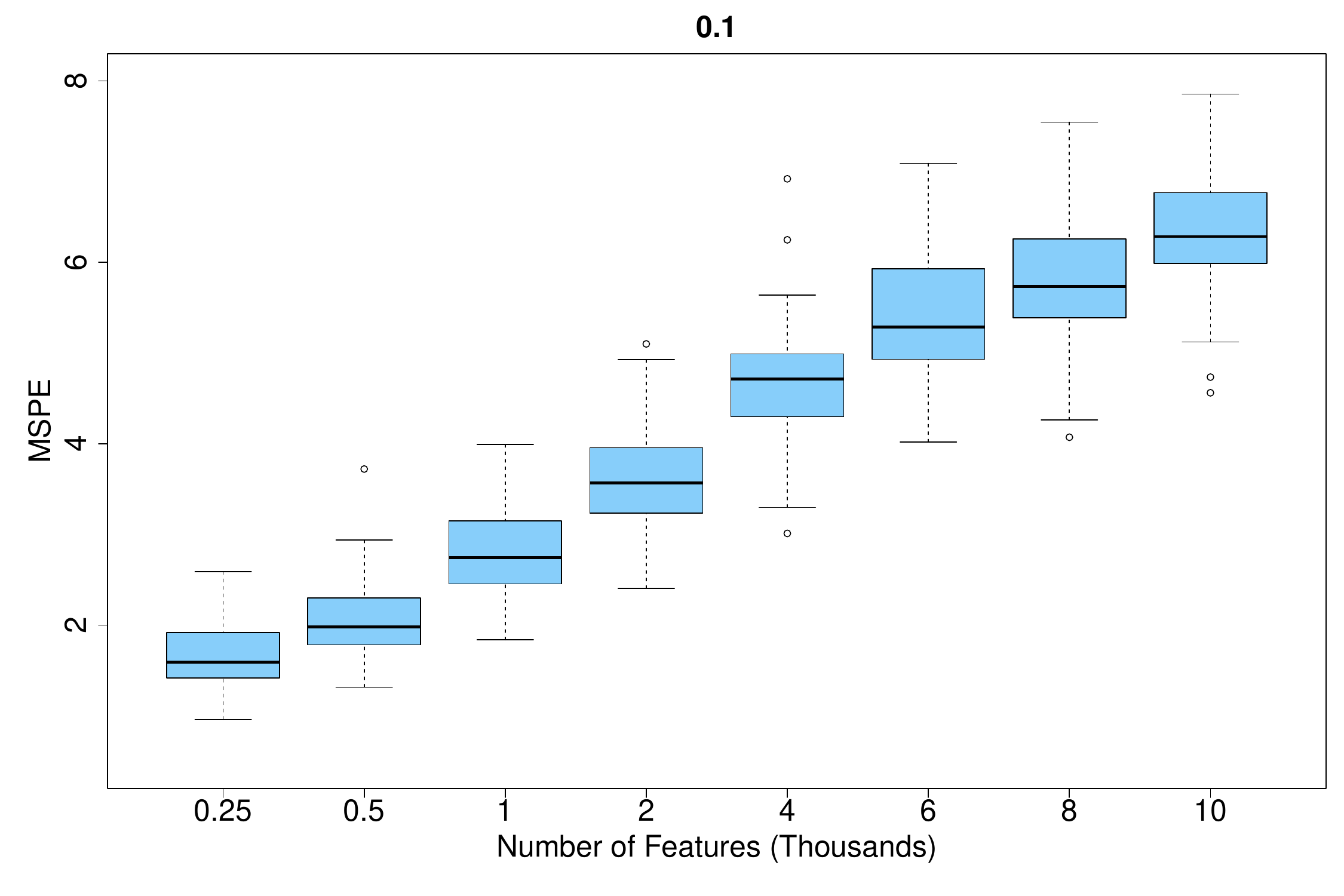}}
\caption{We show the number of included predictors $\mathcal{I}$ through the nonparametric independent screening analysis vis-a-vis predictive accuracy. The left and right panels show plots corresponding to $\tau^2=0.01$ and $\tau^2=0.1$, respectively. The plots are presented for the swiss roll example.}\label{fig:screening_full}
\end{figure}

We investigate the impact of the number of predictors $|\mathcal{I}|$ selected based on their marginal association with the response and the sketching matrix dimension $m$ applied to the $|\mathcal{I}|$-dimensional predictor vector. Figure~\ref{fig:screening_full} shows MSPE values for the swiss roll example with varying numbers of included predictors $|\mathcal{I}|$. Due to the low intrinsic dimensionality of the swiss roll manifold and the response’s dependence solely on the manifold, adding redundant predictors leads to performance degradation, as reflected by increasing MSPE, especially in high-noise settings. This confirms that estimating the regression function depends on the manifold’s dimension, not the number of predictors.

\begin{figure}[h!]
    \centering
    \includegraphics[width = 0.95\textwidth]{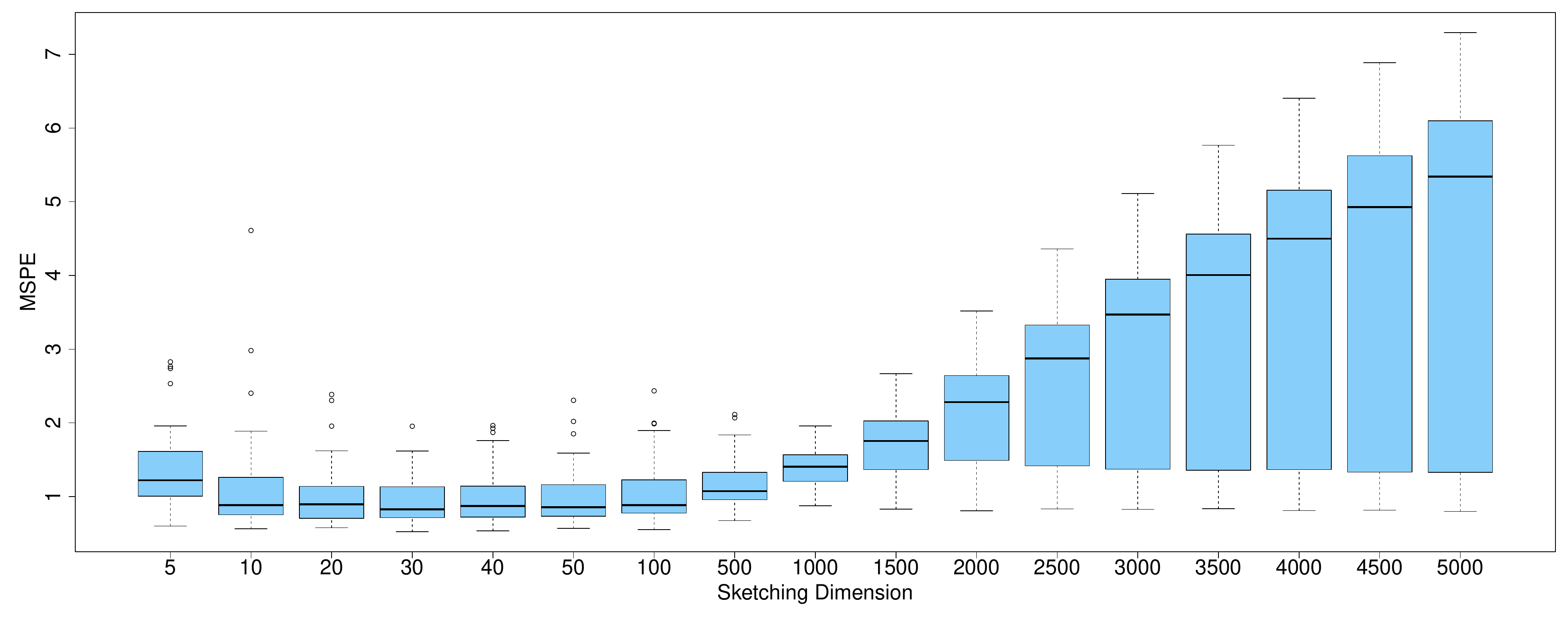}
    \caption{Plot of MSPE as sketching dimension ($m$) increases.}
    \label{fig:sketch_m}
\end{figure}

Figure~\ref{fig:sketch_m} illustrates the effect of sketching dimension $m$ on SkGP performance. As $m$ increases, MSPE decreases until reaching an optimal point, after which performance declines due to the introduction of redundant predictors. A sketching dimension of around $m \sim 50$ performs well across various examples with low-dimensional manifolds.

\begin{figure}[h!]
\centering
\subfloat{\includegraphics[height=4.5cm]{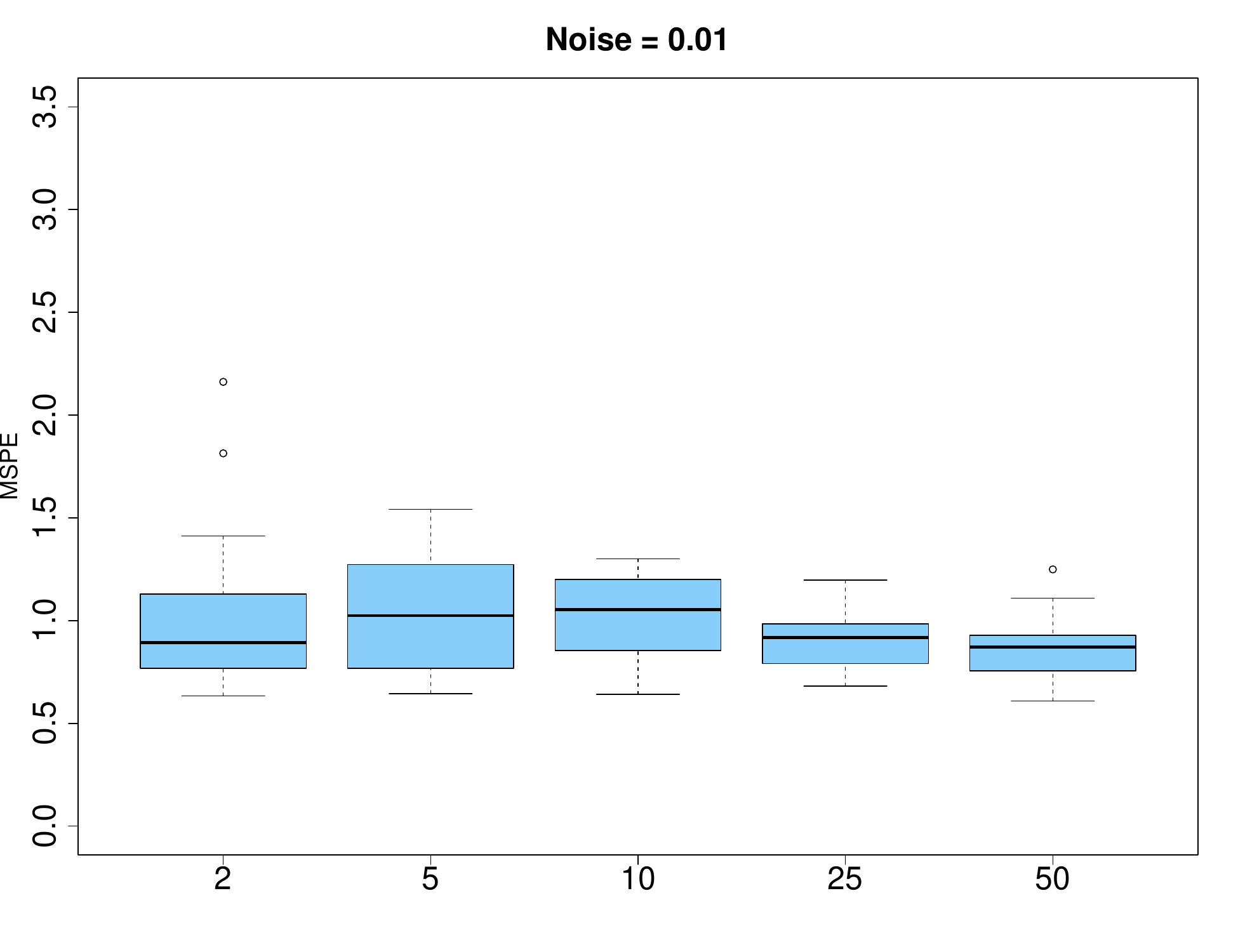}}
\subfloat{\includegraphics[height=4.5cm]{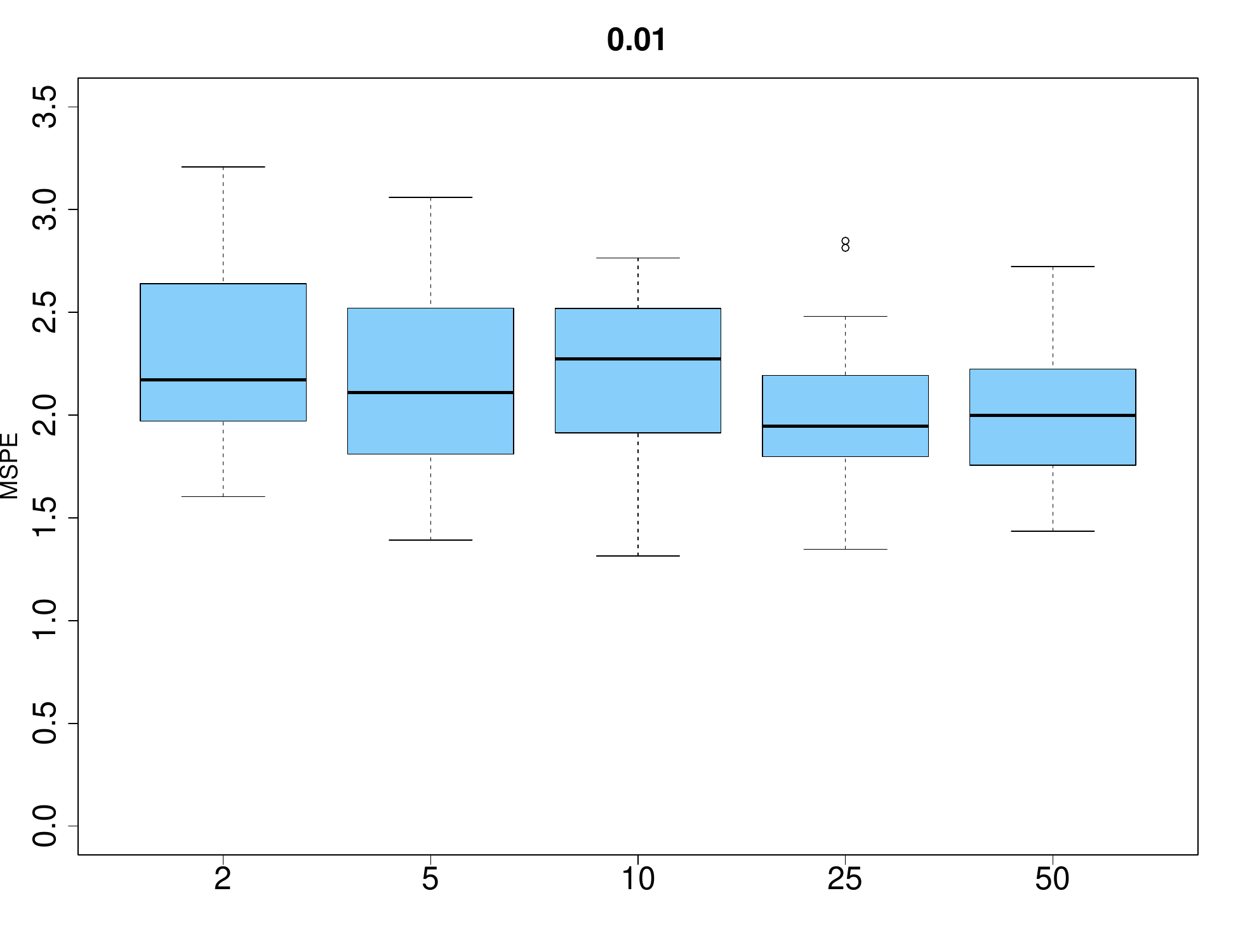}}
\caption{Left panel shows the plot of MSPE with changing number of models $K$ to be stacked together with $\tau^2=0.01$. The right panel shows the same with $\tau^2=0.1$. Increasing the number of models (K) tends to increase predictive accuracy by decreasing the variation in predictions.  }\label{fig:Vary_K}
\end{figure}

Figure~\ref{fig:Vary_K} shows that increasing the number of models $K$ reduces prediction variability, especially in high-noise scenarios. Stacking over many models increases the likelihood of sampling models with higher predictive accuracy.


\section{Analysis of Outdoor Air Pollution Data}\label{sec:real_data}



\begin{figure}[H]
    \centering
    \includegraphics[width = \textwidth]{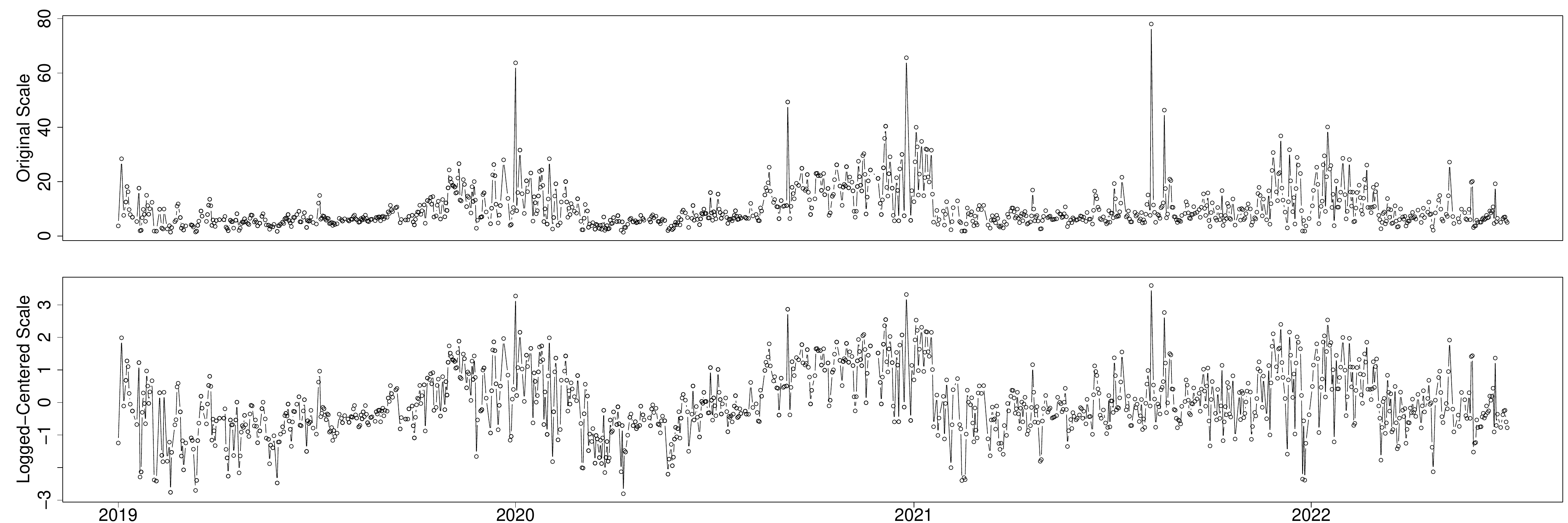}
    \caption{Daily air quality monitor data in Las Vegas. Original (top) and standardized for analysis (bottom). Data before 2019 contains irregular gaps and is excluded.}
    \label{fig:pmfrm}
\end{figure}

We will apply the proposed approach and compare it with relevant competitors in the analysis of air quality using multi-band satellite images over time for 30 different cities in the USA. For each city, the dataset consists of almost daily air quality measurements from EPA federal reference monitors, sometimes with multiple readings in a day, collected between January 2019 and July 2022. We only consider data in this interval, since the data for dates prior to 2019 are sparse in many locations. This results in $1667$ air quality measurements, as depicted in the first row of Figure~\ref{fig:pmfrm}. For each air quality reading, corresponding multi-band satellite images (blue, green, red, and near-infrared) were obtained from Planet’s PlanetScope instrument~\citep{team2017planet}. Since installing air quality monitors is costly, a key goal is to predict air quality using the high-dimensional multi-band images. At each time point, the $128\times 128 = 16384$ pixels from the four bands are vectorized into a $65536$-dimensional vector.


Although the images are high-dimensional, their intrinsic dimension (ID) determined using the two-nearest neighbor (NN) method \citep{facco2017estimating} is much lower. For example, the estimated ID for Las Vegas is $4.18$ with a standard error of $0.1$, suggesting the vectorized image predictors lie on a lower-dimensional manifold, making it ideal for SkGP. Out of the $1667$ samples, every fourth sample is used for testing, leaving $n=1334$ training samples and $n_{new}=333$ test samples.

The raw air quality monitor data in each city displays characteristics such as non-negativity, heavy-tailed distributions, non-stationary patterns, and periodic behavior. To meet the normality assumption for the error in (\ref{eq:fitted_model}), we apply a log transformation and standardize the response, ensuring a mean of zero and a variance of one. As an example, The first and the second row of Figure~\ref{fig:pmfrm} illustrate the original and log-transformed and standardized air quality data for Las Vegas, respectively. Many of the pixels in the satellite images are zero at all times and are included to buffer the image to fit into a square. In our analysis, these zeros are removed. The columns of the image predictor matrix, with zeros removed, 
are pixel-wise standardized to have zero mean and unit variance. For each city, Nonparametric Independent Screening (NIS) \citep{fan2014nonparametric} is used to identify predictors marginally related to air quality, retaining at least $5000$ predictors to avoid discarding useful information.

We evaluated SkGP against all competing methods from our simulation studies, but BART emerged as the strongest alternative. Therefore, we focus our in-depth comparison on BART. Uncompressed GP was omitted because of its high computational cost. While the data contains a temporal aspect, this analysis does not consider time dynamics; we intend to address this in future research.

\begin{figure}[H]
    \centering
    \includegraphics[width = \textwidth]{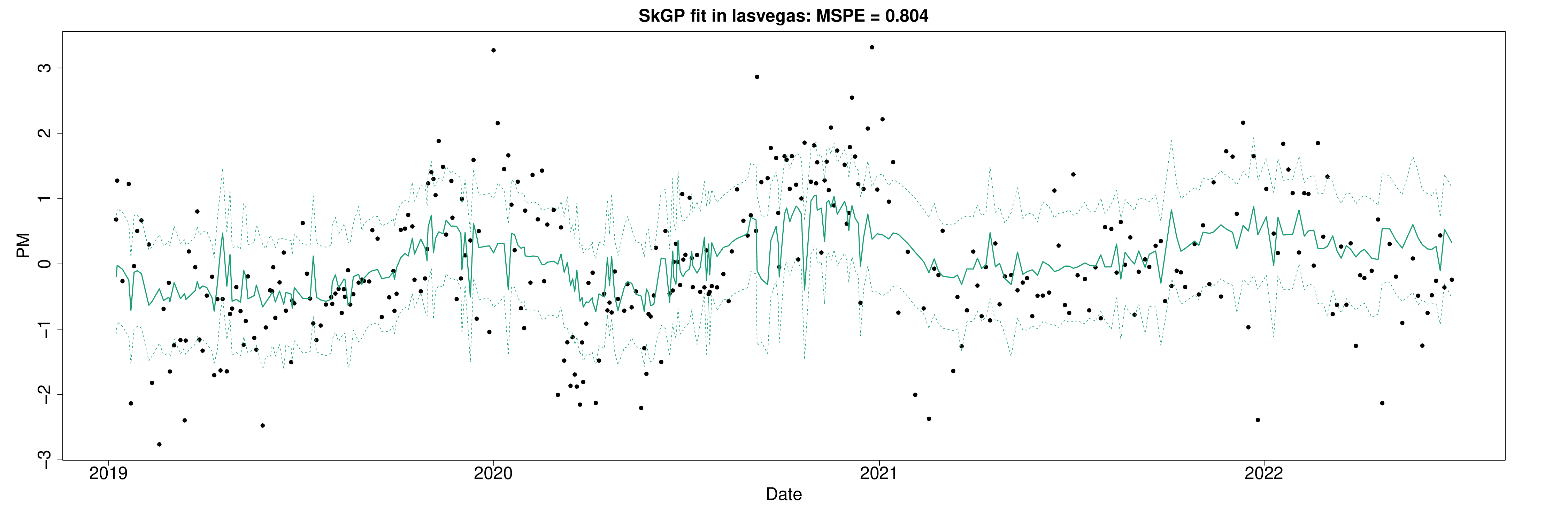}
    \caption{Point prediction and 95\% predictive interval for Las Vegas PM2.5 test points. SkGP accurately recovers the trend with good coverage.}\label{fig:lasvegasnice}
\end{figure}

In fitting SkGP, we use a sketching dimension of $m = 100$ and stack over $K = 20$ models. Table~\ref{tab:Real_Data_Results} shows that SkGP and BART perform similarly in point prediction across $30$ cities. Although all competitors exhibit under-coverage, potentially due to neglecting the time-varying nature of the data, SkGP offers significantly better predictive uncertainty, achieving much higher coverage with slightly wider 95\% predictive intervals. Due to space constraints, we show (see Figure~\ref{fig:lasvegasnice}) predictive point estimates and 95\% predictive interval bands across all hold-out time points for a representative city of Las Vegas. Overall, these findings highlight SkGP's effectiveness in capturing the non-linear regression relationship between air quality and multi-band satellite images.


\begin{table}[h!]
\scriptsize{
\begin{tabular}{|l|lll|lll|}
\hline
Location &      & SkGP     &           &       & BART     & \\ \cline{2-7} 
         & MSPE & Cvg & Length &  MSPE & Cvg & Length \\ \hline
\multicolumn{1}{|l|}{Akron}    & \multicolumn{1}{|l|}{\textbf{0.83}} & \multicolumn{1}{|l|}{0.85} & \multicolumn{1}{|l|}{2.07} & \multicolumn{1}{|l|}{0.93} & \multicolumn{1}{|l|}{0.65} & \multicolumn{1}{|l|}{1.73}\\
\multicolumn{1}{|l|}{Albuquerque}    & \multicolumn{1}{|l|}{0.87} & \multicolumn{1}{|l|}{0.93} & \multicolumn{1}{|l|}{2.10} & \multicolumn{1}{|l|}{\textbf{0.83}} & \multicolumn{1}{|l|}{0.67} & \multicolumn{1}{|l|}{1.59}\\
\multicolumn{1}{|l|}{Atlanta}    & \multicolumn{1}{|l|}{\textbf{0.96}} & \multicolumn{1}{|l|}{0.79} & \multicolumn{1}{|l|}{2.27} & \multicolumn{1}{|l|}{1.01} & \multicolumn{1}{|l|}{0.51} & \multicolumn{1}{|l|}{1.51}\\
\multicolumn{1}{|l|}{Austin}    & \multicolumn{1}{|l|}{\textbf{0.89}} & \multicolumn{1}{|l|}{0.86} & \multicolumn{1}{|l|}{2.23} & \multicolumn{1}{|l|}{0.95} & \multicolumn{1}{|l|}{0.65} & \multicolumn{1}{|l|}{1.78}\\
\multicolumn{1}{|l|}{Birmingham}    & \multicolumn{1}{|l|}{\textbf{0.88}} & \multicolumn{1}{|l|}{0.78} & \multicolumn{1}{|l|}{2.132} & \multicolumn{1}{|l|}{0.93} & \multicolumn{1}{|l|}{0.52} & \multicolumn{1}{|l|}{1.37}\\
\multicolumn{1}{|l|}{Bismark}    & \multicolumn{1}{|l|}{\textbf{0.79}} & \multicolumn{1}{|l|}{0.86} & \multicolumn{1}{|l|}{2.07} & \multicolumn{1}{|l|}{0.87} & \multicolumn{1}{|l|}{0.69} & \multicolumn{1}{|l|}{1.86}\\
\multicolumn{1}{|l|}{Boston}    & \multicolumn{1}{|l|}{\textbf{0.87}} & \multicolumn{1}{|l|}{0.77} & \multicolumn{1}{|l|}{2.03} & \multicolumn{1}{|l|}{0.88} & \multicolumn{1}{|l|}{0.65} & \multicolumn{1}{|l|}{1.73}\\
\multicolumn{1}{|l|}{The Bronx}    & \multicolumn{1}{|l|}{\textbf{0.93}} & \multicolumn{1}{|l|}{0.82} & \multicolumn{1}{|l|}{2.06} & \multicolumn{1}{|l|}{0.97} & \multicolumn{1}{|l|}{0.52} & \multicolumn{1}{|l|}{1.42}\\
\multicolumn{1}{|l|}{Broward}    & \multicolumn{1}{|l|}{\textbf{0.97}} & \multicolumn{1}{|l|}{0.82} & \multicolumn{1}{|l|}{2.03} & \multicolumn{1}{|l|}{1.03} & \multicolumn{1}{|l|}{0.59} & \multicolumn{1}{|l|}{1.72}\\
\multicolumn{1}{|l|}{Cedar Rapids}    & \multicolumn{1}{|l|}{\textbf{0.85}} & \multicolumn{1}{|l|}{0.81} & \multicolumn{1}{|l|}{2.04} & \multicolumn{1}{|l|}{0.93} & \multicolumn{1}{|l|}{0.63} & \multicolumn{1}{|l|}{1.82}\\
\multicolumn{1}{|l|}{Chattanooga}    & \multicolumn{1}{|l|}{0.98} & \multicolumn{1}{|l|}{0.86} & \multicolumn{1}{|l|}{2.19} & \multicolumn{1}{|l|}{\textbf{0.96}} & \multicolumn{1}{|l|}{0.55} & \multicolumn{1}{|l|}{1.55}\\
\multicolumn{1}{|l|}{Chicago}    & \multicolumn{1}{|l|}{\textbf{0.98}} & \multicolumn{1}{|l|}{0.84} & \multicolumn{1}{|l|}{2.29} & \multicolumn{1}{|l|}{1.02} & \multicolumn{1}{|l|}{0.56} & \multicolumn{1}{|l|}{1.67}\\
\multicolumn{1}{|l|}{Dallas}    & \multicolumn{1}{|l|}{\textbf{0.98}} & \multicolumn{1}{|l|}{0.86} & \multicolumn{1}{|l|}{2.32} & \multicolumn{1}{|l|}{1.01} & \multicolumn{1}{|l|}{0.54} & \multicolumn{1}{|l|}{1.44}\\
\multicolumn{1}{|l|}{Fort Collins}    & \multicolumn{1}{|l|}{\textbf{0.64}} & \multicolumn{1}{|l|}{0.90} & \multicolumn{1}{|l|}{1.85} & \multicolumn{1}{|l|}{0.66} & \multicolumn{1}{|l|}{0.69} & \multicolumn{1}{|l|}{1.61}\\
\multicolumn{1}{|l|}{Houston}    & \multicolumn{1}{|l|}{\textbf{0.90}} & \multicolumn{1}{|l|}{0.88} & \multicolumn{1}{|l|}{2.30} & \multicolumn{1}{|l|}{0.93} & \multicolumn{1}{|l|}{0.62} & \multicolumn{1}{|l|}{1.70}\\
\multicolumn{1}{|l|}{Jackson}    & \multicolumn{1}{|l|}{\textbf{0.89}} & \multicolumn{1}{|l|}{0.82} & \multicolumn{1}{|l|}{2.51} & \multicolumn{1}{|l|}{0.90} & \multicolumn{1}{|l|}{0.53} & \multicolumn{1}{|l|}{1.45}\\
\multicolumn{1}{|l|}{Las Vegas}    & \multicolumn{1}{|l|}{\textbf{0.80}} & \multicolumn{1}{|l|}{0.80} & \multicolumn{1}{|l|}{1.89} & \multicolumn{1}{|l|}{0.92} & \multicolumn{1}{|l|}{0.59} & \multicolumn{1}{|l|}{1.40}\\
\multicolumn{1}{|l|}{Los Angeles}    & \multicolumn{1}{|l|}{0.94} & \multicolumn{1}{|l|}{0.84} & \multicolumn{1}{|l|}{2.10} & \multicolumn{1}{|l|}{\textbf{0.92}} & \multicolumn{1}{|l|}{0.70} & \multicolumn{1}{|l|}{1.66}\\
\multicolumn{1}{|l|}{Merced}    & \multicolumn{1}{|l|}{\textbf{0.62}} & \multicolumn{1}{|l|}{0.88} & \multicolumn{1}{|l|}{1.83} & \multicolumn{1}{|l|}{0.67} & \multicolumn{1}{|l|}{0.72} & \multicolumn{1}{|l|}{1.64}\\
\multicolumn{1}{|l|}{Phoenix}    & \multicolumn{1}{|l|}{0.77} & \multicolumn{1}{|l|}{0.92} & \multicolumn{1}{|l|}{2.43} & \multicolumn{1}{|l|}{\textbf{0.73}} & \multicolumn{1}{|l|}{0.77} & \multicolumn{1}{|l|}{2.00}\\
\multicolumn{1}{|l|}{Pittsburgh}    & \multicolumn{1}{|l|}{0.99} & \multicolumn{1}{|l|}{0.83} & \multicolumn{1}{|l|}{2.10} & \multicolumn{1}{|l|}{\textbf{0.97}} & \multicolumn{1}{|l|}{0.52} & \multicolumn{1}{|l|}{1.46}\\
\multicolumn{1}{|l|}{Queens}    & \multicolumn{1}{|l|}{0.87} & \multicolumn{1}{|l|}{0.85} & \multicolumn{1}{|l|}{2.22} & \multicolumn{1}{|l|}{\textbf{0.85}} & \multicolumn{1}{|l|}{0.63} & \multicolumn{1}{|l|}{1.60}\\
\multicolumn{1}{|l|}{Raleigh}    & \multicolumn{1}{|l|}{\textbf{0.79}} & \multicolumn{1}{|l|}{0.87} & \multicolumn{1}{|l|}{2.33} & \multicolumn{1}{|l|}{0.80} & \multicolumn{1}{|l|}{0.63} & \multicolumn{1}{|l|}{1.66}\\
\multicolumn{1}{|l|}{Salt Lake City}    & \multicolumn{1}{|l|}{0.86} & \multicolumn{1}{|l|}{0.91} & \multicolumn{1}{|l|}{2.46} & \multicolumn{1}{|l|}{\textbf{0.78}} & \multicolumn{1}{|l|}{0.58} & \multicolumn{1}{|l|}{1.43}\\
\multicolumn{1}{|l|}{Saint Louis}    & \multicolumn{1}{|l|}{0.96} & \multicolumn{1}{|l|}{0.95} & \multicolumn{1}{|l|}{3.12} & \multicolumn{1}{|l|}{\textbf{0.92}} & \multicolumn{1}{|l|}{0.48} & \multicolumn{1}{|l|}{1.39}\\
\multicolumn{1}{|l|}{Saint Paul}    & \multicolumn{1}{|l|}{\textbf{0.93}} & \multicolumn{1}{|l|}{0.87} & \multicolumn{1}{|l|}{2.42} & \multicolumn{1}{|l|}{0.94} & \multicolumn{1}{|l|}{0.65} & \multicolumn{1}{|l|}{1.94}\\
\multicolumn{1}{|l|}{Terre Haute}    & \multicolumn{1}{|l|}{\textbf{1.10}} & \multicolumn{1}{|l|}{0.89} & \multicolumn{1}{|l|}{1.86} & \multicolumn{1}{|l|}{1.11} & \multicolumn{1}{|l|}{0.61} & \multicolumn{1}{|l|}{1.56}\\
\multicolumn{1}{|l|}{Timonium}    & \multicolumn{1}{|l|}{\textbf{1.05}} & \multicolumn{1}{|l|}{0.91} & \multicolumn{1}{|l|}{2.03} & \multicolumn{1}{|l|}{1.07} & \multicolumn{1}{|l|}{0.66} & \multicolumn{1}{|l|}{1.68}\\
\multicolumn{1}{|l|}{Tulsa}    & \multicolumn{1}{|l|}{\textbf{0.92}} & \multicolumn{1}{|l|}{0.91} & \multicolumn{1}{|l|}{2.39} & \multicolumn{1}{|l|}{0.93} & \multicolumn{1}{|l|}{0.64} & \multicolumn{1}{|l|}{1.70}\\
\multicolumn{1}{|l|}{Yuma}    & \multicolumn{1}{|l|}{\textbf{0.84}} & \multicolumn{1}{|l|}{0.91} & \multicolumn{1}{|l|}{2.41} & \multicolumn{1}{|l|}{0.92} & \multicolumn{1}{|l|}{0.66} & \multicolumn{1}{|l|}{1.69}\\ \hline
\end{tabular}
\caption{Results for 30 cities across the US. SkGP and BART yield comparable predictive accuracy. SkGP outperforms BART for uncertainty quantification in terms of coverage of 95\% predictive intervals.}\label{tab:Real_Data_Results}
}
\end{table}

\subsection{Effect of Stacking}
\begin{figure}[H]
    \centering
    \includegraphics[width = \textwidth]{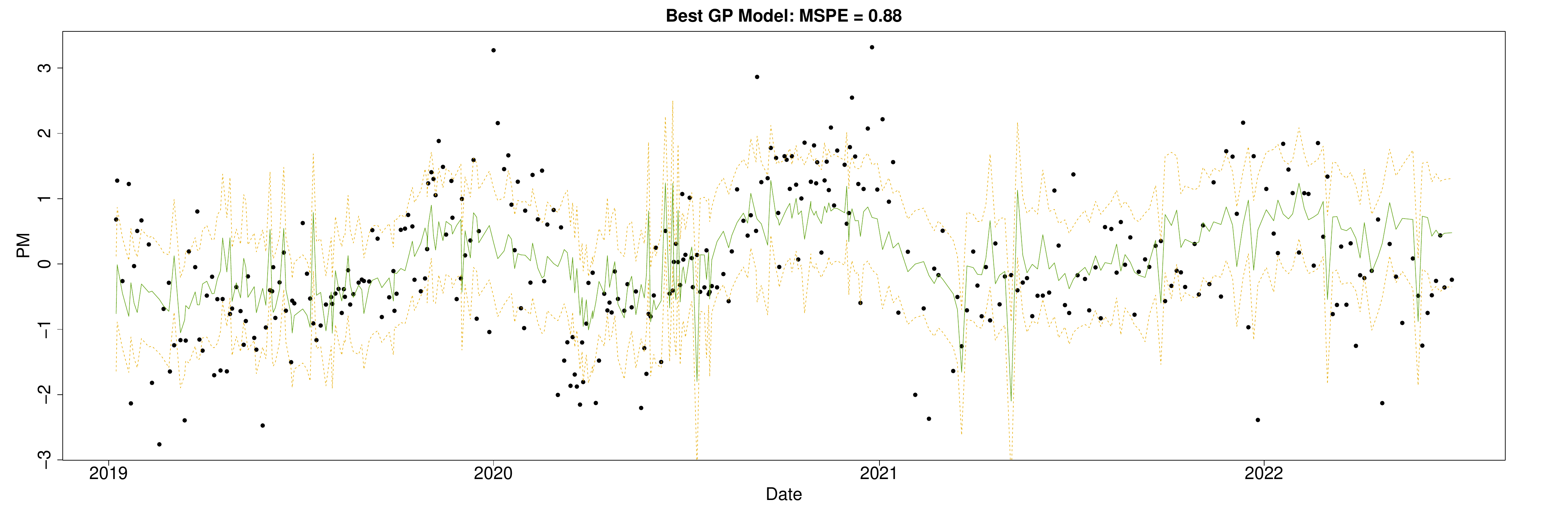}
    \caption{Point prediction and 95\% predictive interval for Las Vegas PM2.5 test points for the best performing model included in the stacked model. The associated model weight is 0.24.}\label{fig:lasvegasbest}
\end{figure}

\begin{figure}[H]
    \centering
    \includegraphics[width = \textwidth]{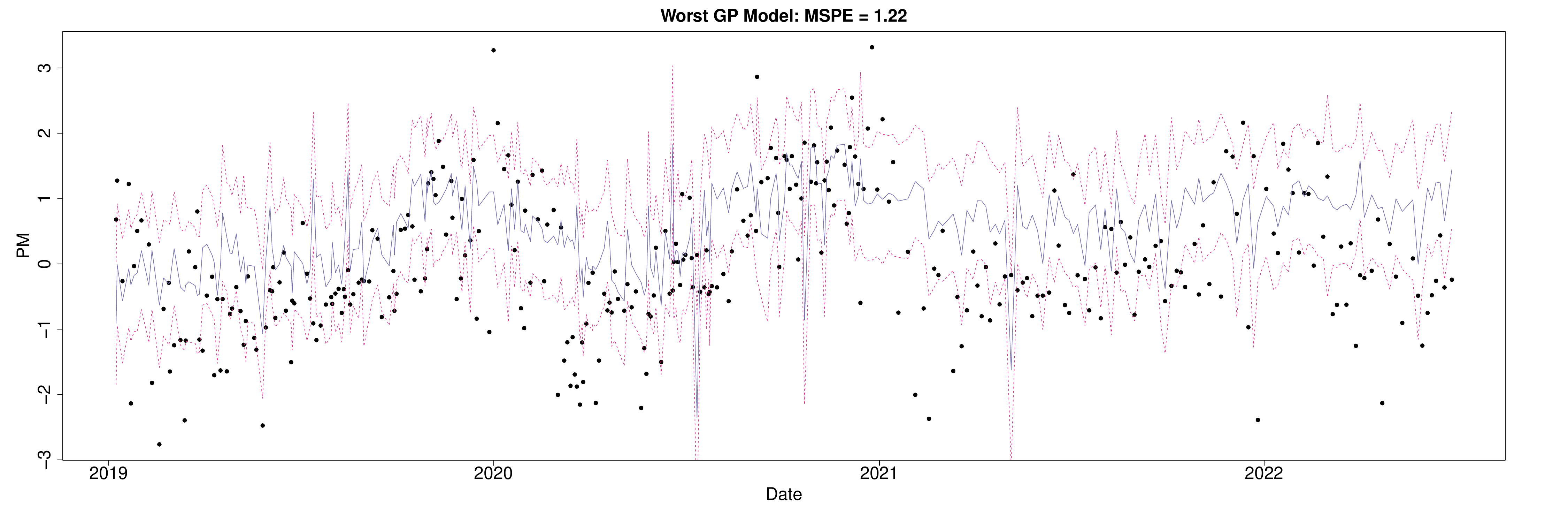}
    \caption{Point prediction and 95\% predictive interval for Las Vegas PM2.5 test points for the worst performing model included in the stacked model. The associated model weight is 0.06.}\label{fig:lasvegasworst}
\end{figure}

We now examine the impact of stacking and demonstrate its clear advantages for model performance. Stacking offers two main benefits: it down-weights poorly performing models and it consistently yields predictions superior to any individual constituent model. To demonstrate this, we consider air pollution data on Las Vegas a representative city. Here single-sketched GP models, each using one sketching matrix, somewhat vary in predictive accuracy. Figures~\ref{fig:lasvegasbest} and \ref{fig:lasvegasworst} illustrate the performance of the worst and the best individual models in terms of MSPE, showing  a spread in the performance. While the best individual model outperforms BART, $14$ of $20$ constituent sketched models actually perform worse than BART. This highlights the risk in relying on a single sketched GP, as results may be somewhat sensitive to the choice of sketching matrix and parameters.

Stacking directly addresses this issue by combining predictions from the individual models in a principled, weighted manner. Table~\ref{tab:SkGP_Model_Comp} confirms that the stacked model (SkGP) achieves the lowest MSPE, with most stacking weight assigned to the top-performing constituent model (corresponding to $k=20$). The majority of the constituent models receive weights below 0.05. In summary, while individual sketched models have theoretical guarantees (Section 2.3), stacking provides a practical improvement in predictive accuracy, especially with limited sample sizes.

\begin{table}[ht]
\centering
\begin{tabular}{lrr@{\qquad}lrr}
\hline
\multicolumn{6}{c}{\bf SkGP: MSPE = 0.80} \\
\hline
Model & MSPE & Weight & Model & MSPE & Weight \\
\hline
$k=1$  & 0.99 & 0.00 & $k=11$ & 1.02 & 0.03 \\
$k=2$  & 1.02 & 0.05 & $k=12$ & 0.89 & 0.00 \\
$k=3$  & 0.87 & 0.05 & $k=13$ & 1.16 & 0.13 \\
$k=4$  & 0.90 & 0.05 & $k=14$ & 1.06 & 0.00 \\
$k=5$  & 1.04 & 0.10 & $k=15$ & 1.00 & 0.00 \\
$k=6$  & 1.01 & 0.04 & $k=16$ & 1.16 & 0.06 \\
$k=7$  & 1.05 & 0.13 & $k=17$ & 1.12 & 0.04 \\
$k=8$  & 0.94 & 0.05 & $k=18$ & 0.90 & 0.00 \\
$k=9$  & 0.89 & 0.01 & $k=19$ & 0.94 & 0.03 \\
$k=10$ & 1.15 & 0.00 & $k=20$ & 0.84 & 0.24 \\
\hline
\end{tabular}
\caption{Results for SkGP and each constituent GP model. SkGP outperforms each model in terms of prediction. Sketching weights, shown in the final column, demonstrates the best performing model (corresponding to $k=20$) receives the highest stacking weight.}\label{tab:SkGP_Model_Comp}
\end{table}




\section{Conclusion and Future Work}\label{sec:conclusion}
This article is the first to present a novel Bayesian approach for predictive inference of outdoor air quality using high-resolution satellite images, when these images lie on a low-dimensional noise-corrupted manifold.
Our methodology exploits two powerful ideas, data sketching and stacking, to eliminate the necessity for computationally demanding manifold structure estimation, providing accurate point predictions and predictive uncertainties. The computation of the posterior predictive distribution does not rely on MCMC sampling, and our framework is amenable to parallel implementation, resulting in substantial reductions in computation and storage costs. Empirical findings underscore the significantly improved point prediction and predictive uncertainty of our approach compared to existing methods. 

As a next step, we plan to extend our framework to jointly predict outdoor air quality using remote-sensing imagery across 30 cities. This application will naturally involve both high-dimensional predictors ($p$) and a large number of observations ($n$). Given this paper’s emphasis on handling large $p$, we have kept $n$ moderate here. For future work with large $n$, we will explore scalable GP methodologies such as the Vecchia approximation \cite{pan2025block} and distributed GP approaches \citep{guhaniyogi2022distributed, guhaniyogi2023distributed}, which enable efficient GP regression at scale. For a comprehensive overview of state-of-the-art scalable GP techniques, see \cite{heaton2019case}. Additionally, exploring applications to non-Gaussian or multivariate outcomes, and simultaneous estimation of the intrinsic dimensionality of the manifold alongside predictive inference for the outcome are also part of our future research directions.

\section*{Appendix}
\underline{\textbf{Proof of Lemma 2.1}}

For any $\zeta>0$,
\begin{align}
\Pi(\Delta_\phi(f_0)>\zeta|\mbox{Data})&\leq \Pi(\sum_{k=1}^K\Delta_\phi^{(k)}(f_0)>\zeta/\sum_{k=1}^Kw_k^2|\mbox{Data})\nonumber\\
&\leq \sum_{k=1}^K\Pi(\Delta_\phi^{(k)}(f_0)>\zeta/K\sum_{k=1}^Kw_k^2|\mbox{Data}),
\end{align}
where the second inequality follows by Cauchy-Schwartz theorem.

We write $T:\mathbb R^p \to \mathbb R^m$ for the linear map $T(\bx)=\bP_n \bx$ induced by the random projection matrix $\bP_n$. Assume that $\mathcal O\subset \mathbb R^p$ is a compact smooth $d$-dimensional manifold satisfying the regularity conditions of Theorem 3.1 in \cite{baraniuk2009random}. Then there exist sequences $\varepsilon_n\in(0,1)$ and $\alpha_n\in(0,1)$ with $\alpha_n\to 1$ such that, if
\[
\mathcal B_n=\Bigl\{
(1-\varepsilon_n)\|\bu-\bv\| \le \|T(\bu)-T(\bv)\| \le (1+\varepsilon_n)\|\bu-\bv\|
\text{ for all }\bu,\bv\in\mathcal O
\Bigr\},
\]
then $\mathbb P(\mathcal B_n)\ge \alpha_n$.

\underline{Claim:} On the event $\mathcal B_n$, the restriction $T|_{\mathcal O}$ is injective. 

Indeed, if $\bx,\by\in\mathcal O$ satisfy $T(\bx)=T(\by)$, then
\[
0=\|T(\bx)-T(\by)\|\ge (1-\varepsilon_n)\|\bx-\by\|,
\]
and since $\varepsilon_n<1$, it follows that $\bx=\by$. Moreover, since $T:\mathbb R^p\to\mathbb R^m$ is linear, the restriction $T|_{\mathcal O}$ is smooth. We next show that, on $\mathcal B_n$, the differential
\[
D(T|_{\mathcal O})_\bx:T_\bx\mathcal O\to\mathbb R^m
\]
is injective for every $\bx\in\mathcal O$. Fix $\bx\in\mathcal O$ and let
$\bv\in T_\bx\mathcal O$ satisfy $D(T|_{\mathcal O})_\bx(\bv)=0$. Choose a $C^1$ curve $\gamma:(-\delta,\delta)\to\mathcal O$ such that $\gamma(0)=\bx$ and $\gamma'(0)=\bv$. Since $T$ is linear,
\[
D(T|_{\mathcal O})_\bx(\bv)
=
\left.\frac{d}{dt}\right|_{t=0}T(\gamma(t)).
\]
By the defining property of $\mathcal B_n$,
\[
\|T(\gamma(t))-T(\bx)\|
\ge (1-\varepsilon_n)\|\gamma(t)-\bx\|
\qquad\text{for all }t.
\]
Dividing by $|t|$ and letting $t\to 0$, we obtain
\[
\|D(T|_{\mathcal O})_\bx(\bv)\|
\ge (1-\varepsilon_n)\|\bv\|.
\]
Since $D(T|_{\mathcal O})_\bx(\bv)=0$ and $\varepsilon_n<1$, it follows that $\bv=0$. Therefore $D(T|_{\mathcal O})_\bx$ is injective for every $\bx\in\mathcal O$, and hence $T|_{\mathcal O}$ is a smooth injective immersion. Since $\mathcal O$ is compact, Proposition 4.22(c) of \cite{lee2012smooth} implies that $T|_{\mathcal O}$ is a smooth embedding. Consequently,
\[
T|_{\mathcal O}:\mathcal O\to T(\mathcal O)
\]
is a diffeomorphism onto its image. Hence, on the event $\mathcal B_n$, the projected predictors lie on the smooth $d$-dimensional manifold $T(\mathcal O)$, and the projected regression problem satisfies the assumptions of Theorem 2.4 of \cite{yang2016bayesian}.

Hence,
\begin{align}
\Pi(\Delta_\phi^{(k)}(f_0)>\zeta/K\sum_{k=1}^Kw_k^2|\mbox{Data})
\leq \Pi(\Delta_\phi^{(k)}(f_0)>\zeta/K\sum_{k=1}^Kw_k^2|\mbox{Data},\mathcal{B}_n)\mathbb P(\mathcal{B}_n) + \mathbb P(\mathcal{B}_n^c).
\end{align}
$\mathbb P(\mathcal{B}_n^c)=1-\alpha_n\rightarrow 0$, as $n\rightarrow\infty$. Considering 
all $w_k$'s are bounded and the function $T$ is a diffeomorphism in $\mathcal{B}_n,$
$\Pi(\Delta_\phi^{(k)}(f_0)>\zeta/K\sum_{k=1}^Kw_k^2|\mbox{Data},\mathcal{B}_n)\rightarrow 0$, as $n\rightarrow\infty$, following Theorem 2.3 of \cite{yang2016bayesian}.

{\noindent\underline{\textbf{Proof of Lemma 2.2}}}

The random compression matrix can be written as
\[
\bP_n^{(k)} = \big[\bP_{\mathcal I}^{(k)} \;\; {\boldsymbol 0}\big]
\]
where $\bP_{\mathcal I}^{(k)}\in\mathbb R^{k\times |\mathcal I|}$ acts on the selected coordinates and the remaining block is zero. Hence, for each observation,
\[
\bP_n^{(k)}\bx_i
=
\bP_n^{(k)}\bphi(\bo_i) + \bP_n^{(k)}\bet_i
=
\bP_{\mathcal I}^{(k)}\bphi_{\mathcal I}(\bo_i) + \bP_{\mathcal I}^{(k)}\bet_{i,\mathcal I}.
\]
Thus, the noise contribution from the coordinates outside $\mathcal I$ is removed by construction. 

We now show that the retained noise component is attenuated by the random projection. Let $\bp_{j}^{(k)}$ denote the $j$th row of $\bP_{\mathcal I}^{(k)}$. The matrix $\bP_{\mathcal I}^{(k)}$ is obtained by first generating a Gaussian matrix $\bG_{\mathcal I}^{(k)}\in\mathbb R^{m\times |\mathcal I|}$ with independent $N(0,1)$ entries and then orthonormalizing its rows using Gram--Schmidt. We first note that this construction yields random orthonormal directions. Indeed, for any orthogonal matrix $\bQ\in\mathbb R^{|\mathcal I|\times |\mathcal I|}$,
\[
\bG_{\mathcal I}^{(k)}\bQ \overset{d}{=} \bG_{\mathcal I}^{(k)},
\]
by rotational invariance of the Gaussian distribution. Moreover, since Gram--Schmidt depends only on inner products and norms, and these are preserved under multiplication by $\bQ$,
\[
\operatorname{GS}\!\left(\bG_{\mathcal I}^{(k)}\bQ\right)
=
\operatorname{GS}\!\left(\bG_{\mathcal I}^{(k)}\right)\bQ,
\]
where $\operatorname{GS}(\cdot)$ refers to the Gram-Schmidt orthogonalization.
Therefore,
\[
\bP_{\mathcal I}^{(k)}\bQ
=
\operatorname{GS}\!\left(\bG_{\mathcal I}^{(k)}\right)\bQ
=
\operatorname{GS}\!\left(\bG_{\mathcal I}^{(k)}\bQ\right)
\overset{d}{=}
\operatorname{GS}\!\left(\bG_{\mathcal I}^{(k)}\right)
=
\bP_{\mathcal I}^{(k)}.
\]
Hence each row $\bp_{j}^{(k)}$ is a random unit vector whose distribution is invariant under rotations in $\mathbb R^{|\mathcal I|}$. 
Now, consider the matrix $\mathbb{E}\!\left[\bp_{j}^{(k)} (\bp_{j}^{(k)})^\top\right]$. By symmetry under permutations of the coordinates, all diagonal entries of this matrix are equal. Since $\|\bp_{j}^{(k)}\|^2=1$, the trace of this matrix is one, so each diagonal entry equals $1/|\mathcal I|$. Moreover, for $\ell\neq \ell'$, the distribution of $\bp_{j}^{(k)}$ is unchanged under the sign change $(p_{j,\ell}^{(k)},p_{j,\ell'}^{(k)}) \mapsto (p_{j,\ell}^{(k)}, -p_{j,\ell'}^{(k)})$, whereas the product $p_{j,\ell}^{(k)}p_{j,\ell'}^{(k)}$ changes sign. Hence, the off-diagonal entries, $\mathbb E[p_{j,\ell}^{(k)}p_{j,\ell'}^{(k)}]~=~0$. Therefore,
\[
\mathbb E\!\left[
\bp_{j}^{(k)}(\bp_{j}^{(k)})^\top
\right]
=
\frac{1}{|\mathcal I|}\bI_{|\mathcal I|}.
\]
Since the random compression matrix is independent of the off-manifold perturbation $\bet_i$, conditional on $\bet_i$,
\[
\mathbb E_{\bP_n}\left[
\left\{(\bp_{j}^{(k)})^\top\bet_{i,\mathcal I}\right\}^2
\mid \bet_i
\right] 
=
\bet_{i,\mathcal I}^\top\; \mathbb E_{\bP_n}\left[\bp_{j}^{(k)}(\bp_{j}^{(k)})^\top \right]\; \bet_{i,\mathcal I}
=
\frac{\|\bet_{i,\mathcal I}\|^2}{|\mathcal I|}.
\]
Therefore,
\[
\bp_{j}^{(k)}\bet_i = O_P\left( \frac{\|\bet_{i,\mathcal I}\|}{\sqrt{|\mathcal I|}} \right).
\]
Now, the bounded off-manifold perturbation assumption $\|\bet_i\|=O_P(1)$, yields
\[
\bp_{j}^{(k)}\bet_i = O_P\left(|\mathcal I|^{-1/2}\right).
\]

{\noindent\underline{\textbf{Proof of Lemma 2.3}}}

Fix \(k\). To simplify notation, write
\[
\bz_i=\bz_i^{(k)},\qquad
\bu_i=\bu_i^{(k)},\qquad
\delta_n=\delta_n^{(k)},
\]
and similarly write \(A_n=A_n^{(k)}\), \(B_n=B_n^{(k)}\),
\(\Pi_Z=\Pi_Z^{(k)}\), and \(\Pi_U=\Pi_U^{(k)}\). Let
\[
\boldm_Z=\psi^2\bC_{ZZ}(\bI+\psi^2\bC_{ZZ})^{-1}\by,
\qquad
\boldm_U=\psi^2\bC_{UU}(\bI+\psi^2\bC_{UU})^{-1}\by
\]
be the two posterior mean vectors, and let
\[
\bV_Z
=
\xi^2\psi^2\bC_{ZZ}
-
\xi^2\psi^4\bC_{ZZ}(\bI+\psi^2\bC_{ZZ})^{-1}\bC_{ZZ},
\]
\[
\bV_U
=
\xi^2\psi^2\bC_{UU}
-
\xi^2\psi^4\bC_{UU}(\bI+\psi^2\bC_{UU})^{-1}\bC_{UU}
\]
be the corresponding posterior covariance matrices. Couple the two Gaussian
posterior vectors by writing
\[
\boldf_Z=\boldm_Z+\bV_Z^{1/2}\be,
\qquad
\boldf_U=\boldm_U+\bV_U^{1/2}\be,
\qquad
\be\sim N(0,\bI_n).
\]
Under the above coupling,
\[
\boldf_Z-\boldf_U
=
(\boldm_Z-\boldm_U)
+
(\bV_Z^{1/2}-\bV_U^{1/2})\be .
\]
Hence,
\[
\begin{aligned}
\mathbb E\left[
\|\boldf_Z-\boldf_U\|_2^2
\,\middle|\,\mathcal D
\right]
&=
\mathbb E\left[
\left\|
(\boldm_Z-\boldm_U)
+
(\bV_Z^{1/2}-\bV_U^{1/2})\be
\right\|_2^2
\,\middle|\,\mathcal D
\right] \\
&=
\|\boldm_Z-\boldm_U\|_2^2
+
2(\boldm_Z-\boldm_U)^\top
(\bV_Z^{1/2}-\bV_U^{1/2})
\mathbb E(\be) \\
&\qquad
+
\mathbb E\left[
\be^\top
(\bV_Z^{1/2}-\bV_U^{1/2})^\top
(\bV_Z^{1/2}-\bV_U^{1/2})
\be
\right].
\end{aligned}
\]
Since \(\mathbb E(\be)=0\) and \(\mathbb E(\be\be^\top)=\bI_n\), the last display becomes
\[
\mathbb E\left[
\|\boldf_Z-\boldf_U\|_2^2
\,\middle|\,\mathcal D
\right]
=
\|\boldm_Z-\boldm_U\|_2^2
+
\operatorname{tr}
\left[
(\bV_Z^{1/2}-\bV_U^{1/2})^\top
(\bV_Z^{1/2}-\bV_U^{1/2})
\right].
\]
Therefore,
\[
\mathbb E\left[
\|\boldf_Z-\boldf_U\|_2^2
\,\middle|\,\mathcal D
\right]
=
\|\boldm_Z-\boldm_U\|_2^2
+
\|\bV_Z^{1/2}-\bV_U^{1/2}\|_F^2.
\]
Dividing by \(n\) gives
\[
\frac1n
\mathbb E
\left[
\|\boldf_Z-\boldf_U\|_2^2
\,\middle|\,
\mathcal D
\right]
=
\frac1n\|\boldm_Z-\boldm_U\|_2^2
+
\frac1n
\|\bV_Z^{1/2}-\bV_U^{1/2}\|_F^2.
\]
We control the two terms separately.

First, define
\[
T_1=\boldm_Z-\boldm_U.
\]
Then
\[
T_1
=
\psi^2\bC_{ZZ}(\bI+\psi^2\bC_{ZZ})^{-1}\by
-
\psi^2\bC_{UU}(\bI+\psi^2\bC_{UU})^{-1}\by.
\]
Using
\[
\psi^2\bC(\bI+\psi^2\bC)^{-1}
=
\bI-(\bI+\psi^2\bC)^{-1}.
\]
we get
\[
T_1
=
\left[
(\bI+\psi^2\bC_{UU})^{-1}
-
(\bI+\psi^2\bC_{ZZ})^{-1}
\right]\by.
\]
Using the matrix identity $A^{-1}-B^{-1}=A^{-1}(B-A)B^{-1}$, for invertible matrices \(A\) and \(B\), we get
\[
(\bI+\psi^2\bC_{UU})^{-1}
-
(\bI+\psi^2\bC_{ZZ})^{-1}
=
\psi^2
(\bI+\psi^2\bC_{UU})^{-1}
(\bC_{ZZ}-\bC_{UU})
(\bI+\psi^2\bC_{ZZ})^{-1}.
\]
Therefore,
\[
T_1
=
\psi^2
(\bI+\psi^2\bC_{UU})^{-1}
(\bC_{ZZ}-\bC_{UU})
(\bI+\psi^2\bC_{ZZ})^{-1}\by.
\]
Since \(\bC_{ZZ}\) and \(\bC_{UU}\) are positive semidefinite,
\[
\|(\bI+\psi^2\bC_{ZZ})^{-1}\|_{\mathrm{op}}\le 1,
\qquad
\|(\bI+\psi^2\bC_{UU})^{-1}\|_{\mathrm{op}}\le 1.\]
Hence
\[
\frac1n\|T_1\|_2^2
\le
\psi^4
\frac{\|\by\|_2^2}{n}
\|\bC_{ZZ}-\bC_{UU}\|_{\mathrm{op}}^2.
\]
Thus, if
\[
\|\bC_{ZZ}-\bC_{UU}\|_{\mathrm{op}}=o_P(1),
\]
then, using \(\|\by\|_2^2/n=O_P(1)\),
\[
\frac1n\|T_1\|_2^2=o_P(1).
\]
Next, we control the covariance term. For any positive semidefinite matrix
\(\bC\),
\[
\psi^2\bC-\psi^4\bC(\bI+\psi^2\bC)^{-1}\bC
=
\bI-(\bI+\psi^2\bC)^{-1}.
\]
Therefore,
\[
\bV_Z
=
\xi^2\left\{\bI-(\bI+\psi^2\bC_{ZZ})^{-1}\right\},
\qquad
\bV_U
=
\xi^2\left\{\bI-(\bI+\psi^2\bC_{UU})^{-1}\right\}.
\]
Thus
\[
\bV_Z-\bV_U
=
\xi^2
\left[
(\bI+\psi^2\bC_{UU})^{-1}
-
(\bI+\psi^2\bC_{ZZ})^{-1}
\right].
\]
Applying the matrix identity $A^{-1}-B^{-1}=A^{-1}(B-A)B^{-1}$ again,
\[
\bV_Z-\bV_U
=
\xi^2\psi^2
(\bI+\psi^2\bC_{UU})^{-1}
(\bC_{ZZ}-\bC_{UU})
(\bI+\psi^2\bC_{ZZ})^{-1}.
\]
Consequently,
\[
\|\bV_Z-\bV_U\|_{\mathrm{op}}
\le
\xi^2\psi^2
\|\bC_{ZZ}-\bC_{UU}\|_{\mathrm{op}}.
\]
Using the matrix square-root inequality for positive semidefinite matrices,
\[
\|\bV_Z^{1/2}-\bV_U^{1/2}\|_F^2
\le
\|\bV_Z-\bV_U\|_*,
\]
and the bound
\[
\|\bV_Z-\bV_U\|_*
\le
n\|\bV_Z-\bV_U\|_{\mathrm{op}},
\]
we obtain
\[
\frac1n
\|\bV_Z^{1/2}-\bV_U^{1/2}\|_F^2
\le
\|\bV_Z-\bV_U\|_{\mathrm{op}}
\le
\xi^2\psi^2
\|\bC_{ZZ}-\bC_{UU}\|_{\mathrm{op}}.
\]
Therefore, the covariance contribution is also \(o_P(1)\), provided
\[
\|\bC_{ZZ}-\bC_{UU}\|_{\mathrm{op}}=o_P(1).
\]
It remains to prove this operator-norm perturbation bound. Let
\[
\bDelta_C=\bC_{ZZ}-\bC_{UU}.
\]
For the squared exponential kernel,
\[
(\bC_{ZZ})_{ij}
=
e^{-\theta\|\bz_i-\bz_j\|^2},
\qquad
(\bC_{UU})_{ij}
=
e^{-\theta\|\bu_i-\bu_j\|^2}.
\]
Define
\[
d_{ij}
=
\|\bz_i-\bz_j\|^2-\|\bu_i-\bu_j\|^2.
\]
Since \(\bz_i=\bu_i+\br_i\),
\[
d_{ij}
=
2\langle \bu_i-\bu_j,\br_i-\br_j\rangle
+
\|\br_i-\br_j\|^2.
\]
Using \(\|\br_i-\br_j\|\le 2\delta_n\), we get
\[
|d_{ij}|
\le
2\|\bu_i-\bu_j\|\|\br_i-\br_j\|
+
\|\br_i-\br_j\|^2
\le
4\delta_n\|\bu_i-\bu_j\|+4\delta_n^2.
\]
Moreover,
\[
(\bC_{ZZ})_{ij}
=
(\bC_{UU})_{ij}e^{-\theta d_{ij}},
\]
and hence
\[
(\bDelta_C)_{ij}
=
(\bC_{UU})_{ij}\{e^{-\theta d_{ij}}-1\}.
\]
Using \(|e^{-x}-1|\le e^{|x|}-1\), we obtain
\[
|(\bDelta_C)_{ij}|
\le
e^{-\theta\|\bu_i-\bu_j\|^2}
\left[
\exp\left\{
\theta(4\delta_n\|\bu_i-\bu_j\|+4\delta_n^2)
\right\}
-1
\right].
\]
Using \(e^x-1\le xe^x\) for \(x\ge0\), we obtain
\[
\begin{aligned}
|(\bDelta_C)_{ij}|
&\le
C
\left(
\delta_n\|\bu_i-\bu_j\|+\delta_n^2
\right)
e^{-\theta\|\bu_i-\bu_j\|^2}
e^{4\theta\delta_n\|\bu_i-\bu_j\|}.
\end{aligned}
\]
Next, for \(t\ge 0\), using \(2ab\le a^2+b^2\),
\[
4\theta\delta_n t
\le
\frac{\theta}{2}t^2+8\theta\delta_n^2
\quad\Longrightarrow\quad
-\theta t^2+4\theta\delta_n t
\le
-\frac{\theta}{2}t^2+8\theta\delta_n^2.
\]
Therefore,
\[
e^{-\theta\|\bu_i-\bu_j\|^2}
e^{4\theta\delta_n\|\bu_i-\bu_j\|}
\le
e^{8\theta\delta_n^2}
e^{-(\theta/2)\|\bu_i-\bu_j\|^2}.
\]
Since \(\delta_n=o_P(1)\), \(e^{8\theta\delta_n^2}=O_P(1)\). Hence
\[
|(\bDelta_C)_{ij}|
\le
O_P(1)
\left(
\delta_n\|\bu_i-\bu_j\|+\delta_n^2
\right)
e^{-(\theta/2)\|\bu_i-\bu_j\|^2}.
\]
Since \(\bDelta_C\) is symmetric, the row-sum bound gives
\[
\|\bDelta_C\|_{\mathrm{op}}
\le
\max_{1\le i\le n}\sum_{j=1}^n|(\bDelta_C)_{ij}|.
\]
Therefore,
\[
\begin{aligned}
\|\bDelta_C\|_{\mathrm{op}}
&\le
O_P(1)\delta_n
\max_{1\le i\le n}
\sum_{j=1}^n
\|\bu_i-\bu_j\|
e^{-(\theta/2)\|\bu_i-\bu_j\|^2} \\
&\qquad
+
O_P(1)\delta_n^2
\max_{1\le i\le n}
\sum_{j=1}^n
e^{-(\theta/2)\|\bu_i-\bu_j\|^2} \\
&=
O_P(1)\left(\delta_n A_n+\delta_n^2 B_n\right),
\end{aligned}
\]
By assumption, \(A_n=O(1)\), \(B_n=O(1)\), and \(\delta_n=o_P(1)\). Hence
\[
\|\bC_{ZZ}-\bC_{UU}\|_{\mathrm{op}}
=
\|\bDelta_C\|_{\mathrm{op}}
=
o_P(1).
\]
Combining the mean and covariance bounds, we have
\[
\mathbb E
\left[
\frac1n\|\boldf_Z-\boldf_U\|_2^2
\,\middle|\,
\mathcal D
\right]
=o_P(1).
\]
Finally, Markov's inequality yields, for every \(\zeta>0\),
\[
\Pi\left(
\frac1n\|\boldf_Z-\boldf_U\|_2^2>\zeta
\,\middle|\,
\mathcal D
\right)
\le
\frac{1}{\zeta}
\mathbb E
\left[
\frac1n\|\boldf_Z-\boldf_U\|_2^2
\,\middle|\,
\mathcal D
\right]
=o_P(1).
\]

{\noindent\underline{\textbf{Proof of Theorem 2.4}}}

We give the proof for the fixed-design case; the random-design case is analogous. For each \(i\), add and subtract the clean-sketch posterior prediction and the
clean-manifold truth:
\[
\begin{aligned}
\sum_{k=1}^K
w_k f_k\!\left(\bx_i\right)
-
f_0(\bx_i)
&= \sum_{k=1}^K
w_k f_Z^{(k)}\!\left(\bz_i^{(k)}\right)
-
f_0(\bx_i) \\
& =
\sum_{k=1}^K
w_k
\left[
f_Z^{(k)}\!\left(\bz_i^{(k)}\right)
-
f_U^{(k)}\!\left(\bu_i^{(k)}\right)
\right] \\
&\qquad\quad+
\left[
\sum_{k=1}^K
w_k f_U^{(k)}\!\left(\bu_i^{(k)}\right)
-
f_0\!\left(\bphi(\bo_i)\right)
\right] \\
&\qquad\quad+
\left[
f_0\!\left(\bphi(\bo_i)\right)
-
f_0(\bx_i)
\right].
\end{aligned}
\]
Using \((a+b+c)^2\le 3a^2+3b^2+3c^2\), we obtain
\[
\Delta_x(f_0)
\le
3R_n+3\Delta_\phi(f_0)+3S_n,
\]
where
\[
R_n
=
\frac1n\sum_{i=1}^n
\left[
\sum_{k=1}^K
w_k
\left\{
f_Z^{(k)}\!\left(\bz_i^{(k)}\right)
-
f_U^{(k)}\!\left(\bu_i^{(k)}\right)
\right\}
\right]^2
\]
and
\[
S_n
=
\frac1n\sum_{i=1}^n
\left[
f_0(\bx_i)-f_0\!\left(\bphi(\bo_i)\right)
\right]^2 .
\]
By convexity of \(x\mapsto x^2\) and the stacking-weight condition
\(w_k\ge0\), \(\sum_{k=1}^K w_k=1\), Jensen's inequality gives
\[
R_n
\le
\sum_{k=1}^K
w_k
\frac1n\sum_{i=1}^n
\left\{
f_Z^{(k)}\!\left(\bz_i^{(k)}\right)
-
f_U^{(k)}\!\left(\bu_i^{(k)}\right)
\right\}^2 .
\]
Therefore, for every \(\zeta>0\),
\[
\{\Delta_x(f_0)>\zeta\}
\subseteq
\{R_n>\zeta/9\}
\cup
\{\Delta_\phi(f_0)>\zeta/9\}
\cup
\{S_n>\zeta/9\}.
\]
Moreover,
\[
\{R_n>\zeta/9\}
\subseteq
\bigcup_{k=1}^K
\left\{
\frac1n\sum_{i=1}^n
\left\{
f_Z^{(k)}\!\left(\bz_i^{(k)}\right)
-
f_U^{(k)}\!\left(\bu_i^{(k)}\right)
\right\}^2
>
\frac{\zeta}{9}
\right\}.
\]
Taking posterior probabilities and applying the union bound gives
\[
\begin{aligned}
\Pi\left(
\Delta_x(f_0)>\zeta
\,\middle|\,
\mathcal D
\right)
&\le
\sum_{k=1}^K
\Pi_{\mathrm{coup}}^{(k)}
\left(
\frac1n\sum_{i=1}^n
\left\{
f_Z^{(k)}\!\left(\bz_i^{(k)}\right)
-
f_U^{(k)}\!\left(\bu_i^{(k)}\right)
\right\}^2
>
\frac{\zeta}{9}
\,\middle|\,
\mathcal D
\right) \\
&\qquad+
\Pi_U\left(
\Delta_\phi(f_0)>\frac{\zeta}{9}
\,\middle|\,
\mathcal D
\right)
+
\mathbf 1\left\{
S_n>\frac{\zeta}{9}
\right\}.
\end{aligned}
\]
The first term converges to zero by Lemma~\ref{lem:posterior-stability} for
each \(k=1,\ldots,K\). Since \(K\) is fixed, the entire sum converges to zero.
The second term converges to zero by Lemma~\ref{lem:clean-manifold}. It remains to control \(S_n\). Since \(\bx_i=\bphi(\bo_i)+\bet_i\) and \(f_0\)
is Lipschitz, there exists \(L<\infty\) such that
\[
\left|
f_0(\bx_i)-f_0\!\left(\bphi(\bo_i)\right)
\right|
\le
L\|\bx_i-\bphi(\bo_i)\|
=
L\|\bet_i\|.
\]
Therefore,
\[
S_n
=
\frac1n\sum_{i=1}^n
\left[
f_0(\bx_i)-f_0\!\left(\bphi(\bo_i)\right)
\right]^2
\le
L^2\max_{1\le i\le n}\|\bet_i\|^2
=
o_P(1).
\]
Hence for every \(\zeta>0\),
\[
P_0\left(
S_n>\frac{\zeta}{9}
\right)
\to 0.
\] 
Combining the three bounds yields
\[
\Pi\left(
\Delta_x(f_0)>\zeta
\,\middle|\,
\mathcal D
\right)
\to0
\]
in \(P_0\)-probability.



\clearpage
\bibliographystyle{ba}
\bibliography{sample}


\clearpage

\setcounter{page}{1}

\begin{frontmatter}

\title{\bf Supplementary Material for\\
Data Sketching and Stacking: A Confluence of Two Strategies for
Predictive Inference in Nonlinear High-Dimensional Regression}

\runtitle{Supplementary Material for SkGP}

\begin{aug}

\author[A]{\fnms{Samuel}~\snm{Gailliot}
\ead[label=e1]{samuel.gailliot@stat.tamu.edu}},
\author[A]{\fnms{Niladri}~\snm{Kal}
\ead[label=e4]{niladrik@tamu.edu}},
\author[B]{\fnms{Roger D.}~\snm{Peng}
\ead[label=e3]{roger.peng@austin.utexas.edu}}
\and
\author[A]{\fnms{Rajarshi}~\snm{Guhaniyogi}
\ead[label=e2]{rajguhaniyogi@stat.tamu.edu}}

\address[A]{Department of Statistics,
Texas A\&M University
\printead[presep={,\ }]{e1,e4,e2}}

\address[B]{Department of Statistics and Data Sciences,
University of Texas at Austin
\printead[presep={,\ }]{e3}}

\runauthor{Gailliot et al.}

\end{aug}

\begin{abstract}
This supplementary material contains additional simulation results.
\end{abstract}

\begin{keyword}
\kwd{Bayesian predictive stacking}
\kwd{predictor sketching}
\kwd{Gaussian processes}
\kwd{high-dimensional predictors}
\kwd{manifold regression}
\kwd{posterior consistency}
\end{keyword}

\end{frontmatter}

\setcounter{section}{0}
\setcounter{figure}{0}
\setcounter{table}{0}
\setcounter{equation}{0}

\renewcommand{\thesection}{S\arabic{section}}
\renewcommand{\thefigure}{S\arabic{figure}}
\renewcommand{\thetable}{S\arabic{table}}
\renewcommand{\theequation}{S\arabic{equation}}

\section{Additional Simulation Results}
\begin{figure}[h!]
\centering
\subfloat[][$\tau^2 = 0.01$]{\includegraphics[height=4.5cm]{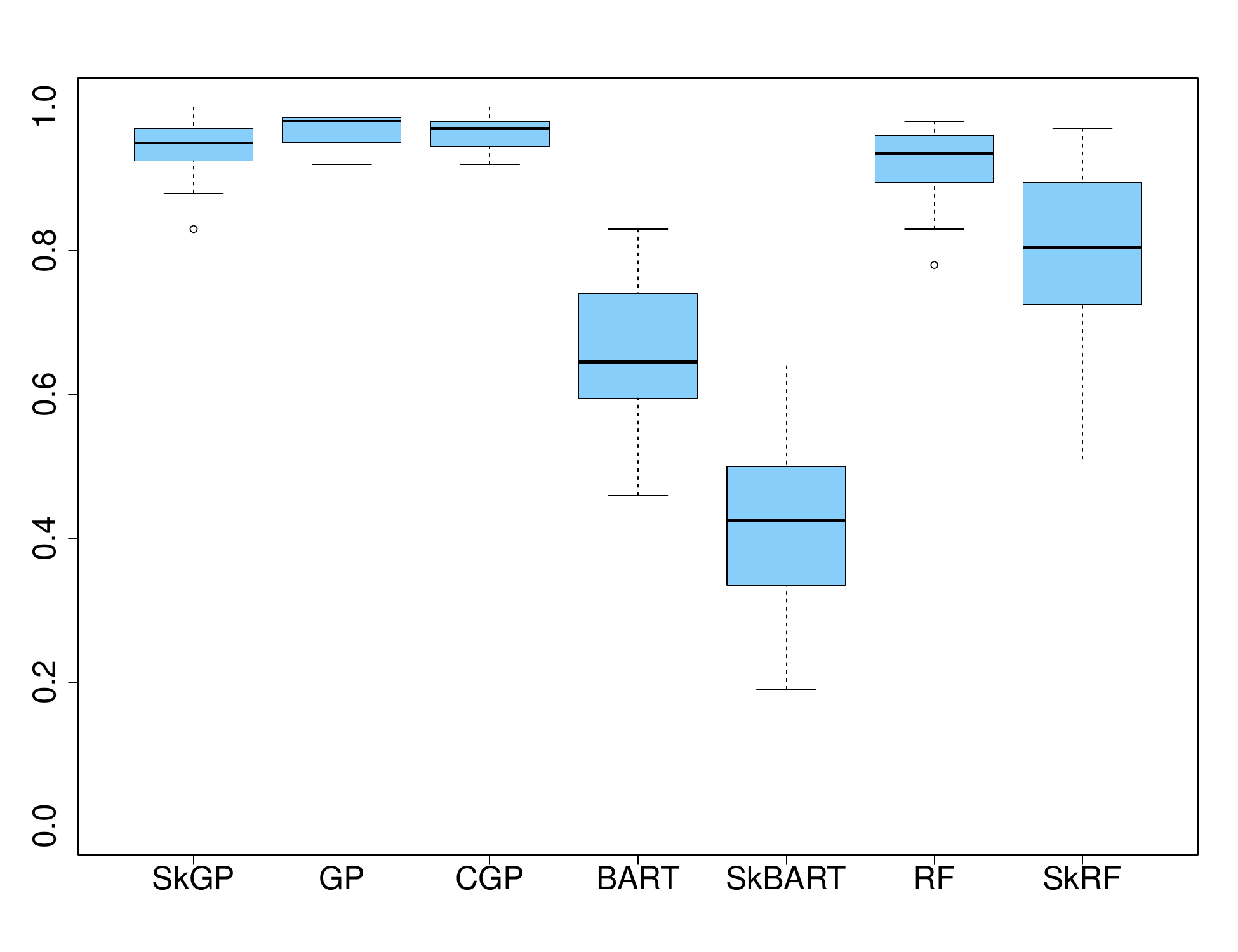}}
\subfloat[][$\tau^2 = 0.03$]{\includegraphics[height=4.5cm]{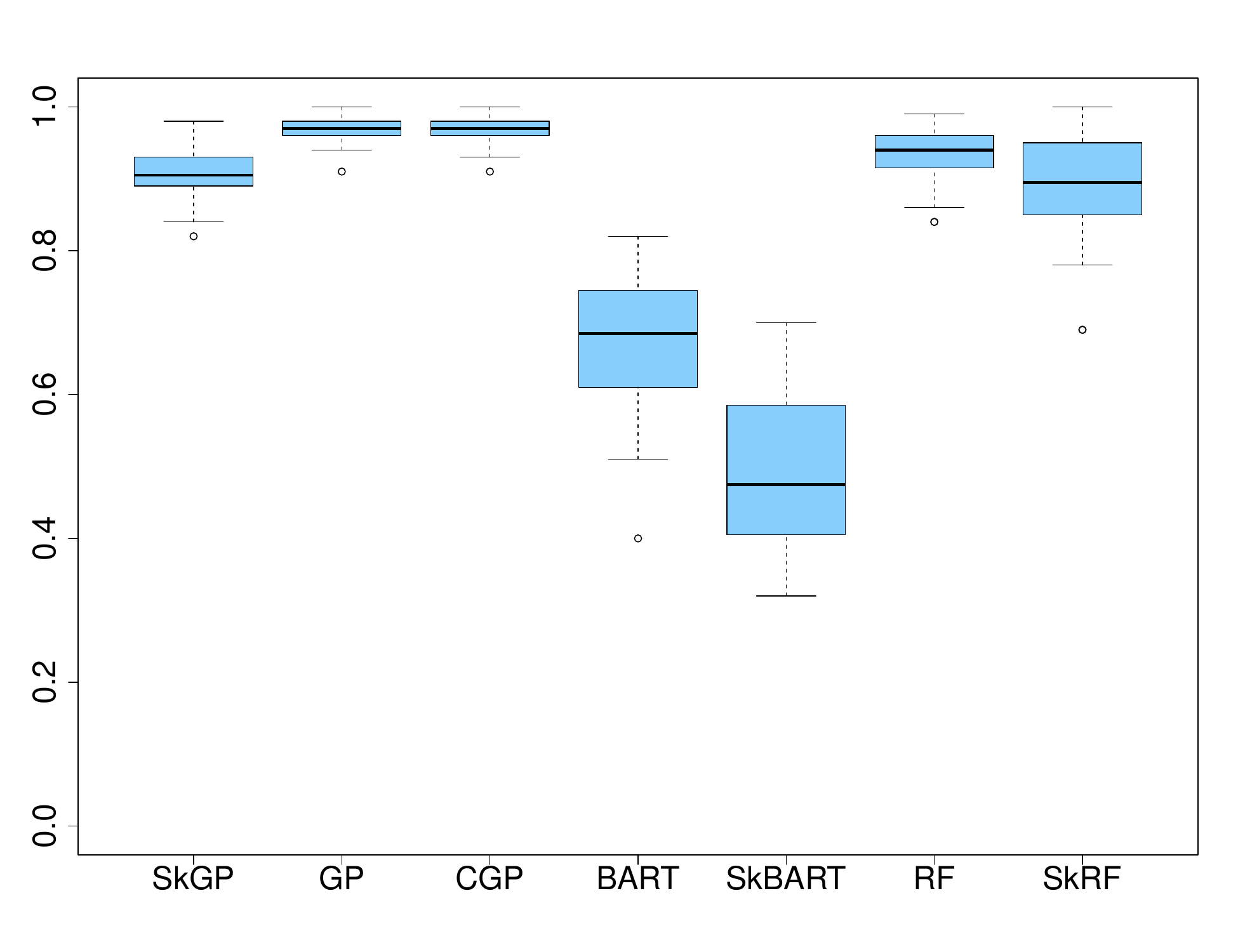}}\par\medskip
\subfloat[][$\tau^2 = 0.05$]{\includegraphics[height=4.5cm]{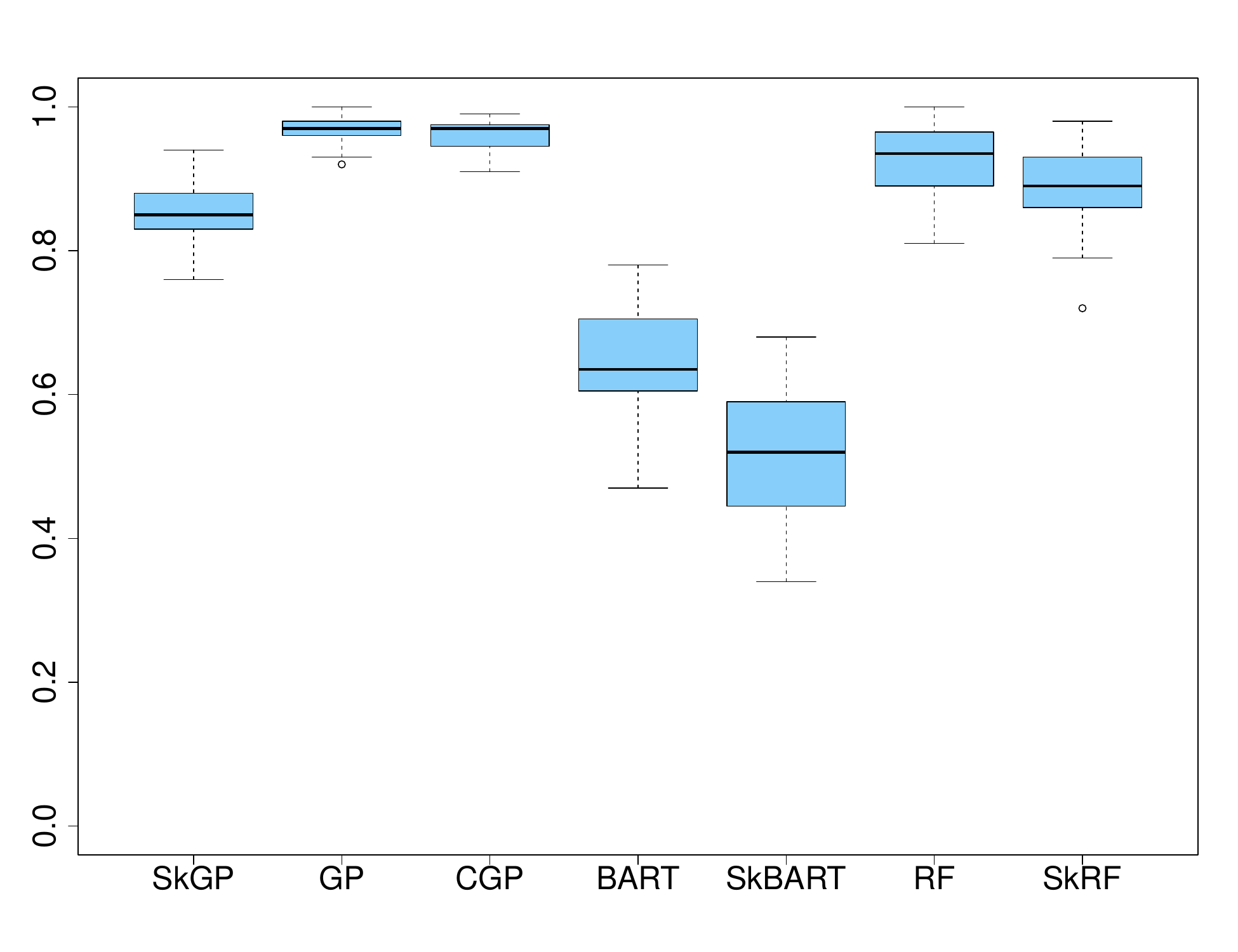}}
\subfloat[][$\tau^2 = 0.1$]{\includegraphics[height = 4.5cm]{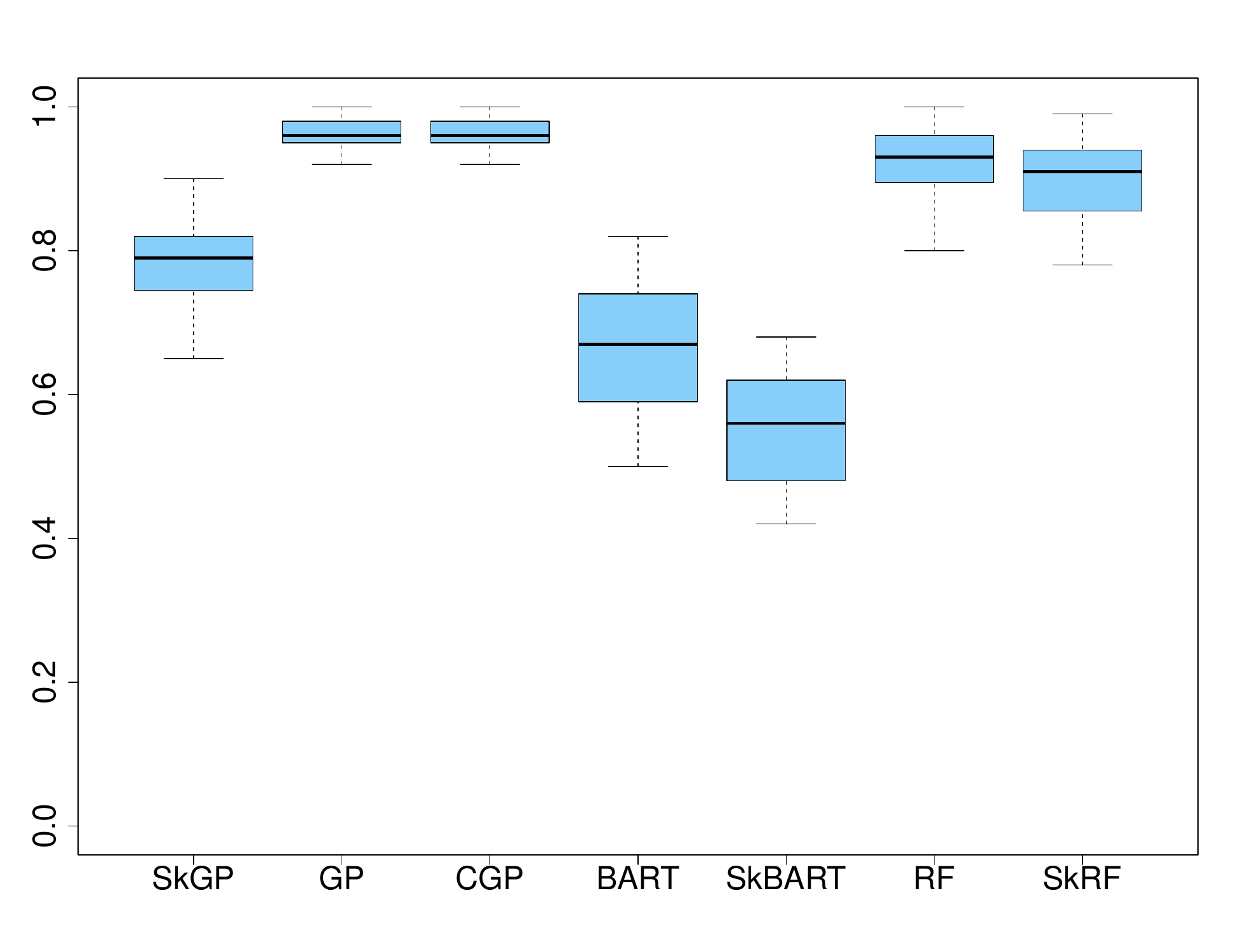}}
\caption{Coverage of 95\% predictive interval for Swiss Roll Simulations for different noise levels $\tau^2$ in the manifold containing predictors in the case of $p=10000$.}\label{fig:swiss_roll_coverage_10K}
\end{figure}

\begin{figure}[h!]
\centering
\subfloat[][$\tau^2 = 0.01$]{\includegraphics[height=4cm]{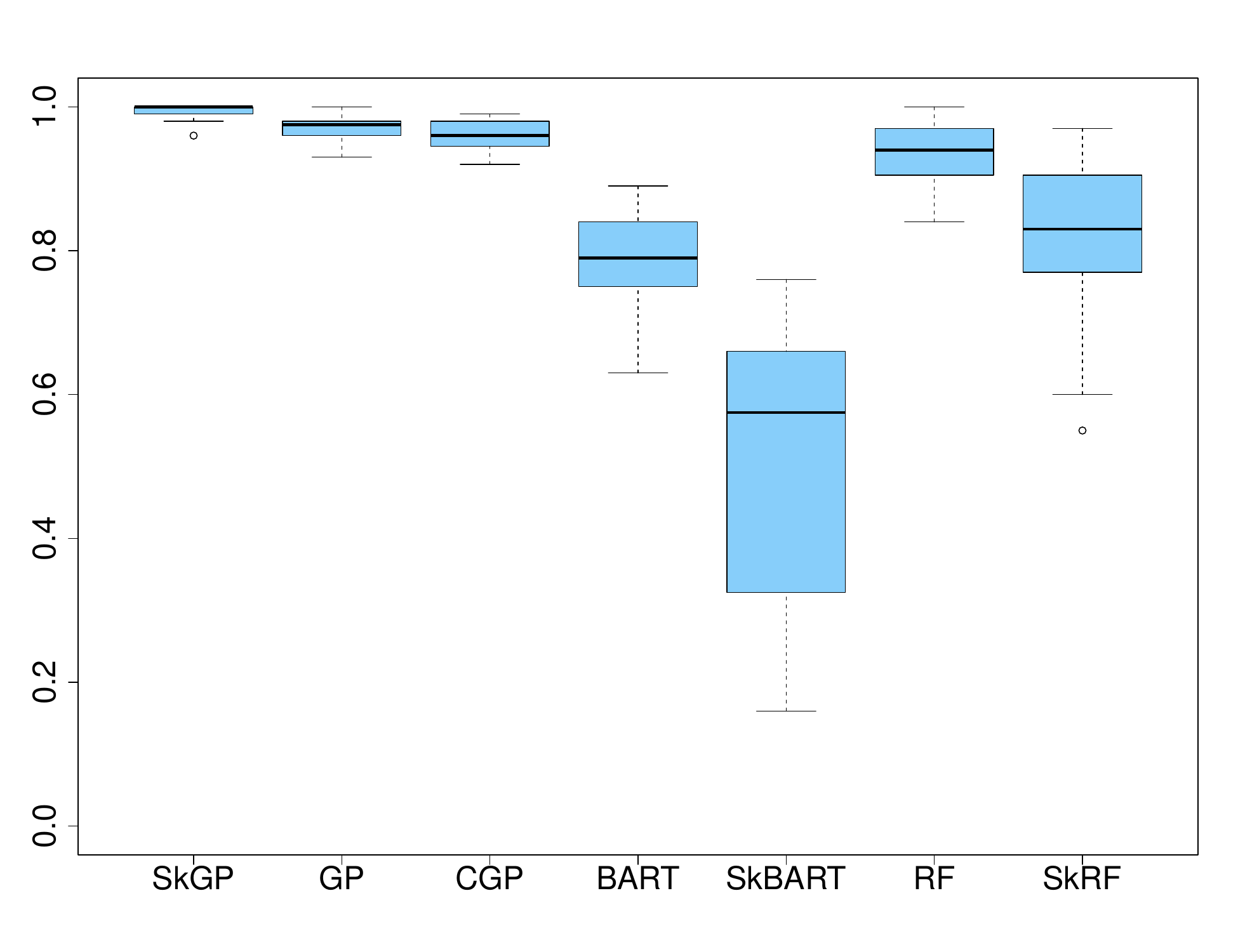}}
\subfloat[][$\tau^2 = 0.03$]{\includegraphics[height=4cm]{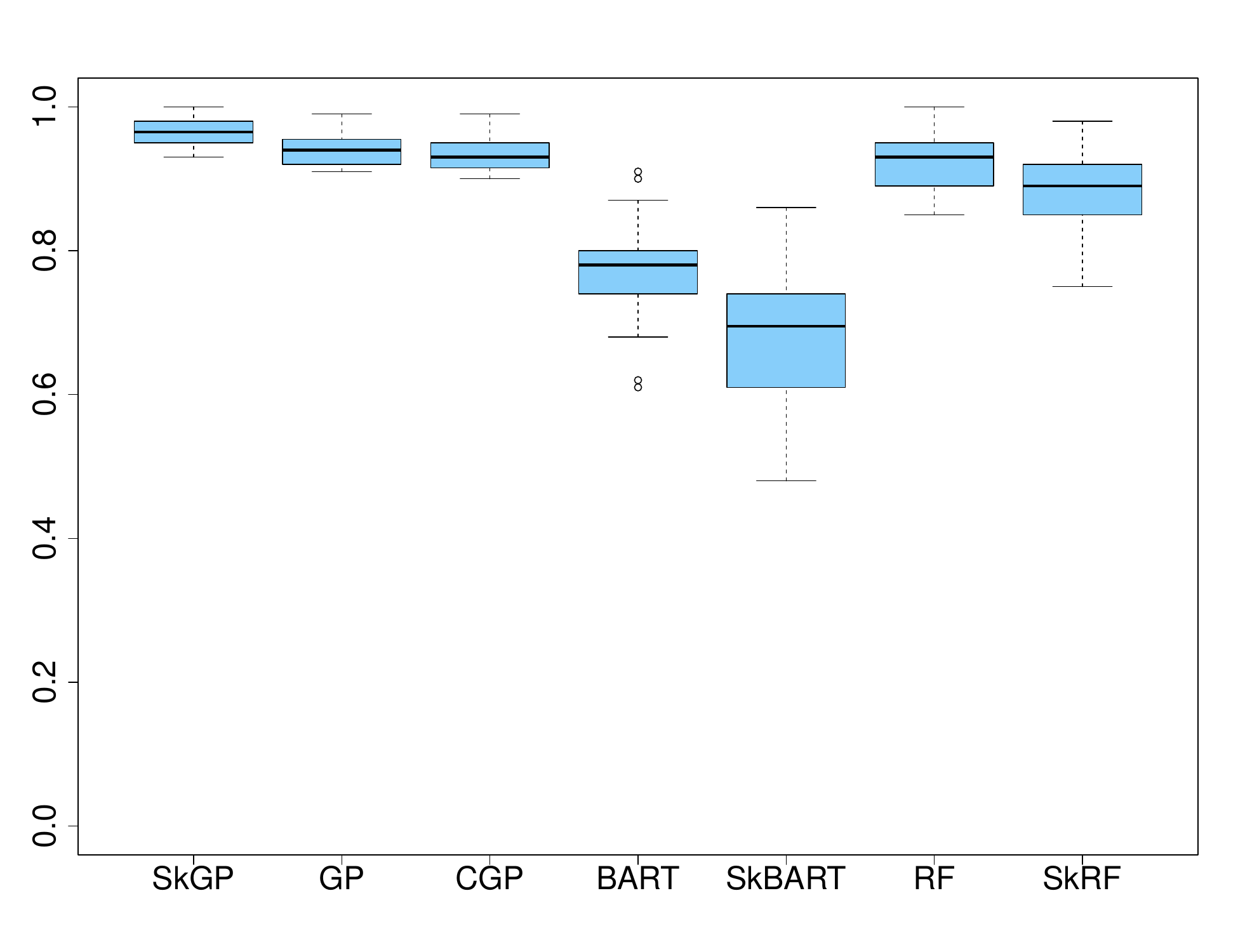}}\par\medskip
\subfloat[][$\tau^2 = 0.05$]{\includegraphics[height=4cm]{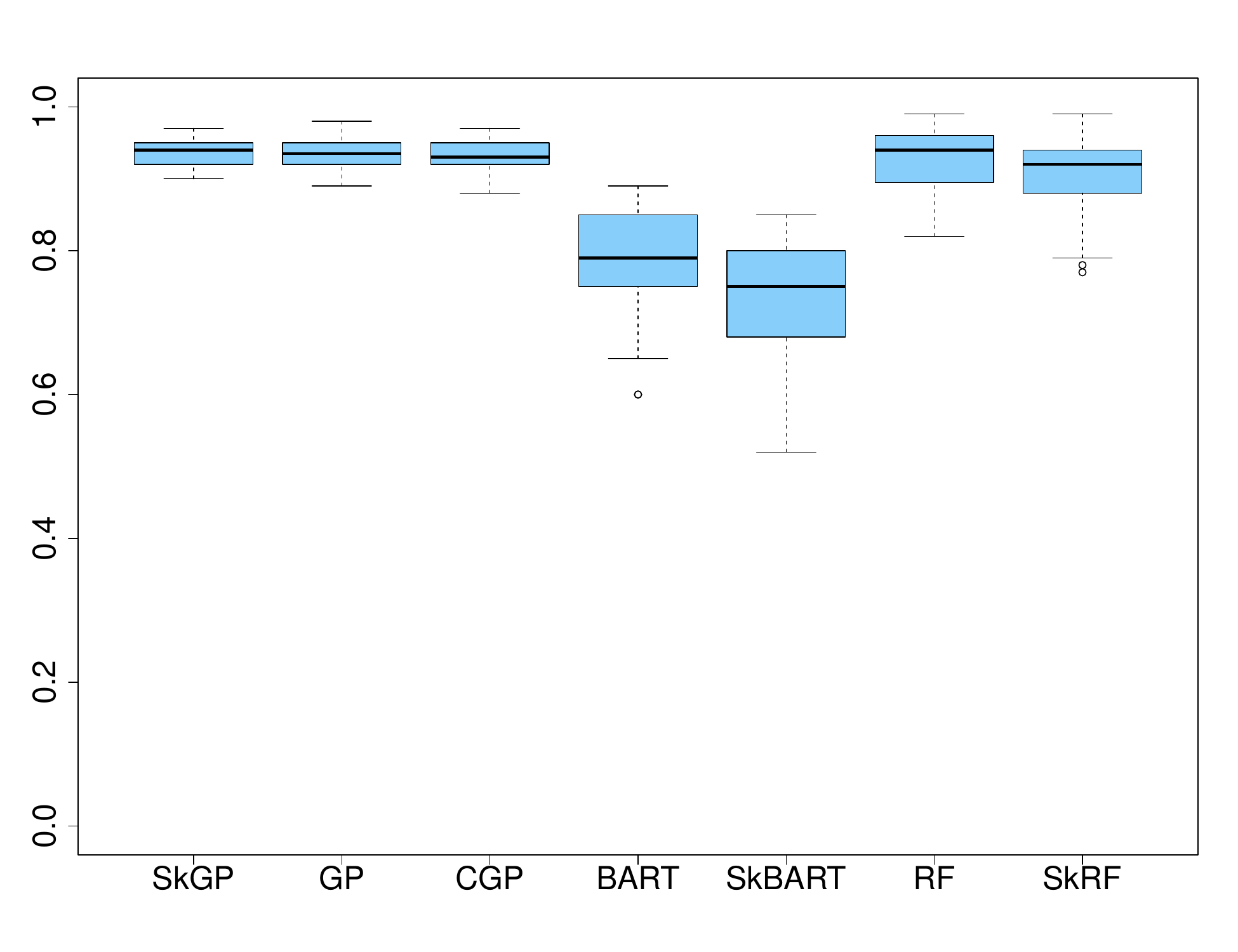}}
\subfloat[][$\tau^2 = 0.1$]{\includegraphics[height = 4cm]{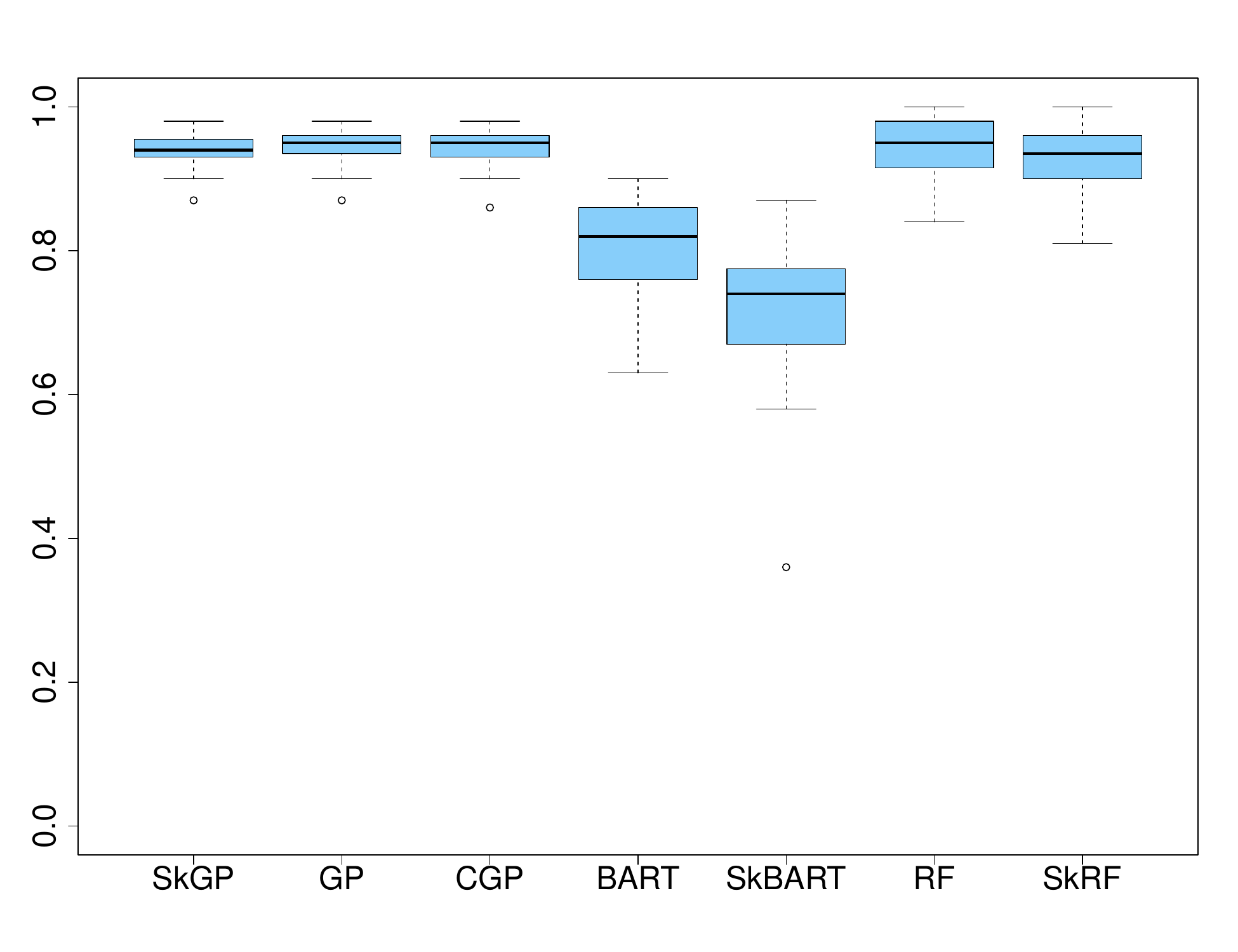}}
\caption{Coverage of 95\% predictive interval for Torus Simulations for different noise levels $\tau^2$ in the manifold containing predictors in the case of $p=10000$.}\label{fig:torus_coverage_10K}
\end{figure}

\begin{figure}[h!]
  \subcaptionbox{Swiss Roll, $p = 10000$}{\includegraphics[width=0.42\textwidth]{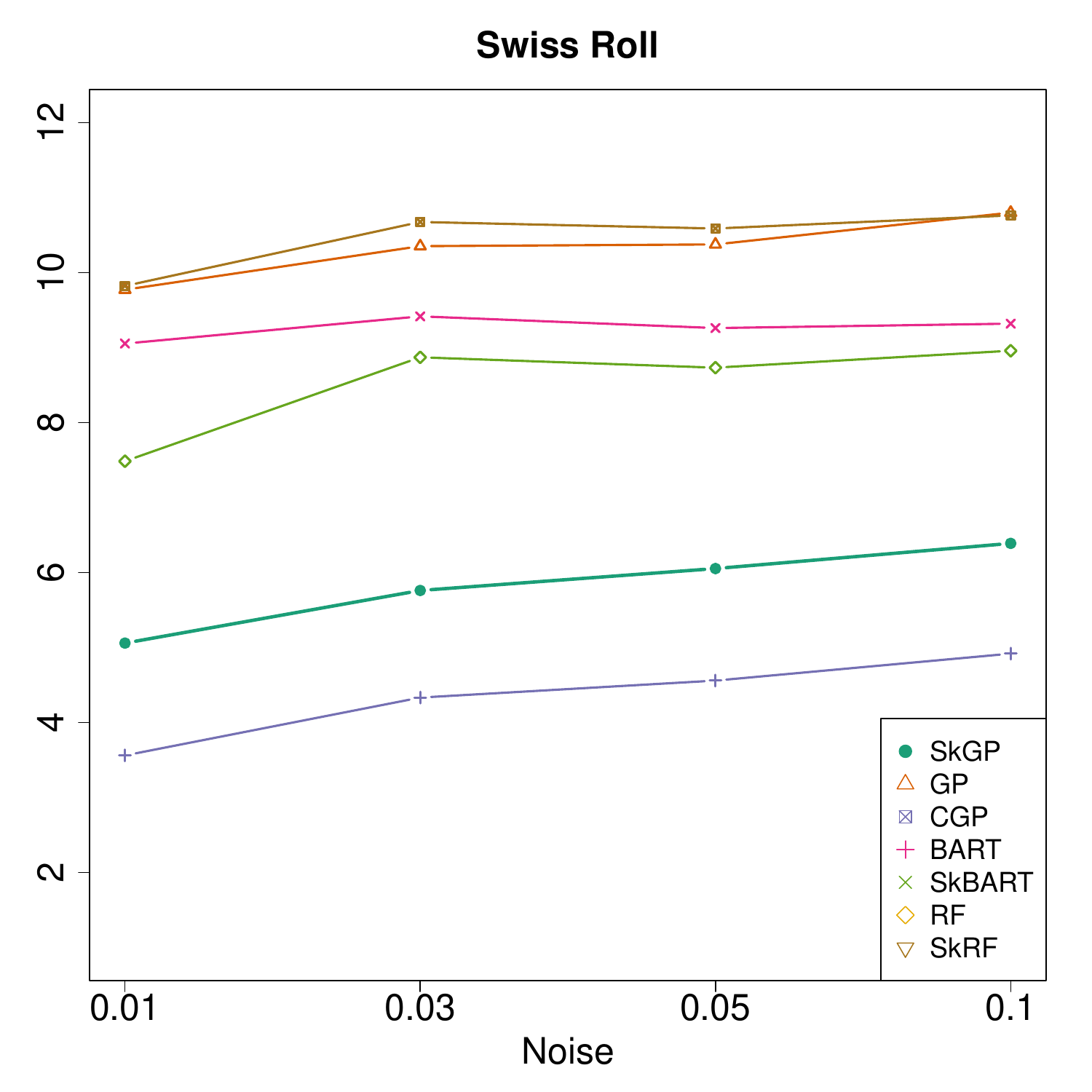}}\hfill%
  \subcaptionbox{Torus, $p = 10000$}{\includegraphics[width=0.42\textwidth]{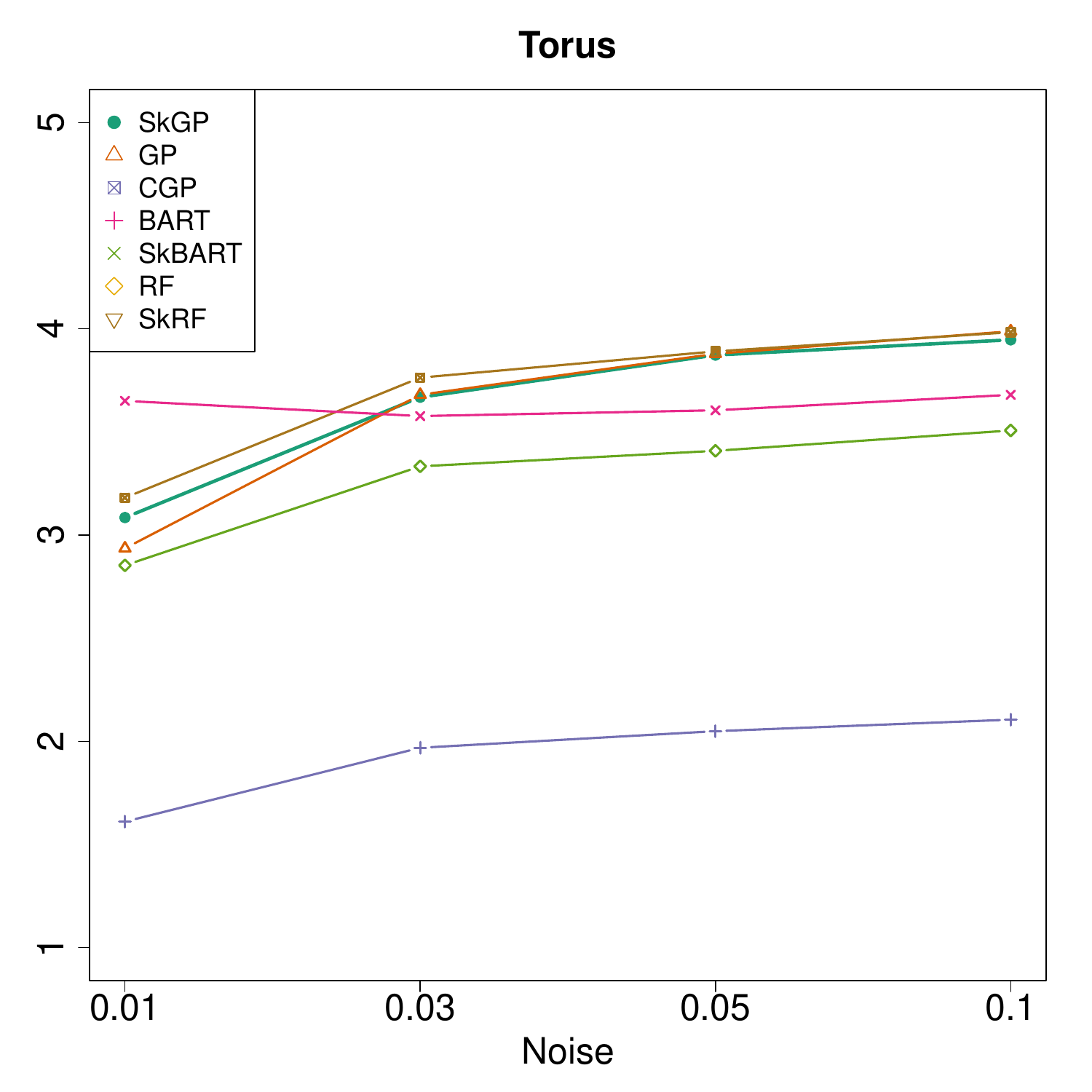}} \\
  \caption{Length of 95\% predictive intervals for all competitors in both swiss roll and torus simulations for $p=10000$.}\label{fig:length_PI_10K}
\end{figure}

To evaluate quality of predictive uncertainty, we calculate the coverage and length of 95\% predictive intervals (PI) for SkGP and other competitors. This section presents coverage probability boxplots over $50$ replications for $p=10000$ in the swiss roll example and torus example in Figures~\ref{fig:swiss_roll_coverage_10K} and \ref{fig:torus_coverage_10K}, respectively.
Figure~\ref{fig:length_PI_10K} displays the median lengths of the 95\% PI for all competitors for $p=10000$ in both the swiss roll and torus examples.
Results for predictive uncertainties for $p=2000$ are presented in Section 3.3 of the main article.

Across all simulations, SkGP's 95\% PI coverage is near the nominal level, with intervals widening as noise increases (higher $\tau^2$). BART and SkBART show poor coverage with narrower PIs. RF and SkRF also exhibit undercoverage ($\sim$ 80\%) with wider PIs than SkGP. In the swiss roll example, GP's coverage is similar to SkGP but with intervals about twice as wide. In the torus example, for $p=10000$, SkGP produces narrower PIs than GP while maintaining similar coverage as noise increases. Overall, SkGP provides more precise and robust predictive uncertainty compared to its competitors.

\end{document}